\documentclass[twocolumn, onecolappendix]{aastex63}

\usepackage{amsmath}
\usepackage{mathrsfs}
\usepackage{afterpage}
\usepackage{scalerel}
\usepackage{natbib}
\bibpunct[; ]{(}{)}{;}{a}{}{,}
\usepackage{multirow}
\usepackage{graphics}
\usepackage{threeparttable}
\usepackage[varg]{txfonts}
\usepackage{xcolor}
\usepackage{bm}
\hypersetup{colorlinks,linkcolor={blue},citecolor={blue},urlcolor={blue}}
\newcommand\iona[2]{#1$\;${\scshape{#2}}}

\begin{document}
\title{Identification and Characterization of a Large Sample of Distant Active Dwarf Galaxies in XMM-SERVS}

\author[0000-0002-4436-6923]{Fan Zou}
\affiliation{Department of Astronomy and Astrophysics, 525 Davey Lab, The Pennsylvania State University, University Park, PA 16802, USA}
\affiliation{Institute for Gravitation and the Cosmos, The Pennsylvania State University, University Park, PA 16802, USA}

\author[0000-0002-0167-2453]{W. N. Brandt}
\affiliation{Department of Astronomy and Astrophysics, 525 Davey Lab, The Pennsylvania State University, University Park, PA 16802, USA}
\affiliation{Institute for Gravitation and the Cosmos, The Pennsylvania State University, University Park, PA 16802, USA}
\affiliation{Department of Physics, 104 Davey Laboratory, The Pennsylvania State University, University Park, PA 16802, USA}

\author[0000-0002-8577-2717]{Qingling Ni}
\affiliation{Max-Planck-Institut f\"{u}r extraterrestrische Physik (MPE), Gie{\ss}enbachstra{\ss}e 1, D-85748 Garching bei M\"unchen, Germany}

\author[0000-0002-1653-4969]{Shifu Zhu}
\affiliation{Department of Astronomy and Astrophysics, 525 Davey Lab, The Pennsylvania State University, University Park, PA 16802, USA}
\affiliation{Institute for Gravitation and the Cosmos, The Pennsylvania State University, University Park, PA 16802, USA}

\author[0000-0002-5896-6313]{David M. Alexander}
\affiliation{Centre for Extragalactic Astronomy, Department of Physics, Durham University, South Road, Durham, DH1 3LE, UK}

\author[0000-0002-8686-8737]{Franz E. Bauer}
\affiliation{Instituto de Astrof{\'{\i}}sica, Facultad de F{\'{i}}sica, Pontificia Universidad Cat{\'{o}}lica de Chile, Campus San Joaquín, Av. Vicuña Mackenna 4860, Macul Santiago, Chile, 7820436}
\affiliation{Centro de Astroingenier{\'{\i}}a, Facultad de F{\'{i}}sica, Pontificia Universidad Cat{\'{o}}lica de Chile, Campus San Joaquín, Av. Vicuña Mackenna 4860, Macul Santiago, Chile, 7820436}
\affiliation{Millennium Institute of Astrophysics, Nuncio Monse{\~{n}}or S{\'{o}}tero Sanz 100, Of 104, Providencia, Santiago, Chile}
\affiliation{Space Science Institute, 4750 Walnut Street, Suite 205, Boulder, CO 80301, USA}

\author[0000-0002-4945-5079]{Chien-Ting J. Chen}
\affiliation{Science and Technology Institute, Universities Space Research Association, Huntsville, AL 35805, USA}
\affiliation{Astrophysics Office, NASA Marshall Space Flight Center, ST12, Huntsville, AL 35812, USA}

\author[0000-0002-9036-0063]{Bin Luo}
\affiliation{School of Astronomy and Space Science, Nanjing University, Nanjing, Jiangsu 210093, China}
\affiliation{Key Laboratory of Modern Astronomy and Astrophysics (Nanjing University), Ministry of Education, Nanjing 210093, China}

\author[0000-0002-0771-2153]{Mouyuan Sun}
\affiliation{Department of Astronomy, Xiamen University, Xiamen, 361005, Fujian, China}

\author[0000-0002-8853-9611]{Cristian Vignali}
\affiliation{Dipartimento di Fisica e Astronomia ``Augusto Righi'', Alma Mater Studiorum, Universit\`a degli Studi di Bologna, Via Gobetti 93/2, 40129 Bologna, Italy}
\affiliation{INAF -- Osservatorio di Astrofisica e Scienza dello Spazio di Bologna, Via Gobetti 93/3, I-40129 Bologna, Italy}

\author[0000-0003-0680-9305]{Fabio Vito}
\affiliation{INAF -- Osservatorio di Astrofisica e Scienza dello Spazio di Bologna, Via Gobetti 93/3, I-40129 Bologna, Italy}

\author[0000-0002-1935-8104]{Yongquan Xue}
\affiliation{CAS Key Laboratory for Research in Galaxies and Cosmology, Department of Astronomy, University of Science and Technology of China, Hefei 230026, China}
\affiliation{School of Astronomy and Space Sciences, University of Science and Technology of China, Hefei 230026, China}

\author[0000-0001-9519-1812]{Wei Yan}
\affiliation{Department of Astronomy and Astrophysics, 525 Davey Lab, The Pennsylvania State University, University Park, PA 16802, USA}
\affiliation{Institute for Gravitation and the Cosmos, The Pennsylvania State University, University Park, PA 16802, USA}

\email{E-mail: fuz64@psu.edu}

\defcitealias{Zou22}{Z22}

\begin{abstract}
Active dwarf galaxies are important because they contribute to the evolution of dwarf galaxies and can reveal their hosted massive black holes. However, the sample size of such sources beyond the local universe is still highly limited. In this work, we search for active dwarf galaxies in the recently completed XMM-Spitzer Extragalactic Representative Volume Survey (XMM-SERVS). XMM-SERVS is currently the largest medium-depth \mbox{X-ray} survey covering $13~\mathrm{deg^2}$ in three extragalactic fields, which all have well-characterized multi-wavelength information. After considering several factors that may lead to misidentifications, we identify 73 active dwarf galaxies at $z<1$, which constitutes the currently largest X-ray-selected sample beyond the local universe. Our sources are generally less obscured than predictions based on the massive-AGN (active galactic nucleus) \mbox{X-ray} luminosity function and have a low radio-excess fraction. We find that our sources reside in similar environments to inactive dwarf galaxies. We further quantify the accretion distribution of the dwarf-galaxy population after considering various selection effects and find that it decreases with \mbox{X-ray} luminosity, but redshift evolution cannot be statistically confirmed. Depending upon how we define an AGN, the active fraction may or may not show a strong dependence on stellar mass. Their Eddington ratios and \mbox{X-ray} bolometric corrections significantly deviate from the expected relation, which is likely caused by several large underlying systematic biases when estimating the relevant parameters for dwarf galaxies. Throughout this work, we also highlight problems in reliably measuring photometric redshifts and overcoming strong selection effects for distant active dwarf galaxies.
\end{abstract}

\section{Introduction}
\label{sec: intro}
Supermassive black holes (SMBHs) are known to be prevalent in massive galaxies and appear to be fundamentally linked to galaxy evolution (e.g., \citealt{Kormendy13}). However, our knowledge about dwarf galaxies, which are usually defined as galaxies with stellar masses ($M_\star$) comparable to or smaller than that of the Large Magellanic Cloud ($3\times10^9~M_\odot$), and the massive black holes (MBHs; $\gtrsim10^2~M_\odot$, i.e., more massive than stellar-mass black holes) residing in them (if present) is scarcer, primarily because they have lower luminosities and were thus difficult to detect in previous wide surveys. MBHs in dwarf galaxies may include intermediate-mass black holes (IMBHs; $10^2-10^5~M_\odot$, i.e., between stellar-mass black holes and SMBHs; e.g., \citealt{Greene20}) and small SMBHs ($\approx10^5-10^6~M_\odot$). In principle, the term ``MBHs'' can include massive SMBHs, but the latter are explicitly named SMBHs instead of MBHs in most cases. We will thus use MBHs to refer to IMBHs and small SMBHs ($\approx10^2-10^6~M_\odot$) in the following text, as done in previous literature (e.g., \citealt{Reines22}). Note that this is simply a choice of terminology, and the exact black-hole mass ($M_\mathrm{BH}$) is not the focus of this work.\par
Although challenging, understanding MBHs in dwarf galaxies is vital in several respects. First, MBHs may play important roles in dwarf-galaxy evolution (e.g., \citealt{Penny18, Barai19, Koudmani19, Manzano-King19}). Our current understanding of galaxy evolution is mostly based on massive galaxies, but dwarf galaxies are numerically more abundant and are essential for a holistic understanding of galaxy assembly (e.g., \citealt{Calabro17}) and for their contribution to the metal enrichment of the intergalactic medium and the missing baryon problem (e.g., \citealt{Tumlinson17}). Especially, dwarf galaxies and their MBHs are becoming increasingly of interest as upcoming wide surveys are becoming sensitive enough to probe the low-mass regime. Second, MBHs may provide significant insights into the seeds of SMBHs. High-redshift seed MBHs have been involved in the explanation of SMBHs (e.g., \citealt{Inayoshi20} and references therein). It is currently impossible to directly probe the seeds at high redshifts, but some of them presumably did not grow much and eventually evolved into MBHs at lower redshifts (e.g., \citealt{Mezcua17}). Hence, MBHs in dwarf galaxies are important for understanding SMBH seeding scenarios (e.g., \citealt{Burke22a}). Third, MBHs are possible engines of tidal disruption events and also the primary targets of the next-generation gravitational wave detector, the Laser Interferometer Space Antenna.\par
Searches for MBHs in dwarf galaxies have flourished over the past decade. Other than the dynamical searches that are limited to nearby galaxies, evidence of the existence of MBHs is mainly from their active galactic nucleus (AGN) signals (e.g., \citealt{Greene20, Reines22}). Similar to some traditional searching techniques for AGNs in massive galaxies, astronomers have mainly used optical spectra, \mbox{X-ray} observations, and radio observations to search for active dwarf galaxies (i.e., dwarf galaxies with AGN activity). Mid-infrared AGN selection, however, seems to face severe challenges for dwarf galaxies because of source confusion and strong star-formation contamination \citep{Mezcua18, Lupi20}. Variability-based AGN selection of active dwarf galaxies can help identify sources missed by the optical spectroscopic method and is expected to be increasingly important in the upcoming decade as the Vera C. Rubin Observatory Legacy Survey of Space and Time (LSST) becomes available, though the variability method has not been well-explored beyond the local universe and often faces selection biases that are hard to characterize (e.g., \citealt{Baldassare18, Baldassare20b, Burke22b, Ward22}). Future surveys that are both deep and wide, such as LSST, will enable detailed studies of distant dwarf galaxies in the future (e.g., \citealt{LSST09}).\par
Systematic searches for active dwarf galaxies beyond the local universe had not begun until roughly a decade ago and are mainly driven by deep \mbox{X-ray} \citep{Schramm13, Mezcua16, Pardo16, Aird18, Mezcua18} and sometimes radio \citep{Mezcua19, Davis22} observations. Population analyses of distant active dwarf galaxies beyond the local universe are still mainly limited by the number of known sources. The current largest sample is probably from the Cosmic Evolution Survey (COSMOS) field in \citet{Mezcua18}, which is based on the medium-depth Chandra COSMOS Legacy Survey \citep{Civano16} and contains 40 sources, 12 of which are above $z=0.5$. \citet{Aird18} also analyzed $\approx40-50$ Chandra-detected active dwarf galaxies compiled from several fields. However, as our analyses will reveal, these sources should only be considered as candidates because several measurement problems were not recognized, and the number of real active dwarf galaxies in these samples may be even smaller.\par
In this work, we search for active dwarf galaxies in a recently finished XMM-Newton survey, the XMM-Spitzer Extragalactic Representative Volume Survey (XMM-SERVS; \citealt{Chen18, Ni21}). It includes a total of 5.4~Ms of flare-filtered XMM-Newton observations in three fields -- Wide Chandra Deep Field-South (W-CDF-S; $4.6~\mathrm{deg^2}$), European Large-Area Infrared Space Observatory Survey-S1 (ELAIS-S1; $3.2~\mathrm{deg^2}$), and XMM-Newton Large Scale Structure (XMM-LSS; $5.3~\mathrm{deg^2}$). The survey provides a roughly uniform 50~ks exposure across the fields, reaching a flux limit of $\approx10^{-15}-10^{-14}~\mathrm{erg~cm^{-2}~s^{-1}}$ in the $0.5-10~\mathrm{keV}$ band, with more than ten thousand AGNs detected. This is currently the largest medium-depth \mbox{X-ray} survey and covers an area about six times larger than the Chandra COSMOS Legacy Survey, though XMM-SERVS is slightly shallower, with flux limits differing by $\approx0.2-0.5$~dex. Therefore, XMM-SERVS is expected to provide a larger sample of distant active dwarf galaxies than that in \citet{Mezcua18}. Similar to COSMOS, the three XMM-SERVS fields have extensive multi-wavelength observations, which enable good source characterization (e.g., \citealt{Zou22}; hereafter \citetalias{Zou22}). Furthermore, COSMOS and the three XMM-SERVS fields have been chosen as LSST Deep-Drilling Fields (DDFs) and also will be targeted by many other facilities, as summarized in \citetalias{Zou22}. The upcoming LSST DDF time-domain observations are expected to find many variability-selected active dwarf galaxies (e.g., \citealt{Baldassare18, Baldassare20b, Burke22a, Burke22b}), and thus our \mbox{X-ray} selections in the same fields will provide further insights in the LSST era.\par
This paper is structured as follows. Section~\ref{sec: data_sample} describes the data and our selection of active dwarf galaxies. Section~\ref{sec: analyses} presents the population analyses of our sample and relevant discussions. Section~\ref{sec: summary} summarizes this work. We adopt a flat $\Lambda\mathrm{CDM}$ cosmology with $H_0=70~\mathrm{km~s^{-1}~Mpc^{-1}}$, $\Omega_\Lambda=0.7$, and $\Omega_M=0.3$.

\section{Data and Sample}
\label{sec: data_sample}
Our active dwarf galaxies are selected from XMM-SERVS. As per Section~\ref{sec: intro}, this survey has superb multi-wavelength photometric data from the \mbox{X-ray} to radio. Furthermore, its optical-to-near-infrared photometry has been refined using a forced-photometry technique \citep{Zou21a, Nyland23}, which minimizes source confusion in low-resolution images, ensures consistency among different bands, and improves photometric redshifts (photo-$z$s). The redshifts have been compiled from several spectroscopic campaigns and, when spectroscopic redshifts (spec-$z$s) are unavailable, photo-$z$s presented in \citet[for XMM-LSS]{Chen18} and \citet[for W-CDF-S and ELAIS-S1]{Zou21b}\footnote{\citet{Ni21} derived photo-$z$s for broad-line AGNs to complement \citet{Zou21b}, but such sources will be excluded from our sample in Section~\ref{sec: zphot_reliability}.} are used. The number of bands used in the photo-$z$ estimations is around $10-15$ down to an $i$-band magnitude of $\approx24$. \citetalias{Zou22} further measured host-galaxy properties (e.g., $M_\star$ and SFR) in the regions with good multi-wavelength coverage in the XMM-SERVS fields by fitting source spectral energy distributions (SEDs) covering the \mbox{X-ray} to far-infrared (FIR) using \texttt{CIGALE} \citep{Yang22}, where the AGN emission has been appropriately considered. We refer readers to \citetalias{Zou22} for more details on the SED fitting and the related validation and analyses. We limit our analyses to the footprints cataloged by \citetalias{Zou22}, i.e., covered by the VISTA Deep Extragalactic Observations survey (VIDEO; \citealt{Jarvis13}), because quality multi-wavelength data are essential for detecting and characterizing dwarf galaxies. The areas are slightly smaller (mainly for XMM-LSS) than the whole XMM-SERVS area, and they cover 4.6, 3.2, and 4.7~$\mathrm{deg}^2$ in W-CDF-S, ELAIS-S1, and XMM-LSS, respectively.\par

\subsection{Selection of X-ray-Detected Dwarf Galaxies}
\label{sec: select_goodsample}
We select non-stellar \mbox{X-ray} sources with $M_\star<3\times10^9~M_\odot$ as active dwarf galaxy candidates, where $M_\star$ is from the AGN-template SED fitting in \citetalias{Zou22}. One may think of using a more conservative criterion with uncertainties included, such as $M_\star+2\mathrm{Err}(M_\star)<3\times10^9~M_\odot$, where $\mathrm{Err}(M_\star)$ is the uncertainty of $M_\star$. When adopting the cataloged errors in \citetalias{Zou22} as $\mathrm{Err}(M_\star)$, most sources (86\%) in our final sample would satisfy this criterion, and the largest $M_\star+2\mathrm{Err}(M_\star)$ value would only be 0.3~dex above $3\times10^9~M_\odot$. Our $\mathrm{Err}(M_\star)$ is generally small, with a median value of 0.11~dex, and it is far from being the dominant uncertainty compared to other selection effects that will be discussed later. We thus still adopt the standard criterion of $M_\star<3\times10^9~M_\odot$.\par
We also require the best-fit reduced chi-square ($\chi^2_r$) of the SED fitting to be smaller than five to remove poor fits, as adopted in \citetalias{Zou22}. The $\chi^2_r$ distribution peaks at $\approx1$ with a light high-$\chi^2_r$ tail. Only 6\% of sources are removed, and there are 353 sources left in total, including 105 sources in W-CDF-S, 78 sources in ELAIS-S1, and 170 sources in XMM-LSS. However, we found that these requirements are far from sufficient to ensure reliability, and we will further apply stricter cuts in the following subsections. We summarize our sample sizes after adding several criteria in Table~\ref{tbl_samplesize}.\par

\begin{table*}
\caption{Sample sizes}
\label{tbl_samplesize}
\centering
\begin{threeparttable}
\begin{tabular}{ccccc}
\hline
\hline
& Initial & Photo-$z$ & $M_\star^\mathrm{gal}$ & \mbox{X-ray} excess\\
\hline
W-CDF-S & 105 (78 + 27) & 34 (7 + 27) & 26 (7 + 19) & 22 (4 + 18)\\
ELAIS-S1 & 78 (59 + 19) & 26 (7 + 19) & 20 (7 + 13) & 13 (3 + 10)\\
XMM-LSS & 170 (123 + 47) & 62 (15 + 47) & 49 (15 + 34) & 38 (12 + 26)\\
Total & 353 (260 + 93) & 122 (29 + 93) & 95 (29 + 66) & 73 (19 + 54)\\
\hline
\hline
\end{tabular}
\begin{tablenotes}
\item
\textit{Notes.} This table summarizes the sample sizes as the selection criteria are progressively applied. The parentheses list the numbers of photo-$z$ sources + spec-$z$ sources. The second column, ``Initial'', refers to the criterion in the first paragraph of Section~\ref{sec: select_goodsample}. The third column, ``Photo-$z$'', records the sample sizes after applying the photo-$z$ quality cut in Section~\ref{sec: zphot_reliability}. The fourth column, ``$M_\star^\mathrm{gal}$'', shows the sample sizes with the $M_\star^\mathrm{gal}$ cut in Section~\ref{sec: mstar_reliability} added further. The fifth column, ``\mbox{X-ray} excess'', shows our final sample sizes after applying the criterion in Section~\ref{sec: select_finalsample}. We note that most sources fail both the photo-$z$ and $M_\star^\mathrm{gal}$ cuts, and the drastic decreases in the sample sizes from the second column to the third column would still exist when switching the sequence of the photo-$z$ and $M_\star^\mathrm{gal}$ cuts.
\end{tablenotes}
\end{threeparttable}
\end{table*}

\subsubsection{Photo-$z$ Reliability}
\label{sec: zphot_reliability}
Although the photo-$z$s in \citet{Chen18} and \citet{Zou21b} have been proven to be generally accurate, they are expected to be much less reliable for active dwarf galaxies. The problems are two-fold, caused by both the dwarf nature and the active nature of our sources. First, the Balmer break in galaxy spectra is an important feature for measuring photo-$z$s, but it becomes weak as the stellar age and metallicity decrease (see, e.g., Figure~4 in \citealt{Paulino-Afonso20}). Dwarf galaxies generally have young light-weighted stellar ages (i.e., with recent star formation) and low metallicities (e.g., \citealt{Gallazzi05}); thus, the corresponding galaxy SEDs may be close to power-laws, causing strong challenges to their photo-$z$ measurements. Second, AGN contributions were not considered when deriving photo-$z$s in \citet{Chen18} and \citet{Zou21b}, but our sources may have considerable AGN contributions.\par
\begin{figure*}
\centering
\resizebox{\hsize}{!}{\includegraphics{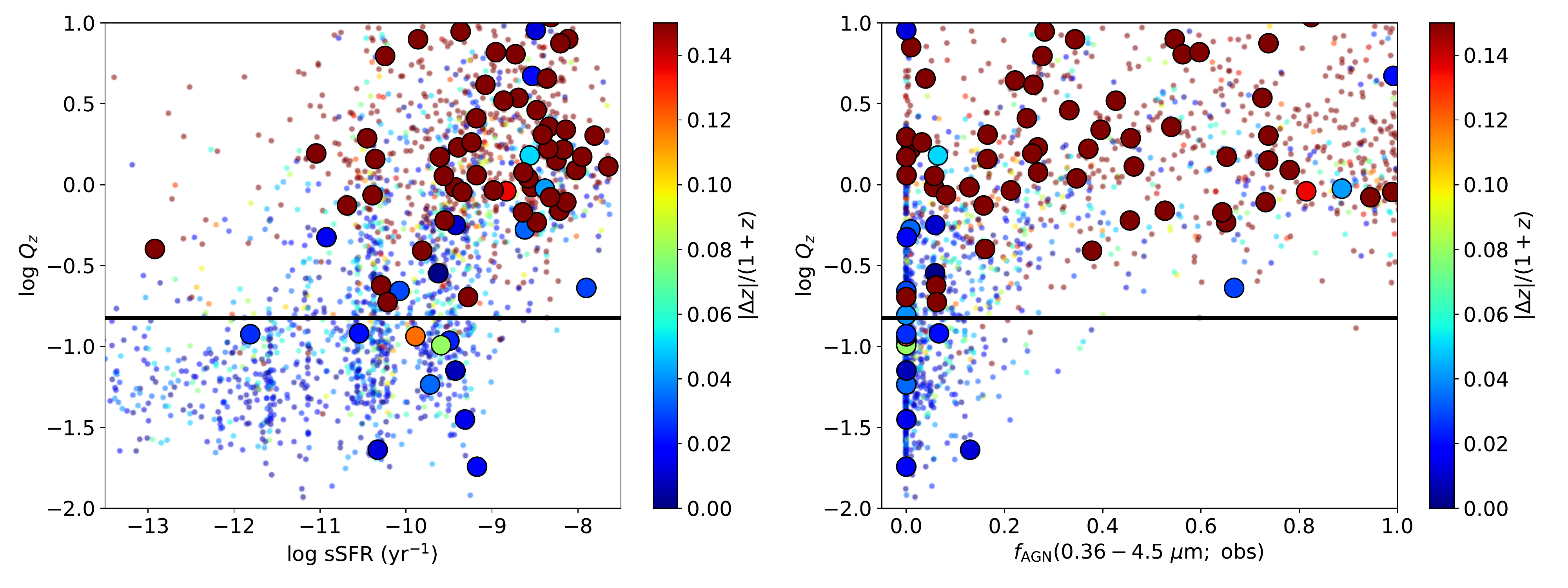}}
\caption{$\log Q_z$ versus $\log\mathrm{sSFR}$ (left) and $f_\mathrm{AGN}(0.36-4.5~\mu\mathrm{m}; \mathrm{obs})$ (right) for all the \mbox{X-ray} sources in XMM-SERVS with spec-$z$s, color-coded by $|\Delta z|/(1+z)$. Galaxies with cataloged $M_\star<3\times10^9~M_\odot$ are marked with large points. The horizontal black lines are $Q_z=0.15$, our threshold for reliable photo-$z$s. Many of our photo-$z$s are unreliable because of high sSFR and $f_\mathrm{AGN}(0.36-4.5~\mu\mathrm{m}; \mathrm{obs})$.}
\label{fig_Qzplot}
\end{figure*}

To illustrate this, we use the specific SFR ($\mathrm{sSFR=SFR/}M_\star$) to represent the Balmer break and the best-fit fractional AGN contribution in the observed-frame $0.36-4.5~\mu\mathrm{m}$ band, $f_\mathrm{AGN}(0.36-4.5~\mu\mathrm{m}; \mathrm{obs})$, to quantify the impact of the AGN emission on the photo-$z$ estimations. We show sSFR and $f_\mathrm{AGN}(0.36-4.5~\mu\mathrm{m}; \mathrm{obs})$ versus $Q_z$, the photo-$z$ quality indicator in \citet{Chen18} and \citet{Zou21b}, in Figure~\ref{fig_Qzplot} for all the \mbox{X-ray} sources in XMM-SERVS with spec-$z$s, color-coded by $|\Delta z|/(1+z)=|z_\mathrm{phot}-z_\mathrm{spec}|/(1+z_\mathrm{spec})$. $Q_z$ is an empirical parameter defined in \citet{Brammer08} and combines several kinds of information -- best-fit chi-square, confidence interval width, and the fraction of the total photo-$z$ probability within a given redshift range. Small $Q_z$ indicates high reliability. Sources with $|\Delta z|/(1+z)\geq0.15$ are catastrophic outliers and correspond to the darkest brown points in the figure. The figure indicates that $Q_z$ is positively correlated with sSFR and $f_\mathrm{AGN}(0.36-4.5~\mu\mathrm{m}; \mathrm{obs})$, and $|\Delta z|/(1+z)$ generally strongly increases with $Q_z$. Note that we found that $f_\mathrm{AGN}(0.36-4.5~\mu\mathrm{m}; \mathrm{obs})$ is much more strongly correlated with $Q_z$ than for the more fundamental fractional AGN contribution in the IR, $f_\mathrm{AGN}(\mathrm{IR})$, cataloged in \citetalias{Zou22} because photo-$z$s were derived based on SEDs only limited to the observed-frame $0.36-4.5~\mu\mathrm{m}$. Most of our sources have large sSFR and/or $f_\mathrm{AGN}(0.36-4.5~\mu\mathrm{m}; \mathrm{obs})$ and are also catastrophic photo-$z$ outliers. We found that sources (both the general XMM-SERVS sources and dwarfs) with $Q_z<0.15$ have $|\Delta z|/(1+z)<0.15$ in nearly all the cases, and thus we empirically adopt $Q_z=0.15$ as the threshold for reliable photo-$z$ measurements for our sources. This threshold is stricter than the nominal high-quality photo-$z$ threshold, $Q_z=1$, in \citet{Chen18} and \citet{Zou21b}. We further require the best-fit photo-$z$ to be between its 68\% lower and upper limits; otherwise, the photo-$z$ probability distribution may have multiple peaks or be highly skewed.\par
For sources with significant characteristics that are similar to spectroscopic broad-line AGNs (e.g., with AGN-dominated SEDs), \citet{Ni21} have derived their photo-$z$s using a dedicated method different from that in \citet{Zou21b}, and nearly all of them have $Q_z>1$ in \citet{Zou21b}. We are unable to calibrate these photo-$z$s from \citet{Ni21} for our dwarf sample because only one of them has a spec-$z$ after the $M_\star^\mathrm{gal}$ cut (see Section~\ref{sec: mstar_reliability} for more details) and turns out to have a photo-$z$ inconsistent with its spec-$z$. Besides, the photo-$z$s and $M_\star$ of broad-line AGNs are generally expected to be less reliable. We hence further exclude these broad-line AGNs with photo-$z$s from \citet{Ni21}. Even if we include them without any photo-$z$ quality cut, our final sample size would only increase by 9\%.\par

\begin{figure}
\centering
\resizebox{\hsize}{!}{\includegraphics{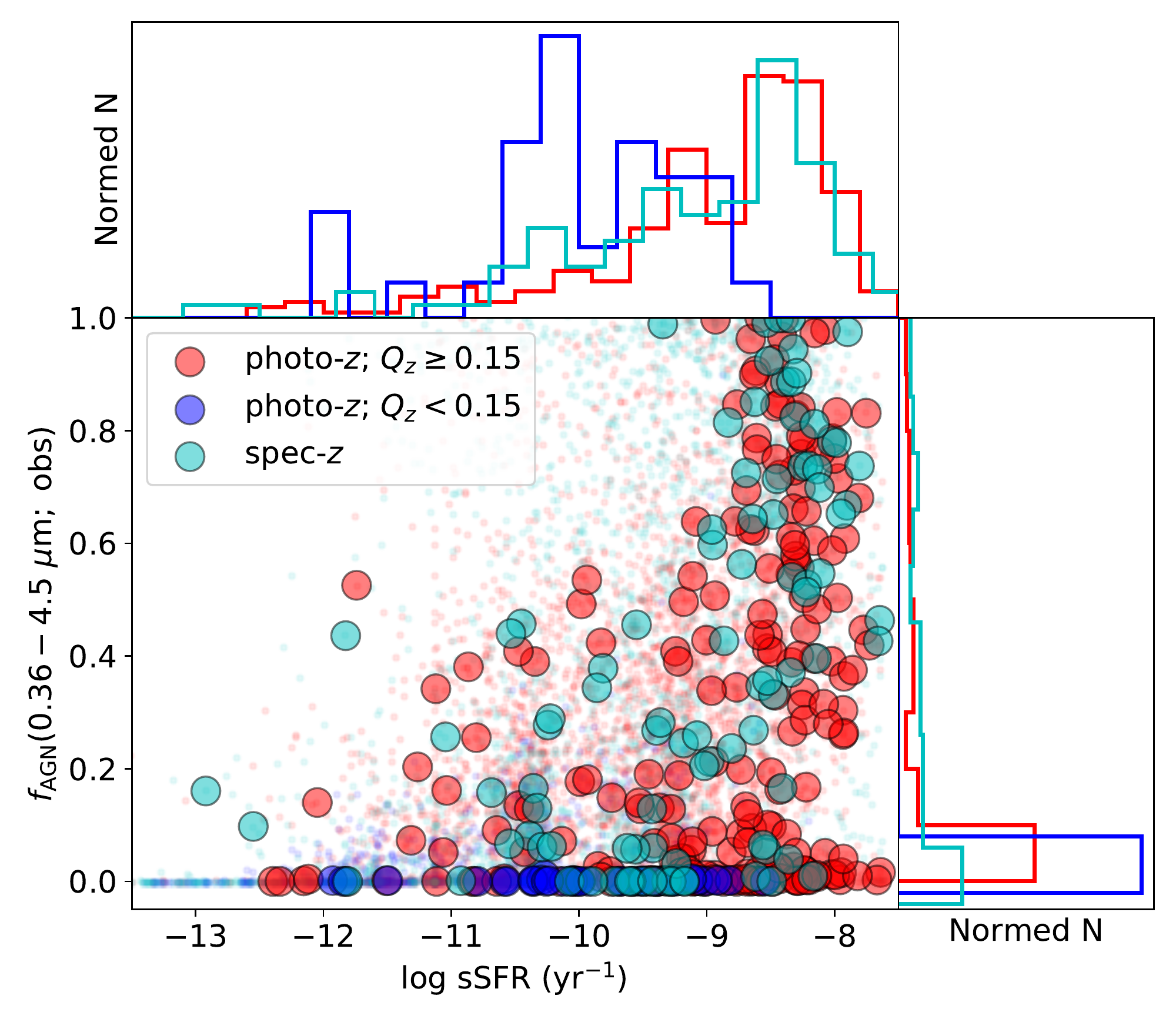}}
\caption{$\log\mathrm{sSFR}$ versus $f_\mathrm{AGN}(0.36-4.5~\mu\mathrm{m}; \mathrm{obs})$ for all the XMM-SERVS sources. Galaxies with cataloged $M_\star<3\times10^9~M_\odot$ are marked with large points. Distributions of $\log\mathrm{sSFR}$ and $f_\mathrm{AGN}(0.36-4.5~\mu\mathrm{m}; \mathrm{obs})$ are shown in the top and right panels, respectively. Sources whose redshifts are photo-$z$s with $Q_z\geq0.15$ (i.e., the large red points) are removed from our sample. The photo-$z$ quality cut only retains sources with small sSFR and near-zero $f_\mathrm{AGN}(0.36-4.5~\mu\mathrm{m}; \mathrm{obs})$.}
\label{fig_Qzdropout}
\end{figure}

The photo-$z$ quality cut is only applied to those candidates without spec-$z$s and removes 231 sources. 122 sources are left, including 34, 26, and 62 sources in W-CDF-S, ELAIS-S1, and XMM-LSS, respectively. We show sources in the $f_\mathrm{AGN}(0.36-4.5~\mu\mathrm{m}; \mathrm{obs})-\mathrm{sSFR}$ plane in Figure~\ref{fig_Qzdropout}. For sources without spec-$z$s, the quality cut only retains those with $\mathrm{sSFR\lesssim10^{-9}~yr^{-1}}$ and $f_\mathrm{AGN}(0.36-4.5~\mu\mathrm{m}; \mathrm{obs})\approx0$. The sources with spec-$z$s, instead, are more scattered in Figure~\ref{fig_Qzdropout}. This highlights the importance of obtaining deep spectroscopic observations in these fields; otherwise, the photo-$z$ quality cut will exert strong selection effects on the active dwarf galaxy sample. The cut also tends to remove high-redshift sources, which generally require higher $f_\mathrm{AGN}(0.36-4.5~\mu\mathrm{m}; \mathrm{obs})$ to be detectable in the \mbox{X-ray} and also have higher sSFR, as the star-forming galaxy main sequence (SFMS) increases with redshift. Although it is inevitable that sources with unreliable redshifts may have biased $f_\mathrm{AGN}(0.36-4.5~\mu\mathrm{m}; \mathrm{obs})$ and sSFR, which may undermine Figure~\ref{fig_Qzdropout}, our adopted cut is independent of these two parameters.\par

We emphasize that the difficulty of deriving reliable photo-$z$s discussed in this section is not unique to our fields. We have checked the active dwarf galaxy sample in \citet{Mezcua18} in COSMOS and compared their spec-$z$s and photo-$z$s cataloged in \citet{Marchesi16}. 9/21 are catastrophic photo-$z$ outliers, indicating that the same problem likely also exists in COSMOS. As far as we know, this problem has not been noted before, and thus extra caution should be taken when analyzing previous active dwarf galaxies with only photo-$z$s beyond the local universe.\par
It is unclear to us how to practically refine the COSMOS sample with photo-$z$s because the photo-$z$ methodologies in COSMOS are technically different. We also found small systematic offsets between some COSMOS SED-fitting results (e.g., \citealt{Laigle16}) and the results in \citetalias{Zou22} -- their SFMSs may differ by $\approx0.1-0.3$~dex, possibly because of their different SED-fitting methods. Such a systematic factor-of-two difference generally exists among different SED-fitting results. To ensure consistency, we do not include the COSMOS sample in our analyses. Besides, even if we do include the COSMOS sample, its sample size is too small to have a large impact on our results. As a rough estimation, the number of reliable active dwarf galaxies in \citet{Mezcua18} should be $\lesssim30$, while our final sample size (Section~\ref{sec: finalsample}) in XMM-SERVS is two to three times larger.

\subsubsection{$M_\star$ Reliability}
\label{sec: mstar_reliability}
\begin{figure*}
\centering
\resizebox{\hsize}{!}{
\includegraphics{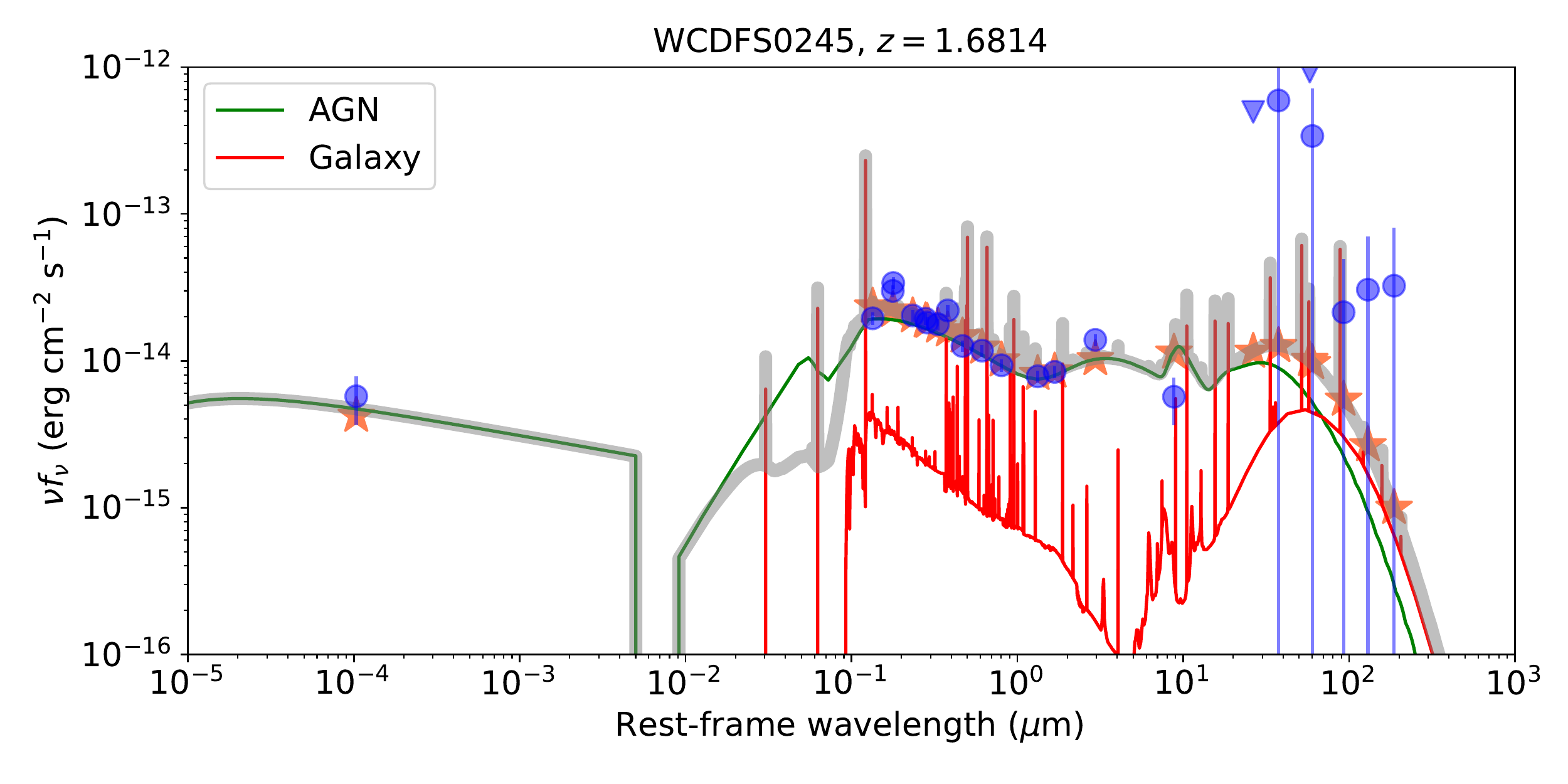}
\includegraphics{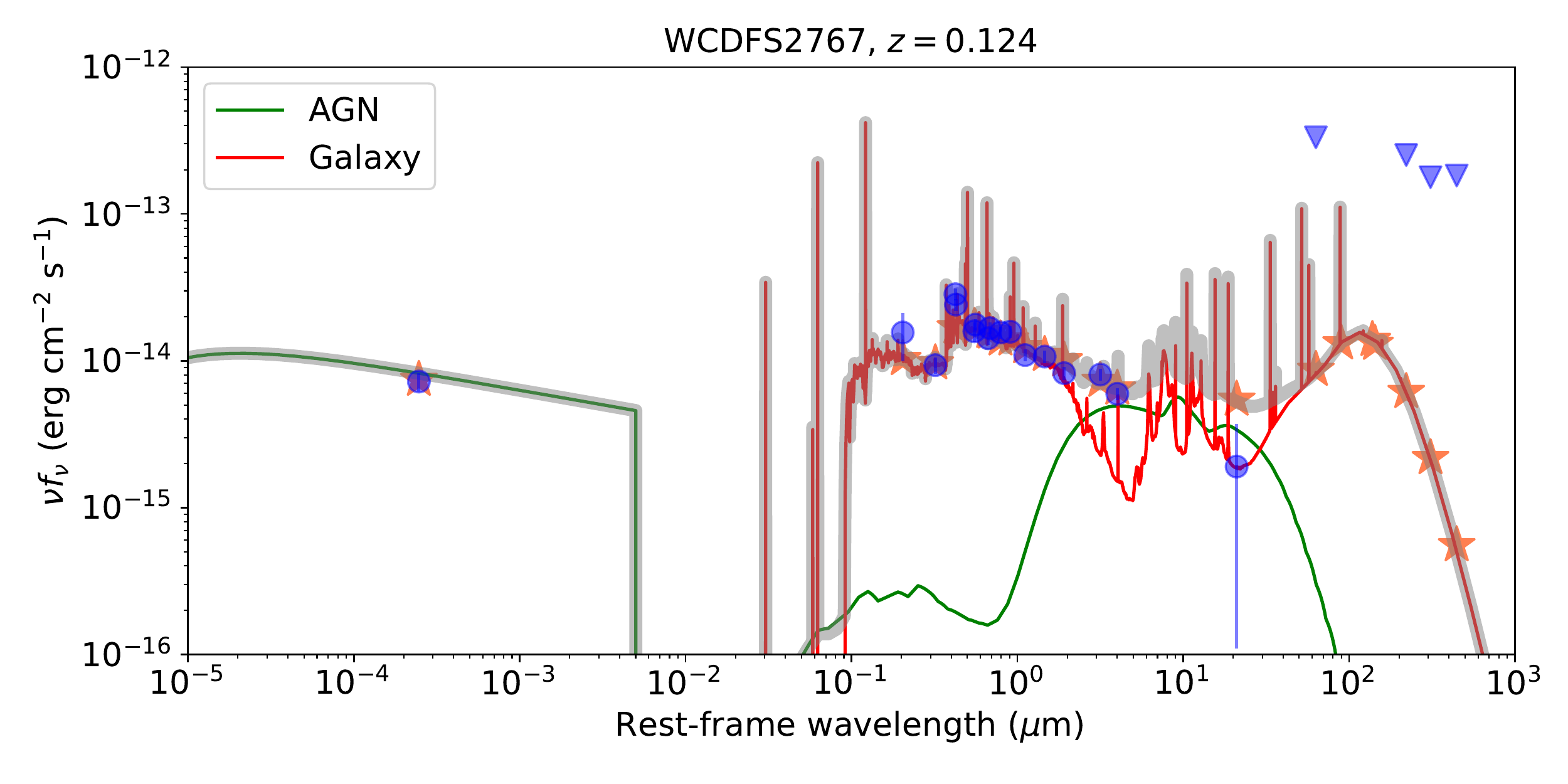}
}
\caption{Example rest-frame SEDs from \citetalias{Zou22} of high-redshift \mbox{X-ray}-detected dwarf galaxies with $M_\star<3\times10^9~M_\odot$ (left) and bona-fide active dwarf galaxies (right). The source XIDs and redshifts are listed as the panel titles. The blue points and downward triangles are the observed photometry and upper limits, respectively. The orange stars are the best-fit modeled photometry in the given bands, and the thick grey lines represent the best-fit models. The SEDs are decomposed into AGN components (green) and galaxy components (red). In contrast to the right panel, the AGN component in the left panel dominates the optical-to-NIR SED, and thus the host $M_\star$ cannot be measured reliably.}
\label{fig_example_sed}
\end{figure*}

We found that some best-fit SEDs are dominated by type~1 AGN emission, especially for sources with spec-$z$s~$\gtrsim1$. Figure~\ref{fig_example_sed} presents a high-redshift example and a bona-fide active dwarf galaxy. For the high-redshift source, its galaxy emission makes little contribution to its optical-to-NIR (near-infrared) SED. For such sources, the $M_\star$ measurements are usually unreliable, and the SED fitting may arbitrarily return small $M_\star$ values, which may range below $3\times10^9~M_\odot$, regardless of the real $M_\star$.\par
\begin{figure*}
\centering
\resizebox{\hsize}{!}{
\includegraphics{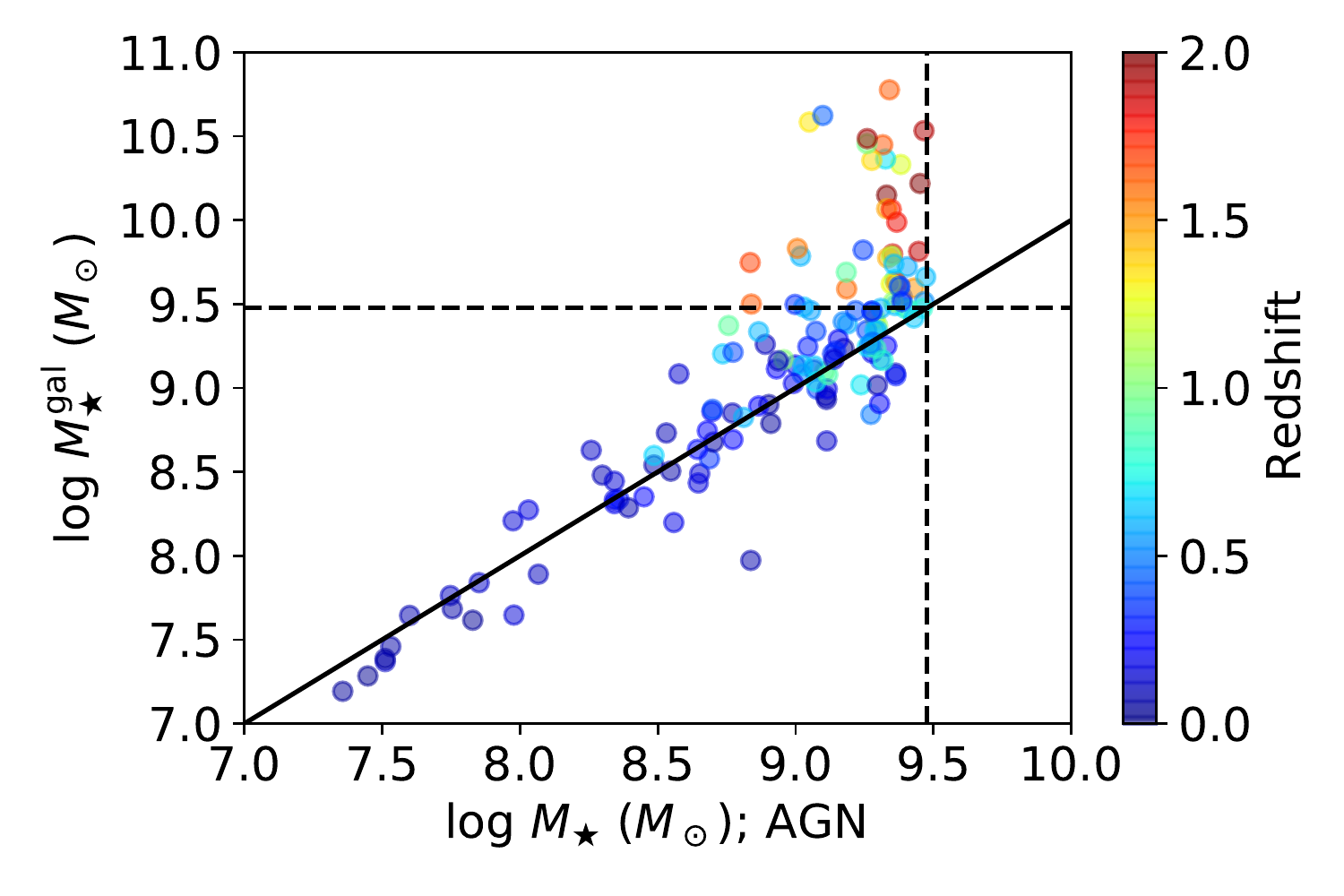}
\includegraphics{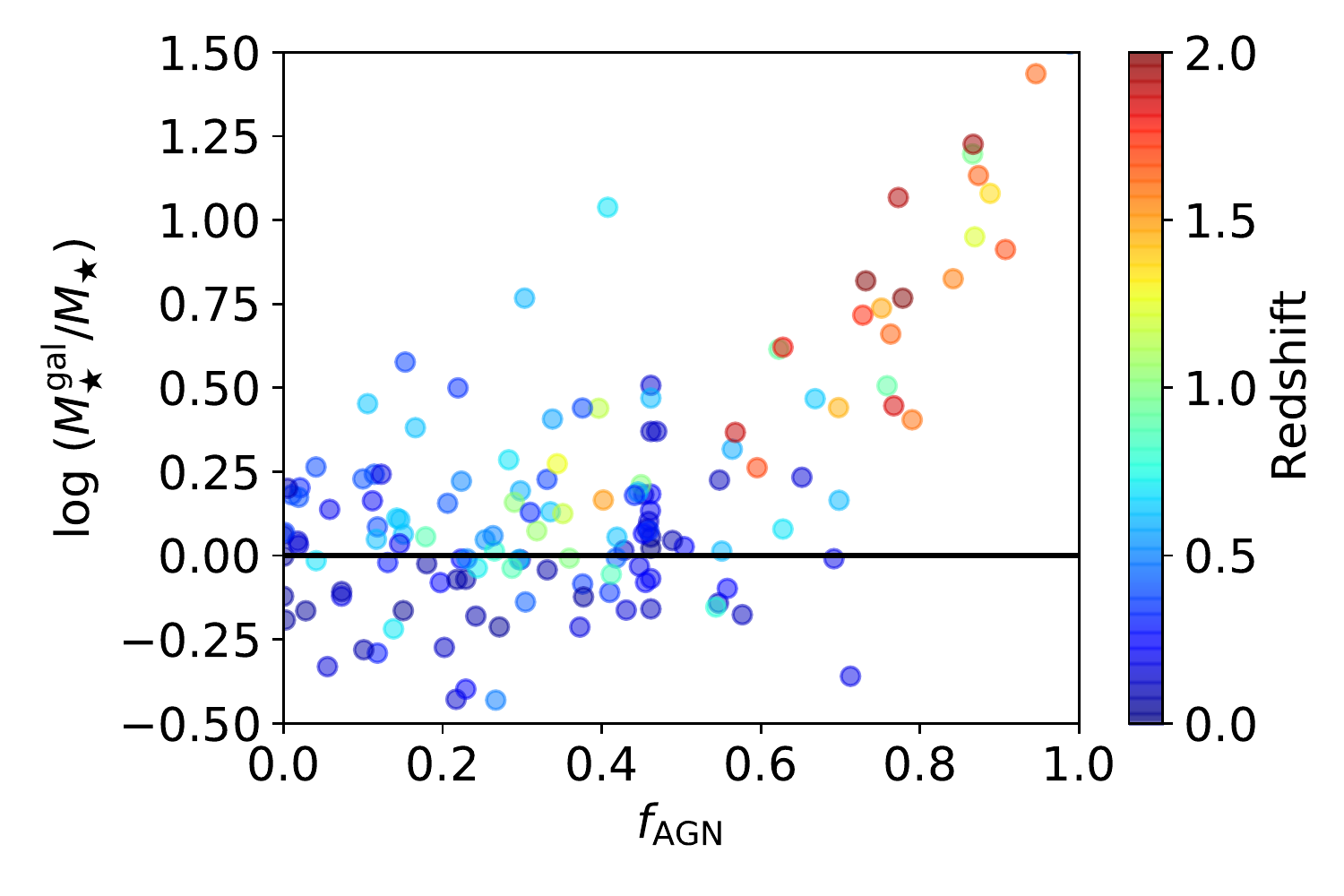}
}
\caption{Left: the comparison between $M_\star^\mathrm{gal}$ and the adopted AGN-template-based $M_\star$ values, color-coded by redshift. The black solid line represents a one-to-one relationship, and the black dashed lines are our adopted mass threshold ($3\times10^9~M_\odot$) for both $M_\star^\mathrm{gal}$ and $M_\star$. Sources above $M_\star^\mathrm{gal}=3\times10^9~M_\odot$ are not considered. Right: the difference between $M_\star^\mathrm{gal}$ and $M_\star$ versus $f_\mathrm{AGN}(\mathrm{IR})$, color-coded by the redshift. The black solid line represents zero difference. The difference between $M_\star^\mathrm{gal}$ and $M_\star$ generally increases with $f_\mathrm{AGN}(\mathrm{IR})$, and high-redshift sources tend to have larger $f_\mathrm{AGN}(\mathrm{IR})$ and often exceed the threshold of $M_\star^\mathrm{gal}$.}
\label{fig_mstarcut}
\end{figure*}

To quantify this effect, we first denote $M_\star^\mathrm{gal}$ as the fitted $M_\star$ based on normal-galaxy templates (i.e., without AGN components) in \citetalias{Zou22}. Since the total emission from both the AGN and galaxy components is assigned only to the galaxy when deriving $M_\star^\mathrm{gal}$, this parameter is expected to be close to the AGN-template-based $M_\star$ when the AGN contribution is small and gives a soft upper limit for the actual $M_\star$ when the AGN contribution is non-negligible. We compare $M_\star^\mathrm{gal}$ with the adopted AGN-template-based $M_\star$ for our candidates in the left panel of Figure~\ref{fig_mstarcut}, and high-$z$ sources tend to have much larger $M_\star^\mathrm{gal}$ than the dwarf-galaxy $M_\star$ threshold. The right panel of Figure~\ref{fig_mstarcut} shows the difference between the two $M_\star$ measurements versus $f_\mathrm{AGN}(\mathrm{IR})$ in \citetalias{Zou22}, where the difference increases with $f_\mathrm{AGN}(\mathrm{IR})$, as expected, and high-$z$ sources also generally have high $f_\mathrm{AGN}(\mathrm{IR})$ values. These indicate that it is challenging to confirm the dwarf nature of high-$z$ active dwarf-galaxy candidates because their strong AGN emission outshines their hosts. To avoid this SED issue, we remove 27 sources with $M_\star^\mathrm{gal}\ge3\times10^9~M_\odot$, and this requirement provides a conservative dwarf-galaxy criterion. All of our candidates with $z>1.1$ fail the $M_\star^\mathrm{gal}$ criterion, and 95 sources are left, including 26, 20, and 49 in W-CDF-S, ELAIS-S1, and XMM-LSS, respectively. Also note that all the photo-$z$ sources surviving the photo-$z$ quality cut in Section~\ref{sec: zphot_reliability} pass the $M_\star^\mathrm{gal}$ cut, as expected.\par

The above $M_\star^\mathrm{gal}$ criterion only addresses the problem that the stellar emission may be outshined and thus ``hidden'' by the AGN emission. However, old stars can also be ``hidden'' by young stars, and neglecting such old stars may cause underestimations of $M_\star$, especially for starburst galaxies (e.g., \citealt{Papovich01}). This is related to the adopted star-formation history (SFH). The normal-galaxy $M_\star^\mathrm{gal}$ in \citetalias{Zou22} is based on delayed SFHs, which can provide good characterizations for general galaxies, whose main stellar populations are either young or old, but cannot describe the starburst case with both very young and old stellar populations. Fortunately, our sources are generally not starburst galaxies (cf., Figure~\ref{fig_Qzdropout}), and only two sources in our final sample (see Section~\ref{sec: finalsample}) have $\mathrm{sSFR>10^{-8}~yr^{-1}}$. More conservatively, we further check the $M_\star^\mathrm{gal}$ based on ``bursting or quenching'' (BQ) SFHs in \citetalias{Zou22} (denoted as $M_\star^\mathrm{bqgal}$),\footnote{We adopt their BQ results based on their Table~3 instead of their cataloged results from their Table~6 because the latter are not run for most sources.} which allow the existence of an old stellar population besides a young population and thus generally return larger $M_\star$. Such $M_\star^\mathrm{bqgal}$ values should be even more conservative upper limits for the real $M_\star$, and we found that only 15\% of our final sample would then have $M_\star^\mathrm{bqgal}>3\times10^9~M_\odot$. Only one source (XID = XMM02399), which also turns out to have interesting properties, exceeds $10^{10}~M_\odot$, and we will discuss this source in detail in Appendix~\ref{sec: xmm02399} (see also Section~\ref{sec: hr}). The normal-galaxy $M_\star^\mathrm{gal}$ criterion could be adjusted to the one based on $M_\star^\mathrm{bqgal}$. However, it should be noted that we are not really improving anything by changing SFHs, but instead trying to identify underlying plausible systematic uncertainties. It is well-known that SED fitting has an inherent factor-of-two uncertainty that can hardly be narrowed down (e.g., \citealt{Conroy13, Leja19, Pacifici23}; \citetalias{Zou22}) because of various factors, including the choice of SFH. Besides, AGN studies rarely adopt BQ SFHs or similar complex ones because of much heavier computational requirements and strong degeneracies with the AGN emission (e.g., see Section~4.5 of \citetalias{Zou22}), and thus adopting normal-galaxy SFHs ensures a general consistency with the literature. We thus still adopt the original normal-galaxy $M_\star^\mathrm{gal}$ criterion, but the inevitable uncertainty in $M_\star$ discussed in this paragraph should be kept in mind.

\subsection{Selection of Active Dwarf Galaxies}
\label{sec: select_finalsample}
We then assess if the detected \mbox{X-rays} are sufficiently bright to indicate the presence of AGNs residing in these dwarf galaxies. We directly compare their counts instead of fluxes for better accuracy. Due to the non-negligible point-spread function (PSF) size of XMM-Newton, nearby galaxies close to the dwarf of interest may also contribute to the observed emission. Therefore, the observed counts are from both the surrounding sources and the dwarf galaxies themselves.\par
For a given dwarf of interest, we select sources in \citetalias{Zou22} within one arcmin, a sufficiently large radius, around this dwarf as its nearby sources. If a nearby source is cataloged in XMM-SERVS, we directly adopt its observed counts as the contribution. For the others, some of them may be AGNs as well, and we select AGNs as those being identified as mid-IR or reliable SED AGNs in Section~3.2 of \citetalias{Zou22} and adopt their expected \mbox{X-ray} emission as the predicted values through SED fitting in Section~3.2.2 of \citetalias{Zou22}. Note that these predictions are intrinsic \mbox{X-ray} emission before absorption by the intrinsic obscuration and thus may overestimate the fluxes. However, this is acceptable because we want to be conservative. These neighboring galaxies are assumed to be from the \textit{general} galaxy population (i.e., not subpopulations with special properties such as our active dwarf galaxies), and thus their redshifts and galaxy properties should be reliable, as justified in detail in \citet{Chen18}, \citet{Zou21b}, and \citetalias{Zou22}.\par
For non-AGN normal galaxies, their \mbox{X-ray} emission is mainly from \mbox{X-ray} binaries (XRBs) and hot gas, where the hot-gas emission mainly contributes in the soft \mbox{X-rays}. We estimate their \mbox{X-ray} fluxes following similar procedures as \citet{Basu-Zych20}. Note that we correct the $M_\star$ and SFR differences among the literature caused by different initial mass functions (IMFs) following \citet{Speagle14} and \citet{Madau14}:
\begin{align}
&M_\star^C=0.94M_\star^K=0.58M_\star^S,\\
&\mathrm{SFR}^C=0.94\mathrm{SFR}^K=0.63\mathrm{SFR}^S,
\end{align}
where the superscripts ``C'', ``K'', and ``S'' represent the Chabrier \citep{Chabrier03}, Kroupa \citep{Kroupa01}, and Salpeter \citep{Salpeter55} IMFs, respectively. The $M_\star$ and SFR differences from the Chabrier and Kroupa IMFs are generally negligible, but the Salpeter IMF can cause noticeable differences. We will always use the Chabrier IMF $M_\star$ and SFR, as adopted in \citetalias{Zou22}, in the following text.\par
We adopt the scaling relation in \citet{Lehmer16} for the XRB emission, where the total XRB emission is further separated into the contributions from low-mass XRBs (LMXBs) and high-mass XRBs (HMXBs). The LMXB and HMXB emission scales with $M_\star$ and SFR, respectively. Their relation gives
\begin{align}
L_\mathrm{2-10~keV}^\mathrm{XRB}=&L_\mathrm{2-10~keV}^\mathrm{LMXB}+L_\mathrm{2-10~keV}^\mathrm{HMXB}\nonumber\\
=&10^{29.30}(1+z)^{2.19}M_\star^K+10^{39.40}(1+z)^{1.02}\mathrm{SFR}^K,
\end{align}
where the luminosity, $M_\star$, and SFR are in $\mathrm{erg~s^{-1}}$, $M_\odot$, and $M_\odot~\mathrm{yr^{-1}}$, respectively. We adopt a power law with a photon index of $\Gamma=1.8$ as the XRB spectrum.\par
For the hot-gas emission, mass-dominated galaxies and galaxies dominated by star formation (SF) have different scaling relations, where the hot-gas emission from the former and the latter mainly scales with $M_\star$ and SFR, respectively. Following \citet{Basu-Zych20}, we regard a galaxy as mass-dominated or SF-dominated if its sSFR is smaller or larger than $10^{-10.1}~\mathrm{yr^{-1}}$, respectively. We adopt the scaling relation in \citet{Kim15} for mass-dominated galaxies:
\begin{align}
\log(L_\mathrm{0.3-8~keV}^\mathrm{gas}/10^{40}~\mathrm{erg~s^{-1}})=2.98\log(L_K/10^{11}L_{K\odot})-0.25,
\end{align}
where $L_K$ is the $K$-band luminosity. We further use $M_\star/M_\odot=0.66L_K/L_{K\odot}$ \citep{Lehmer14} to convert the above luminosity scaling relation to a $M_\star$ scaling relation. Following \citet{Kim15}, we adopt the corresponding hot-gas spectrum as an \texttt{apec} model with the gas temperature $kT$ set by the following equation.
\begin{align}
\log(L_\mathrm{0.3-8~keV}^\mathrm{gas}/10^{40}~\mathrm{erg~s^{-1}})=5.39\log(kT/0.5~\mathrm{keV})+0.16.
\end{align}
For SF-dominated galaxies, we adopt the scaling relation in \citet{Mineo12}:
\begin{align}
L_\mathrm{0.5-2~keV}^\mathrm{gas}=8.3\times10^{38}\mathrm{SFR}^S.
\end{align}
The corresponding spectrum is set to a \texttt{mekal} model with $kT=0.5~\mathrm{keV}$.\par
We follow the same procedures as above to estimate the galaxy emission from the dwarf targets themselves. The above scaling relations may underestimate the galaxy \mbox{X-ray} emission because dwarf galaxies generally have low metallicities and young ages (e.g., \citealt{Gallazzi05}), which both elevate the LMXB and HMXB emission (e.g., \citealt{Fragos13b, Prestwich13, Lehmer21}). However, this almost does not cause any problem because the resulting expected galaxy emission from the dwarf targets is one order of magnitude smaller than that from their nearby sources. For example, we have tried using the metallicity- and age-dependent scaling relations in \citet{Fragos13b} and \citet{Lehmer21}, where we estimate metallicities based on the fundamental metallicity relation in \citet{Curti20} and stellar ages from the best-fit SFHs in \citetalias{Zou22}, and the resulting final sample size in Section~\ref{sec: finalsample} only decreases by at most one.\par
Given the luminosity and the appropriate spectral models above, we can calculate the total expected \mbox{X-ray} flux in any desired \mbox{X-ray} band by applying corresponding K corrections using \texttt{Sherpa} \citep{Freeman01, Doe07}. We then convert the expected flux to the expected counts in each camera (EPIC MOS1, MOS2, and PN) by multiplying the ratios between the cataloged net counts in each camera and the flux of the target of interest in the XMM-SERVS catalogs. These expected counts need to be converted to those in a given source aperture of the target by multiplying by the enclosed energy fraction (EEF) within the target aperture, as discussed in the following text. Then, the expected counts within the target aperture are summed over all the cameras.\par
We adopt the target aperture as $3\times3$ pixels (i.e., $12\times12''$) around the dwarf of interest. The EEF is the integration of the PSF within the aperture, denoted as $\mathrm{EEF}(r, \theta)$, where $r$ is the separation between the nearby source and the target, and $\theta$ represents all the other parameters determining this system. We then apply appropriate weightings on $\theta$ and further eliminate $\theta$: $\mathrm{EEF}(r)=E_\theta\{\mathrm{EEF}(r, \theta)\}$ and $\sigma_\mathrm{EEF}^2(r)=\mathrm{Var}_\theta\{\mathrm{EEF}(r, \theta)\}$. $\mathrm{EEF}(r)$ is used to convert the total expected counts from nearby sources to those in the target aperture in each camera and in each band, and $\sigma_\mathrm{EEF}(r)$ is the uncertainty of $\mathrm{EEF}(r)$, mainly driven by the variation of the PSF shape.\par
We adopt the parametrization in \citet{Read11} for the PSF shape, in which the PSF is mainly described as an elliptical King profile plus an elliptical Gaussian core, and the corresponding parameters are stored in XMM-Newton Current Calibration Files. $\theta$ includes the relative angles of the PSF and the target aperture and parameters determining the PSF shape. For the angle parameters, their weights are flat. The PSF-shape parameters mainly include the photon energy and off-axis angle; we use the observed spectrum (i.e., after convolution with typical XMM-Newton response files) of a power-law with a photon index of 1.4 as the photon-energy weight, and the weight of the off-axis angle is the angle itself.\par
We then compare the observed counts within the target aperture ($C_\mathrm{obs}$) with the non-AGN prediction (i.e., the dwarf target is not an AGN). The relevant distributions are
\begin{align}
C_\mathrm{obs}\mid C_\mathrm{non\text{-}AGN}&\sim\mathrm{Poi}\left(B+C_\mathrm{non\text{-}AGN}+C_\mathrm{AGN}\right),\\
C_\mathrm{non\text{-}AGN}&\sim N_+(C_\mathrm{pred}, \sigma_\mathrm{pred}^2),
\end{align}
where $B$ is the expected background counts from the background maps, $C_\mathrm{non\text{-}AGN}$ is the expected counts from nearby sources and the galaxy emission of the target, $C_\mathrm{AGN}$ is the expected AGN counts from the target, $C_\mathrm{pred}$ is our predicted value for $C_\mathrm{non\text{-}AGN}$, $\sigma_\mathrm{pred}$ is the uncertainty of $C_\mathrm{pred}$, and $N_+$ denotes a normal distribution truncated at 0. $\sigma_\mathrm{pred}$ includes $\sigma_\mathrm{EEF}$ the uncertainty of the expected fluxes, and we estimate the latter by propagating the uncertainties of $M_\star$ and SFR and adding 0.3~dex in quadrature for each involved galaxy to account for the typical scatters of the scaling relationships used previously. Strictly speaking, $B$ partially overlaps with $C_\mathrm{non\text{-}AGN}$ because both include the average unresolved source emission, but this component is generally much smaller than $B$ and thus does not cause noticeable problems even when double counted.\par
We then test the null hypothesis that $C_\mathrm{AGN}=0$. From the above distributions, the $p$-value of the hypothesis test is
\begin{align}
p\text{-value}=\sqrt{\frac{2}{\pi}}\frac{\int_0^{+\infty}P_\mathrm{IG}(C_\mathrm{obs}, B+x)\exp\left(-\frac{(x-C_\mathrm{pred})^2}{2\sigma_\mathrm{pred}^2}\right)dx}{\sigma_\mathrm{pred}\left[\mathrm{erf}\left(\frac{C_\mathrm{pred}}{\sqrt{2}\sigma_\mathrm{pred}}\right)+1\right]},
\end{align}
where $P_\mathrm{IG}$ is the regularized lower incomplete gamma function. To mitigate the effects of obscuration, we choose the comparison band as follows. For sources detected in the hard band (HB), the comparison band is the HB; for sources undetected in the HB but detected in the full band (FB), the comparison band is the FB; for the remaining sources that are only detected in the soft band (SB), the comparison band is the SB. For W-CDF-S and ELAIS-S1, the SB, HB, and FB energy ranges are $0.2-2$, $2-12$, and $0.2-12$~keV, respectively \citep{Ni21}; while for XMM-LSS, the energy ranges are $0.5-2$, $2-10$, and $0.5-10$~keV \citep{Chen18}. We regard a source to be an active dwarf galaxy if its $p$-value is smaller than 0.01. This removes 21 sources and leaves 73 sources, including 22, 13, and 38 sources in W-CDF-S, ELAIS-S1, and XMM-LSS, respectively. We found that these hypothesis-test results are not sensitive to $\sigma_\mathrm{pred}$, and only three more sources are added even if we set $\sigma_\mathrm{pred}=0$. This also indicates that it generally does not matter even if $C_\mathrm{non\text{-}AGN}$ does not strictly follow a truncated normal distribution as assumed.\par
\begin{figure}
\centering
\resizebox{\hsize}{!}{\includegraphics{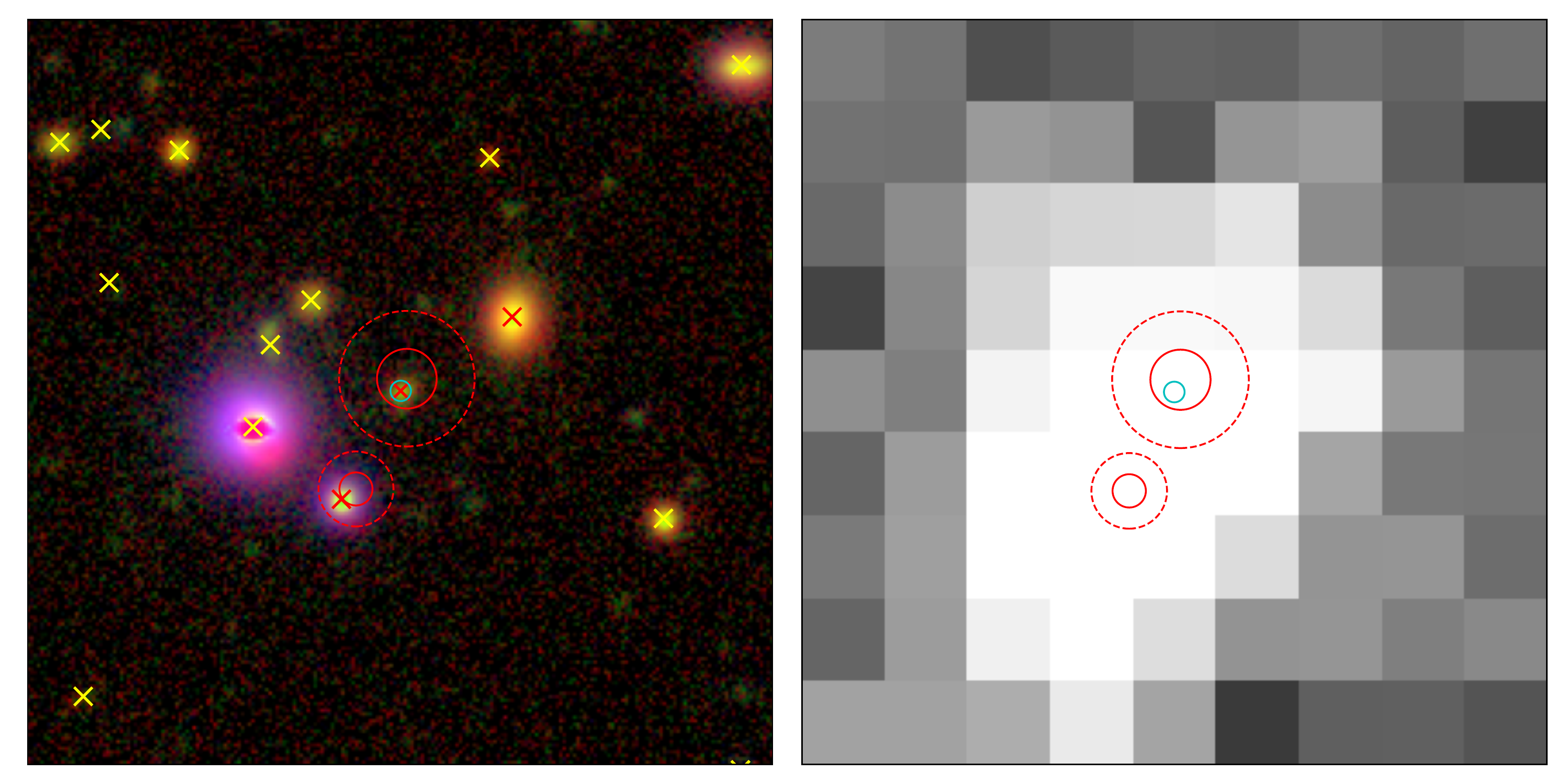}}
\resizebox{\hsize}{!}{\includegraphics{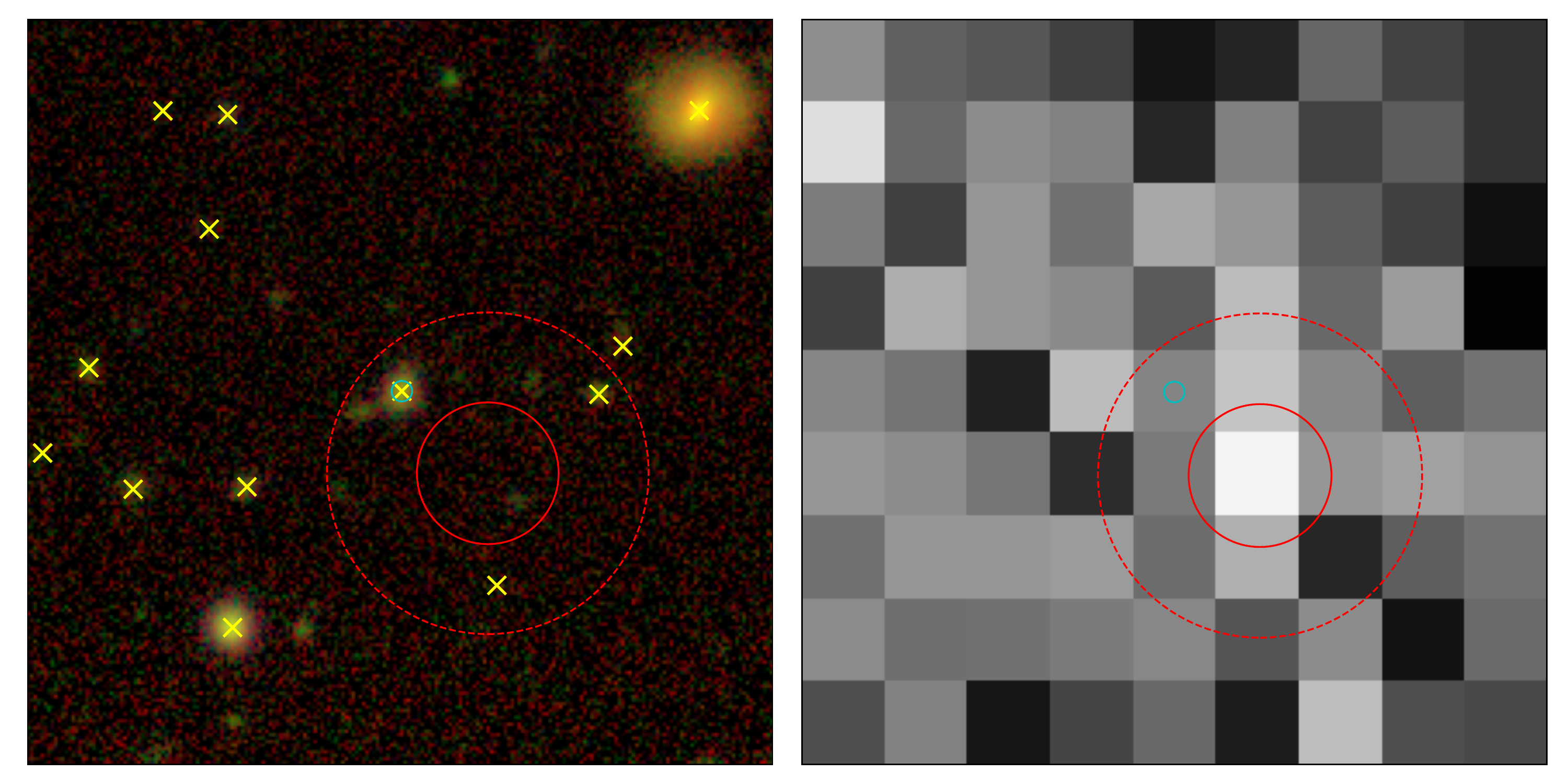}}
\caption{The optical (left) and XMM-Newton $0.2-12$~keV (right) images of example sources failing the \mbox{X-ray} excess criterion. The top panels are for XID = WCDFS4029, and the bottom ones are for XID = WCDFS3998. The three-color optical images are generated from $g$-band (blue), $i$-band (green), and $z$-band (red) images from \citet[top]{Ni19}. The red and yellow crosses mark AGNs and non-AGNs in \citetalias{Zou22}, respectively. The cyan circles are centered at the dwarf targets with radii of $0.5''$. The red solid and dashed circles represent 68\% and 99.73\% \mbox{X-ray} positional uncertainties, respectively, for the \mbox{X-ray} sources.}
\label{fig_exampimg}
\end{figure}

Figure~\ref{fig_exampimg} presents optical and XMM-Newton images for two example sources that are removed, where the optical and XMM-Newton images have been aligned using the \texttt{reproject} package \citep{Robitaille20}, and the optical image is from \citet{Ni19}. In the top panels of Figure~\ref{fig_exampimg} (XID = WCDFS4029), two additional AGNs are found to lie close to the dwarf galaxy, one of which is also included in the XMM-SERVS catalogs. This dwarf is clearly contaminated by the nearby AGNs and can hardly be cleaned reliably. In the bottom panels of Figure~\ref{fig_exampimg} (XID = WCDFS3998), the observed \mbox{X-ray} emission is not sufficiently strong. Therefore, the relatively larger PSF size of XMM-Newton compared to Chandra leads to source confusion and further complexity; due to the same reason, we lack the information of whether the \mbox{X-ray} emission is from the center or the outskirts of the host galaxy and thus can hardly exclude contamination from ultraluminous \mbox{X-ray} sources (ULXs). Some of these sources can reach high $L_\mathrm{X}$ of $10^{42}~\mathrm{erg~s^{-1}}$ (e.g., \citealt{Farrell09}), though such cases are rare. Nevertheless, \citet{Mezcua18} showed that, in COSMOS, the fraction of their active dwarf galaxies whose \mbox{X-ray} emission is actually from ULXs is generally limited ($\lesssim15\%$), and similar conclusions are also drawn in \citet{Birchall20} for nearby active dwarf galaxies selected through XMM-Newton. Thus, we expect that our sample also has limited ULX contamination.\par

\subsection{Resulting Final Sample}
\label{sec: finalsample}
The 73 sources remaining in Section~\ref{sec: select_finalsample} constitute our final active dwarf-galaxy sample. 54 of them (74\%) have spectroscopic redshifts. The evolution of our sample sizes with the selection criteria is summarized in Table~\ref{tbl_samplesize}. We summarize the basic properties of our final sample in Tables~\ref{tbl_release_wcdfs}, \ref{tbl_release_es1}, and \ref{tbl_release_xmmlss} for W-CDF-S, ELAIS-S1, and XMM-LSS, respectively.\par

\begin{table*}
\caption{Source catalog in W-CDF-S}
\label{tbl_release_wcdfs}
\centering
\resizebox{\hsize}{!}{
\begin{threeparttable}
\begin{tabular}{ccccccccccc}
\hline
\hline
XID & Tractor ID & R. A. & Decl. & $z$ & ztype & $\log L_\mathrm{X, obs}$ & $\log\mathrm{SFR}$ & $\log M_\star$ & $\log M_\star^\mathrm{gal}$ & $\log M_\star^\mathrm{bqgal}$\\
& & (degree) & (degree) & & & ($\mathrm{erg~s^{-1}}$) & ($M_\odot~\mathrm{yr^{-1}}$) & ($M_\odot$) & ($M_\odot$) & ($M_\odot$)\\
(1) & (2) & (3) & (4) & (5) & (6) & (7) & (8) & (9) & (10) & (11)\\
\hline
WCDFS0162 & 255877 & 51.905781 & $-28.958321$ & 0.116 & zphot & 41.29 & $-1.48$ & 8.89 & 9.26 & 8.91\\
WCDFS0217 & 364633 & 51.984135 & $-28.414293$ & 0.123 & zspec & 41.36 & $-1.47$ & 7.51 & 7.37 & 7.57\\
WCDFS0761 & 645301 & 51.833397 & $-27.244995$ & 0.044 & zspec & 40.45 & $-0.70$ & 9.18 & 9.24 & 9.18\\
WCDFS0986 & 453469 & 52.293705 & $-28.119442$ & 0.150 & zspec & 41.28 & $-2.39$ & 8.65 & 8.49 & 8.64\\
WCDFS1018 & 395902 & 52.600765 & $-28.227808$ & 0.705 & zspec & 42.65 & $0.76$ & 9.24 & 9.02 & 9.72\\
WCDFS1340 & 288812 & 52.637676 & $-28.778755$ & 0.127 & zspec & 41.39 & $-1.01$ & 7.85 & 7.84 & 8.13\\
WCDFS1346 & 265349 & 52.805256 & $-28.906603$ & 0.645 & zspec & 43.21 & $0.04$ & 8.74 & 9.20 & 9.30\\
WCDFS1355 & 352132 & 52.682285 & $-28.614000$ & 0.916 & zspec & 42.89 & $0.51$ & 8.76 & 9.37 & 9.63\\
WCDFS1394 & 350692 & 52.803371 & $-28.478706$ & 0.042 & zspec & 40.88 & $-1.56$ & 7.83 & 7.62 & 8.03\\
WCDFS1417 & 385839 & 52.793171 & $-28.373316$ & 0.344 & zspec & 42.89 & $0.07$ & 8.69 & 8.58 & 8.46\\
WCDFS1440 & 388690 & 52.752987 & $-28.272453$ & 0.079 & zspec & 40.72 & $-2.99$ & 8.84 & 7.97 & 8.33\\
WCDFS1459 & 402544 & 52.739395 & $-28.202730$ & 0.111 & zspec & 40.91 & $-2.62$ & 8.07 & 7.89 & 7.93\\
WCDFS1770 & 321314 & 53.202751 & $-28.624237$ & 0.140 & zspec & 41.50 & $-0.70$ & 9.11 & 8.68 & 9.06\\
WCDFS2044 & 504127 & 53.058399 & $-27.850225$ & 0.122 & zspec & 42.02 & $-3.56$ & 9.36 & 9.09 & 9.25\\
WCDFS2140 & 579203 & 53.082249 & $-27.483568$ & 0.094 & zphot & 41.02 & $-3.18$ & 8.30 & 8.48 & 8.30\\
WCDFS2759 & 76747 & 53.763737 & $-28.118191$ & 0.364 & zspec & 42.66 & $0.24$ & 9.08 & 8.99 & 9.24\\
WCDFS2767 & 50433 & 53.756840 & $-28.367943$ & 0.124 & zspec & 41.55 & $-1.68$ & 7.75 & 7.76 & 7.80\\
WCDFS2903 & 135853 & 53.713116 & $-27.707296$ & 0.210 & zspec & 41.73 & $-0.48$ & 8.70 & 8.86 & 8.78\\
WCDFS3040 & 14452 & 54.146523 & $-28.566229$ & 0.156 & zspec & 41.35 & $-0.36$ & 7.98 & 7.65 & 7.82\\
WCDFS3049 & 24383 & 54.123714 & $-28.513514$ & 0.201 & zspec & 41.63 & $-0.24$ & 9.00 & 9.14 & 8.93\\
WCDFS3606 & 663441 & 52.205837 & $-27.084970$ & 0.116 & zphot & 41.04 & $-0.97$ & 9.11 & 8.95 & 8.91\\
WCDFS4044 & 190734 & 53.735924 & $-27.263348$ & 0.232 & zphot & 42.96 & $-1.44$ & 8.93 & 9.11 & 9.06\\
\hline
\hline
\end{tabular}
\begin{tablenotes}
\item
\textit{Notes.} This table is sorted in ascending order of (1) XID, the XMM-SERVS source ID in \citet{Ni21}. (2) The source ID in \citetalias{Zou22}. (3) and (4) J2000 coordinates. (5) Redshift. (6) Redshift type. ``zspec'' and ``zphot'' indicate that the redshifts are spectroscopic and photometric redshifts, respectively. (7) Observed $2-10$~keV luminosity. (8) and (9) AGN template-based $M_\star$ and SFR in \citetalias{Zou22}. (10) and (11) Normal-galaxy and BQ-galaxy template-based $M_\star$ in \citetalias{Zou22}, respectively.
\end{tablenotes}
\end{threeparttable}
}
\end{table*}

\begin{table*}
\caption{Source catalog in ELAIS-S1}
\label{tbl_release_es1}
\centering
\resizebox{\hsize}{!}{
\begin{threeparttable}
\begin{tabular}{ccccccccccc}
\hline
\hline
XID & Tractor ID & R. A. & Decl. & $z$ & ztype & $\log L_\mathrm{X, obs}$ & $\log\mathrm{SFR}$ & $\log M_\star$ & $\log M_\star^\mathrm{gal}$ & $\log M_\star^\mathrm{bqgal}$\\
& & (degree) & (degree) & & & ($\mathrm{erg~s^{-1}}$) & ($M_\odot~\mathrm{yr^{-1}}$) & ($M_\odot$) & ($M_\odot$) & ($M_\odot$)\\
(1) & (2) & (3) & (4) & (5) & (6) & (7) & (8) & (9) & (10) & (11)\\
\hline
ES0552 & 644246924653 & 9.443832 & $-43.783956$ & 0.198 & zspec & 41.77 & $-2.05$ & 8.34 & 8.31 & 8.30\\
ES0598 & 644246876754 & 8.917671 & $-43.935414$ & 0.802 & zspec & 42.94 & $0.90$ & 9.46 & 9.47 & 9.51\\
ES0618 & 644246961129 & 8.932128 & $-43.675953$ & 0.078 & zspec & 40.52 & $-1.80$ & 7.75 & 7.68 & 7.75\\
ES0695 & 644246995151 & 8.630936 & $-43.587412$ & 0.052 & zspec & 40.58 & $-1.58$ & 8.90 & 8.90 & 8.79\\
ES0875 & 644247085422 & 9.254822 & $-43.304044$ & 0.220 & zspec & 42.31 & $-0.91$ & 8.45 & 8.35 & 8.40\\
ES0904 & 644247097403 & 9.187931 & $-43.266448$ & 0.058 & zspec & 40.53 & $-2.35$ & 7.51 & 7.39 & 7.54\\
ES0922 & 644247059777 & 8.834723 & $-43.381866$ & 0.627 & zspec & 43.01 & $1.12$ & 9.29 & 9.34 & 9.23\\
ES1525 & 644246565825 & 10.076456 & $-44.636604$ & 0.182 & zspec & 41.37 & $-1.26$ & 9.11 & 8.99 & 9.09\\
ES1692 & 644246507665 & 9.727813 & $-44.838687$ & 0.138 & zphot & 41.49 & $-0.57$ & 9.07 & 9.11 & 9.00\\
ES1906 & 644246984907 & 9.495692 & $-43.606902$ & 0.665 & zspec & 42.61 & $0.68$ & 8.49 & 8.60 & 8.90\\
ES2302 & 644247133625 & 9.481544 & $-43.158952$ & 0.186 & zspec & 41.85 & $-2.11$ & 8.34 & 8.33 & 8.36\\
ES2367 & 644246584145 & 9.235555 & $-44.577481$ & 0.196 & zphot & 41.32 & $-0.94$ & 9.28 & 9.21 & 9.28\\
ES2468 & 644246664997 & 9.588642 & $-44.334487$ & 0.105 & zphot & 40.57 & $-2.04$ & 8.77 & 8.85 & 8.78\\
\hline
\hline
\end{tabular}
\begin{tablenotes}
\item
\textit{Notes.} Same as Table~\ref{tbl_release_wcdfs}, but for ELAIS-S1.
\end{tablenotes}
\end{threeparttable}
}
\end{table*}

\begin{table*}
\caption{Source catalog in XMM-LSS}
\label{tbl_release_xmmlss}
\centering
\resizebox{\hsize}{!}{
\begin{threeparttable}
\begin{tabular}{ccccccccccc}
\hline
\hline
XID & Tractor ID & R. A. & Decl. & $z$ & ztype & $\log L_\mathrm{X, obs}$ & $\log\mathrm{SFR}$ & $\log M_\star$ & $\log M_\star^\mathrm{gal}$ & $\log M_\star^\mathrm{bqgal}$\\
& & (degree) & (degree) & & & ($\mathrm{erg~s^{-1}}$) & ($M_\odot~\mathrm{yr^{-1}}$) & ($M_\odot$) & ($M_\odot$) & ($M_\odot$)\\
(1) & (2) & (3) & (4) & (5) & (6) & (7) & (8) & (9) & (10) & (11)\\
\hline
XMM00235 & 846032 & 34.335213 & $-5.480953$ & 0.018 & zspec & 40.18 & $-1.60$ & 6.89 & 6.91 & 7.00\\
XMM00275 & 1064101 & 34.351822 & $-4.682919$ & 0.621 & zspec & 42.76 & $0.66$ & 9.30 & 9.34 & 9.55\\
XMM00309 & 1197673 & 34.367554 & $-4.229480$ & 0.127 & zphot & 41.17 & $-2.04$ & 8.26 & 8.63 & 8.41\\
XMM00310 & 1049050 & 34.368221 & $-4.717017$ & 0.067 & zspec & 41.00 & $-3.85$ & 8.70 & 8.68 & 8.65\\
XMM00557 & 1047971 & 34.482452 & $-4.736839$ & 0.104 & zspec & 41.86 & $-1.29$ & 8.94 & 9.16 & 8.85\\
XMM00569 & 959133 & 34.487633 & $-5.063463$ & 0.240 & zspec & 42.43 & $-0.68$ & 8.87 & 8.89 & 9.11\\
XMM00637 & 1182069 & 34.520947 & $-4.241128$ & 0.390 & zspec & 42.71 & $0.57$ & 8.77 & 9.21 & 9.50\\
XMM00768 & 1154153 & 34.597561 & $-4.383302$ & 0.564 & zspec & 42.74 & $0.92$ & 9.05 & 9.46 & 9.79\\
XMM00795 & 1197492 & 34.613106 & $-4.230209$ & 0.050 & zspec & 40.25 & $-2.25$ & 7.36 & 7.19 & 7.38\\
XMM00838 & 1057468 & 34.640095 & $-4.683757$ & 0.403 & zphot & 41.95 & $-0.55$ & 8.70 & 8.87 & 8.71\\
XMM00852 & 824637 & 34.644726 & $-5.551417$ & 0.359 & zspec & 42.71 & $-0.04$ & 9.13 & 9.20 & 9.08\\
XMM00893 & 1061574 & 34.671246 & $-4.711613$ & 0.500 & zspec & 42.31 & $-0.40$ & 9.02 & 9.08 & 8.97\\
XMM00929 & 1198658 & 34.688000 & $-4.194396$ & 0.629 & zspec & 43.29 & $0.68$ & 9.03 & 9.13 & 9.36\\
XMM00937 & 1158612 & 34.692986 & $-4.330591$ & 0.405 & zspec & 42.74 & $0.74$ & 9.22 & 9.46 & 9.39\\
XMM01271 & 1094414 & 34.872738 & $-4.567514$ & 0.294 & zspec & 42.63 & $0.18$ & 9.36 & 9.07 & 9.58\\
XMM01487 & 537544 & 34.993084 & $-5.055025$ & 0.094 & zphot & 40.35 & $-1.52$ & 8.53 & 8.73 & 8.57\\
XMM01643 & 618315 & 35.091015 & $-4.655634$ & 0.285 & zspec & 42.03 & $-0.05$ & 8.36 & 8.34 & 8.66\\
XMM01671 & 693786 & 35.107349 & $-4.309465$ & 0.645 & zspec & 42.74 & $0.70$ & 9.43 & 9.42 & 9.85\\
XMM01843 & 530442 & 35.193939 & $-5.092589$ & 0.270 & zspec & 42.45 & $-0.38$ & 8.64 & 8.63 & 8.67\\
XMM01944 & 694600 & 35.253357 & $-4.305611$ & 0.492 & zspec & 43.06 & $0.32$ & 9.27 & 8.84 & 9.48\\
XMM02309 & 483035 & 35.436298 & $-5.421105$ & 0.093 & zspec & 40.86 & $-1.96$ & 8.39 & 8.29 & 8.43\\
XMM02399 & 744798 & 35.485558 & $-4.132769$ & 0.615 & zspec & 43.66 & $1.65$ & 9.31 & 9.47 & 10.46\\
XMM02438 & 538356 & 35.513119 & $-5.053028$ & 0.255 & zspec & 42.39 & $-1.00$ & 9.33 & 9.25 & 9.40\\
XMM02611 & 517637 & 35.609409 & $-5.172750$ & 0.062 & zphot & 40.58 & $-1.88$ & 8.91 & 8.79 & 8.75\\
XMM02667 & 765953 & 35.634304 & $-4.296326$ & 0.033 & zspec & 40.58 & $-1.63$ & 7.45 & 7.28 & 7.52\\
XMM02884 & 741519 & 35.763706 & $-4.090158$ & 0.872 & zphot & 43.59 & $0.30$ & 9.12 & 9.08 & 9.14\\
XMM03004 & 480648 & 35.825909 & $-5.344553$ & 0.082 & zspec & 41.44 & $-4.57$ & 9.30 & 9.02 & 9.21\\
XMM03169 & 590232 & 35.912758 & $-4.797773$ & 0.694 & zphot & 42.70 & $0.72$ & 9.26 & 9.24 & 9.35\\
XMM03297 & 223117 & 35.974476 & $-4.602885$ & 0.091 & zspec & 41.53 & $-1.36$ & 7.60 & 7.64 & 7.90\\
XMM03466 & 80641 & 36.062706 & $-5.137580$ & 0.173 & zphot & 41.15 & $-1.04$ & 8.03 & 8.27 & 8.04\\
XMM03822 & 218471 & 36.261375 & $-4.654545$ & 0.854 & zspec & 43.24 & $0.78$ & 9.29 & 9.24 & 9.68\\
XMM03914 & 383611 & 36.307674 & $-4.018576$ & 0.149 & zphot & 41.47 & $-0.53$ & 9.14 & 9.17 & 9.17\\
XMM04151 & 46389 & 36.458359 & $-5.237194$ & 0.196 & zphot & 41.70 & $-0.52$ & 8.99 & 9.03 & 9.00\\
XMM04366 & 352708 & 36.581959 & $-4.152509$ & 0.504 & zphot & 42.31 & $0.35$ & 9.27 & 9.26 & 9.20\\
XMM04379 & 196099 & 36.588135 & $-4.880373$ & 0.215 & zspec & 41.92 & $-1.23$ & 9.30 & 8.91 & 9.33\\
XMM04396 & 146862 & 36.595638 & $-4.900545$ & 0.781 & zphot & 42.81 & $0.20$ & 9.32 & 9.16 & 9.19\\
XMM04963 & 244870 & 36.905552 & $-4.510224$ & 0.248 & zspec & 42.41 & $-1.44$ & 8.77 & 8.69 & 8.81\\
XMM05115 & 33507 & 37.016174 & $-5.291483$ & 0.321 & zphot & 42.14 & $-0.96$ & 9.28 & 9.46 & 9.24\\
\hline
\hline
\end{tabular}
\begin{tablenotes}
\item
\textit{Notes.} Same as Table~\ref{tbl_release_wcdfs}, but for XMM-LSS, and the XIDs are from \citet{Chen18}.
\end{tablenotes}
\end{threeparttable}
}
\end{table*}

\begin{figure*}
\centering
\resizebox{\hsize}{!}{
\includegraphics{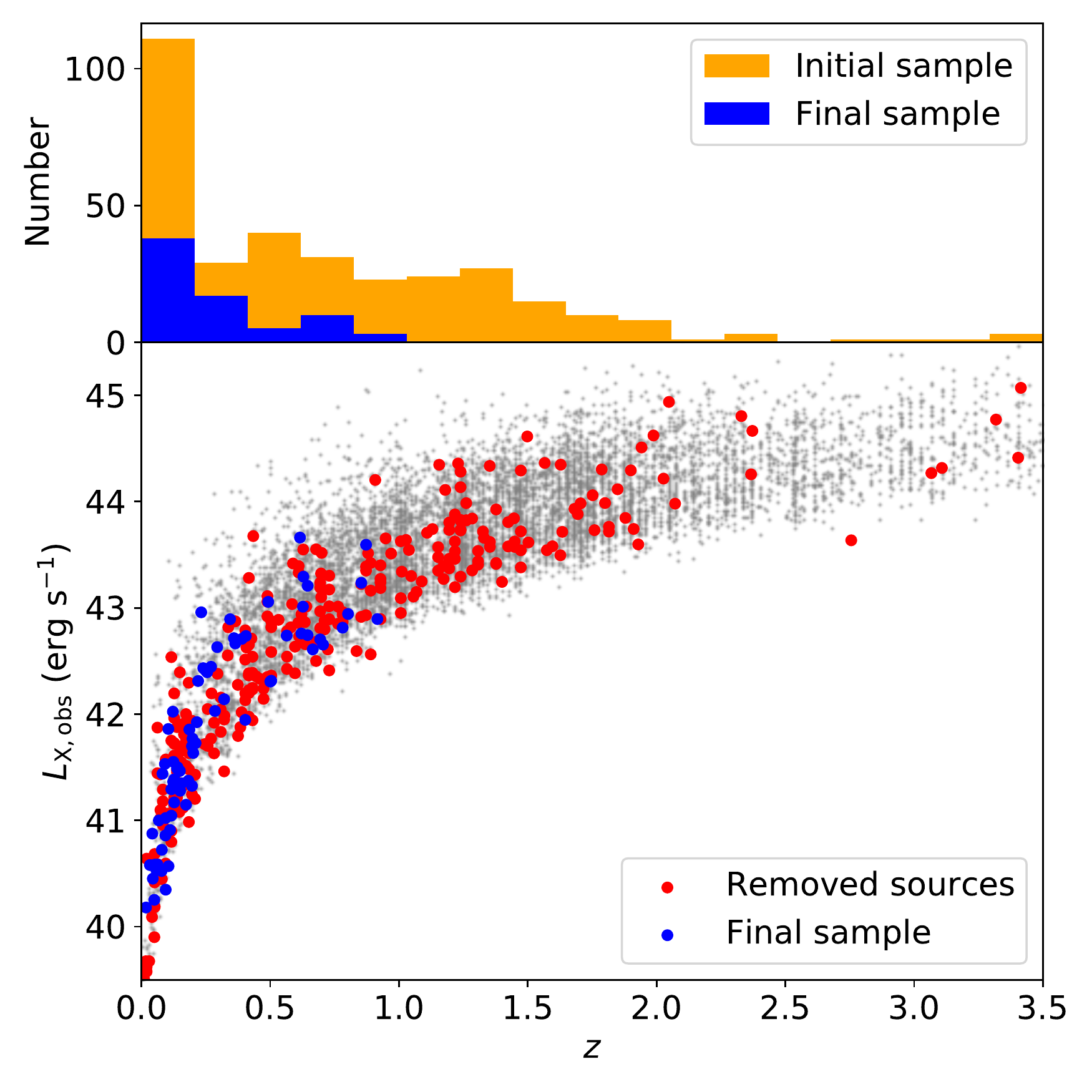}
\includegraphics{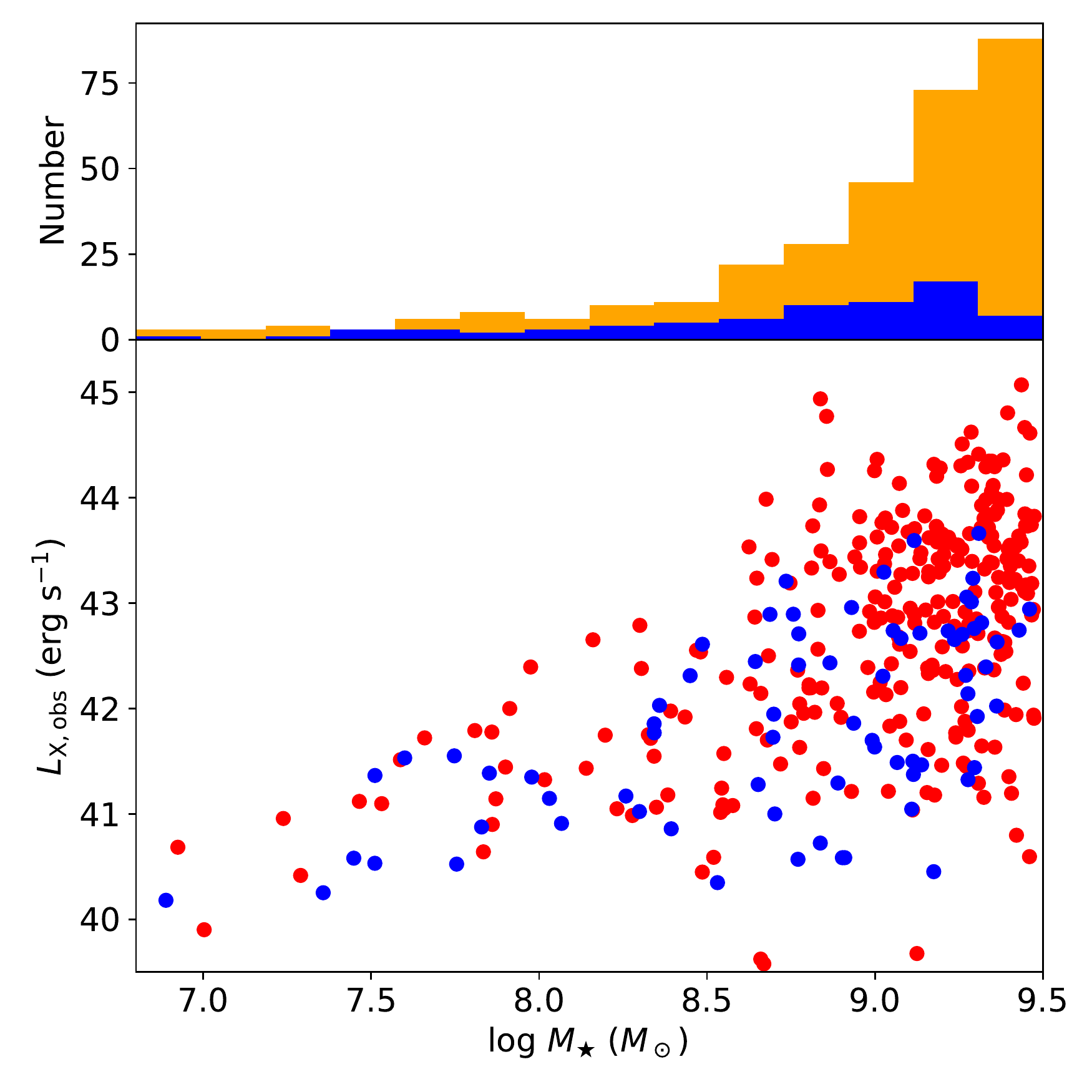}
}
\caption{Left: the $z-L_\mathrm{X, obs}$ plane. The grey small points are all the XMM-SERVS sources. The blue points are our final active dwarf-galaxy sample, and the red ones are those removed from the initial sample. The top histograms show the redshift distributions of the initial (orange) and final (blue) samples. Right: similar to the left panel but for the $M_\star-L_\mathrm{X, obs}$ plane, and its top histograms are for the $M_\star$ distributions. Our final sample size is much smaller than the initial sample size, highlighting the overall challenges of reliably selecting distant active dwarf galaxies. Selection biases inevitably exist, especially at high redshifts.}
\label{fig_zmstarlx}
\end{figure*}

We present our sources in the $z-L_\mathrm{X, obs}$, where $L_\mathrm{X, obs}$ is the observed $2-10$~keV luminosity, and $M_\star-L_\mathrm{X, obs}$ planes in Figure~\ref{fig_zmstarlx}. Table~\ref{tbl_samplesize} and Figure~\ref{fig_zmstarlx} show that only a small fraction of sources can be retained in the final sample, highlighting the challenge of reliably searching for active dwarf galaxies beyond the local universe. Especially, high-redshift sources in the initial sample are less likely to be reliable, and they mainly fail the photo-$z$ quality and $M_\star^\mathrm{gal}$ cuts. We reiterate that our overall selection criteria are designed to be conservative, and it is possible that some of our excluded objects are real active dwarf galaxies, though this is challenging to pin down further, given our current data.\par

We have inspected their \mbox{X-ray} and optical images and did not find any apparent issues. We have also checked the matching between the XMM-SERVS catalogs and optical-to-IR catalogs. For W-CDF-S and ELAIS-S1, \citet{Ni21} presented the false-matching rate as a function of a parameter, $p_\mathrm{any}$. Our final sample has a mean $p_\mathrm{any}$ of 0.87 and 0.82 for W-CDF-S and ELAIS-S1, respectively, which corresponds to a false rate of $\approx3\%$ (see Figure~19 in \citealt{Ni21}). For XMM-LSS, our mean matching reliability is 96\%. Therefore, our final sample should have good matching reliability, and the expected mismatch rate is $\approx3\%$, corresponding to $\approx3$ mismatched sources.\par
\begin{figure}
\centering
\resizebox{\hsize}{!}{\includegraphics{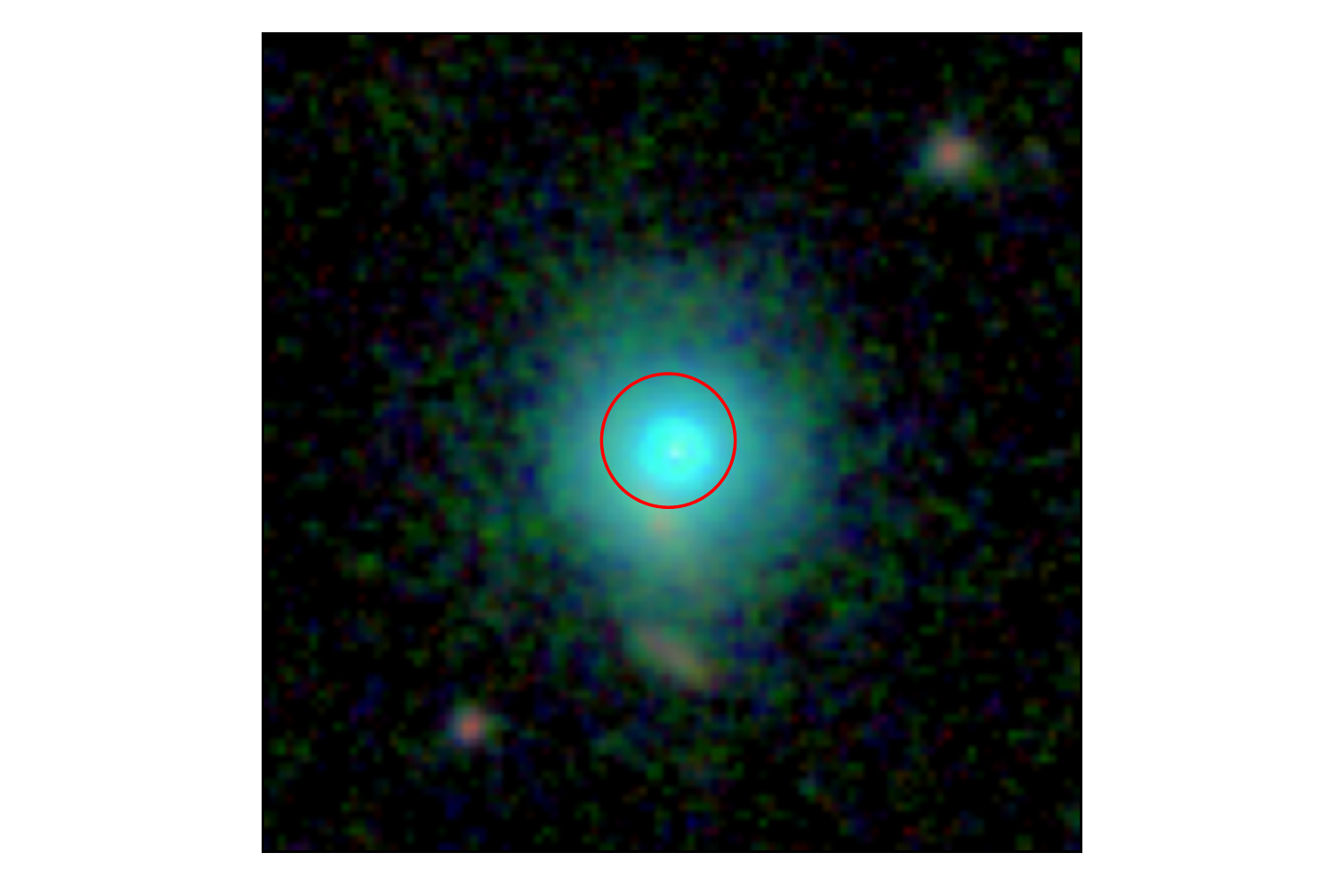}}
\caption{The three-color Hubble image of WCDFS2044 generated from F814W (blue), F125W (green), and F160W (red) images. The red circle is centered at the XMM-Newton centroid with a radius of $0.5''$.}
\label{fig_hstimg_in_hlfgoodss}
\end{figure}

As a representative example, one of our sources (XID = WCDFS2044, R. A. = 03:32:14.02, Decl. = $-$27:51:00.8, and spec-$z$ = 0.122) resides in both the smaller Chandra Deep Field-South (CDF-S; \citealt{Luo17}) and Hubble Legacy Fields GOODS-South \citep{Illingworth16, Whitaker19} and thus has deeper and higher angular-resolution Chandra and Hubble observations. We found apparent Chandra and Hubble counterparts of this source and plot its Hubble image in Figure~\ref{fig_hstimg_in_hlfgoodss}. Its Hubble morphology has a S\'ersic index of 4.2 and a half-light radius of 0.6~kpc \citep{van_der_Wel12}, supporting its SED-fitting results in \citetalias{Zou22} that WCDFS2044 is a quiescent dwarf galaxy ($M_\star=2.3\times10^9~M_\odot$ and $\mathrm{SFR}=2.8\times10^{-4}~M_\odot~\mathrm{yr^{-1}}$).

\section{Analyses and Results}
\label{sec: analyses}
We further investigate several properties of our selected sample in this section. Section~\ref{sec: hr} analyzes \mbox{X-ray} hardness ratios (HRs) and obscuration. Section~\ref{sec: radio} presents the radio properties of our sources. Section~\ref{sec: host} presents the cosmic environments. Section~\ref{sec: agnfrac} derives the accretion distribution and active fraction of the dwarf-galaxy population. Section~\ref{sec: ledd} discusses AGN bolometric luminosities, black-hole masses, and Eddington ratios of our sample.

\subsection{Hardness Ratio}
\label{sec: hr}
The median FB net source counts of our sources is 122, insufficient for detailed \mbox{X-ray} spectral fitting. We thus analyze their HRs for simplicity to probe their spectral shapes. HR is defined as $(\lambda_H-\lambda_S)/(\lambda_H+\lambda_S)$, where $\lambda_S$ and $\lambda_H$ are the SB and HB source count rates, respectively. The cataloged XMM-SERVS HRs are only reliable mainly for sources detected in both the SB and HB, but many of our sources are only detected in the SB because the SB has higher sensitivity. Therefore, we recalculate the HRs of our sources and, to be consistent, all the XMM-SERVS sources for comparison.\par
The classical method of estimating the net count rate by directly subtracting the background fails for undetected bands, and thus we adopt a Bayesian method to correctly account for the Poisson nature. We follow the framework in \citet{Park06} but revise the mathematical and algorithmic implementations, the details of which are presented in Appendix~\ref{sec: myhr}. For each source, we obtain its \mbox{X-ray} image counts within $5\times5$ pixels (i.e., $20''\times20''$) and exposure time, and the expected background intensity is from background maps. The EEF within this given aperture is further absorbed into the exposure time for aperture correction. These are then utilized in our calculations in Appendix~\ref{sec: myhr} to return posterior cumulative distribution functions (CDFs) of HR, where we set the prior parameters as $\psi_{S1}=\psi_{H1}=1$ and $\psi_{S2}=\psi_{H2}=0$ (see Equation~\ref{eq_hrprior} for their definitions). We adopt the HR as the $50^\mathrm{th}$ percentile of the CDF and the associated $1\sigma$ uncertainty range as the $16^\mathrm{th}-84^\mathrm{th}$ percentiles.\par
\begin{figure*}
\centering
\resizebox{\hsize}{!}{\includegraphics{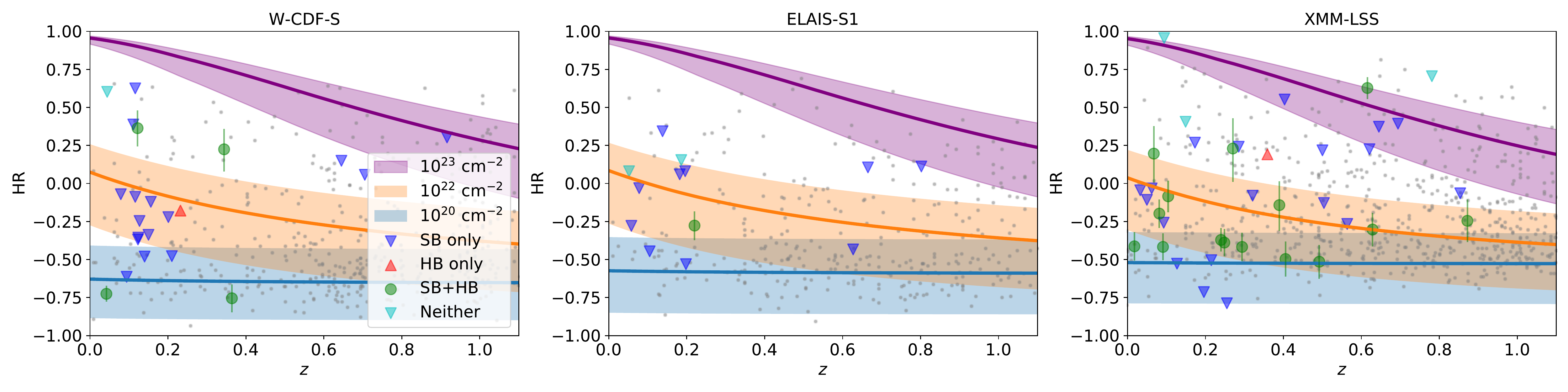}}
\caption{HR versus $z$ in W-CDF-S (left), ELAIS-S1 (middle), and XMM-LSS (right). The background grey points are general XMM-SERVS sources detected in both the SB and HB. The large colored points are our active dwarf galaxies, and their legend in the left panel indicates their detection status in the SB and HB. The blue downward triangles, which are detected in the SB but not the HB, represent 95\% HR upper limits; while the red upward triangles, which are detected in the HB but not the SB, are 95\% HR lower limits. Green points with error bars represent the median HR values with $1\sigma$ uncertainties for sources detected in both bands. The cyan downward triangles are the 95\% HR upper limits for sources only detected in the FB. The shaded regions are the expected relations for redshifted absorbed power-laws, whose photon indices vary between 1.4 and 2.6, with several different $N_\mathrm{H}$ values, as labeled in the legend. Their corresponding solid curves mark the case when the photon index is 1.8. Our sample is generally below the $N_\mathrm{H}=10^{23}~\mathrm{cm^{-2}}$ region and thus not heavily obscured in \mbox{X-rays}.}
\label{fig_zhr}
\end{figure*}

We present our sources in the $z-\mathrm{HR}$ plane in Figure~\ref{fig_zhr}, together with the expected $z-\mathrm{HR}$ curves for redshifted absorbed power-law models with photon indices between 1.4 and 2.6 and the Galactic absorption included. The curves are calculated using the Portable Interactive Multi-Mission Simulator (PIMMS).\footnote{\url{https://heasarc.gsfc.nasa.gov/docs/software/tools/pimms.html}} For a given spectral model, we obtain the corresponding expected total net count rate in a given camera (EPIC PN, MOS1, and MOS2) and a given band, and the predicted HR is calculated as follows.
\begin{align}
\mathrm{HR}&=\frac{\sum_j\left(x_{H,j}\hat{\lambda}_{H,j}^\mathrm{tot}-x_{S,j}\hat{\lambda}_{S,j}^\mathrm{tot}\right)}{\sum_j\left(x_{H,j}\hat{\lambda}_{H,j}^\mathrm{tot}+x_{S,j}\hat{\lambda}_{S,j}^\mathrm{tot}\right)},\\
x_{S,j}&=\mathrm{median}\{t_{\mathrm{exp},j}(\mathrm{SB})/t_\mathrm{exp}(\mathrm{SB})\},\\
x_{H,j}&=\mathrm{median}\{t_{\mathrm{exp},j}(\mathrm{HB})/t_\mathrm{exp}(\mathrm{HB})\},
\end{align}
where $j\in\{\mathrm{PN, MOS1, MOS2}\}$, $t_\mathrm{exp, j}$ is from the single-camera exposure maps across our fields, $t_\mathrm{exp}$ is from the camera-merged exposure maps, and $\hat{\lambda}_{\bullet, i}^\mathrm{tot}$ is the PIMMS-predicted single-camera count rate. Note that it is still appropriate to calculate spectral shapes using photoelectric absorption up to $N_\mathrm{H}=10^{23}~\mathrm{cm^{-2}}$. The ``reflection'' component would only be prominent at much higher $N_\mathrm{H}$ ($>10^{24}~\mathrm{cm^{-2}}$). Although the Compton-scattering losses out of the line of sight may become non-negligible at $N_\mathrm{H}\gtrsim10^{23}~\mathrm{cm^{-2}}$, this effect is nearly energy-independent at the XMM-Newton energy coverage and thus does not change the overall spectral shape. For example, Figure~5 in \citet{Li19} can serve as a clear illustration -- when only using the photoelectric absorption, the inferred $N_\mathrm{H}$ (i.e., the spectral shape) would not be biased up to $N_\mathrm{H}=10^{24}~\mathrm{cm^{-2}}$, but only the intrinsic emission (i.e., the spectral normalization) would be underestimated.\par

Figure~\ref{fig_zhr} shows that our sources are generally not heavily obscured ($N_\mathrm{H}<10^{23}~\mathrm{cm^{-2}}$). We plot the 95\% HR upper limits for sources that are only detected in the FB (i.e., cyan downward triangles in Figure~\ref{fig_zhr}).\footnote{Although these sources are undetected in the SB and HB, we can still provide loose constraints on their HRs. Intuitively, the fact that they are only detected in the FB but not in the SB or HB already provides some information about the underlying spectra -- if their spectra were so hard that their SB counts were fully dominated by noise, it would be impossible to obtain detectable FB counts by summing fully background-dominated SB counts and undetectable (but maybe excessive) HB counts; similarly, their spectra cannot be too soft. Since we are interested in how hard the spectra can be, we only show the HR upper limits for these sources.} Although the overall constraints on their HRs are weak, most of their upper limits are still below the $N_\mathrm{H}=10^{23}~\mathrm{cm^{-2}}$ region and thus disfavor high $N_\mathrm{H}$. As a population, their joint HR posterior gives $\mathrm{HR}=-0.18_{-0.08}^{+0.08}$, also significantly below the $N_\mathrm{H}=10^{23}~\mathrm{cm^{-2}}$ region. We note that many HR upper limits of the sources detected in the SB but not the HB in Figure~\ref{fig_zhr} are above the $N_\mathrm{H}=10^{22}~\mathrm{cm^{-2}}$ region but still below the $N_\mathrm{H}=10^{23}~\mathrm{cm^{-2}}$ region, and the $N_\mathrm{H}=10^{22}~\mathrm{cm^{-2}}$ region significantly overlaps with the $N_\mathrm{H}=10^{20}~\mathrm{cm^{-2}}$ region. Therefore, we do not consider $N_\mathrm{H}$ thresholds lower than $10^{23}~\mathrm{cm^{-2}}$ to mitigate the influence of these uncertainties.\par
This low incidence of heavy obscuration is surprising from some perspectives (e.g., \citealt{Merloni14, Liu17}). We calculate the expected number of sources with $N_\mathrm{H}\geq10^{23}~\mathrm{cm^{-2}}$ in each field as follows:
\begin{align}
\hat{n}(>10^{23})&=\sum_i\frac{\int_{23}^{24}P_{\mathrm{det},i}\mathrm{XLF}_id\log N_\mathrm{H}}{\int_{20}^{24}P_{\mathrm{det},i}\mathrm{XLF}_id\log N_\mathrm{H}},\\
P_{\mathrm{det},i}(N_\mathrm{H})&=P_\mathrm{det}\left(f_\mathrm{X}^\mathrm{FB}(L_{\mathrm{X},i}, z_i)\beta_\mathrm{FB}(N_\mathrm{H}, z_i)\right),\label{eq_Pdet_i}\\
\mathrm{XLF}_i(N_\mathrm{H})&=\mathrm{XLF}(N_\mathrm{H}, L_{\mathrm{X},i}, z_i),\label{eq_XLF_i}
\end{align}
where $L_{\mathrm{X},i}$ is the intrinsic $2-10$~keV luminosity of the $i^\mathrm{th}$ source taken from \citetalias{Zou22}, $z_i$ is the redshift of the $i^\mathrm{th}$ source, the intrinsic source column density $N_\mathrm{H}$ is in $\mathrm{cm^{-2}}$, $P_\mathrm{det}$ is the detection probability as a function of the observed FB flux, $f_\mathrm{X}^\mathrm{FB}(L_\mathrm{X}, z)$ is the intrinsic FB flux for a source with $L_\mathrm{X}$ at redshift $z$ assuming a photon index of 1.8 and is calculated using Equation~A4 in \citetalias{Zou22}, $\beta_\mathrm{FB}$ is the FB absorption factor for a source with a photon index of 1.8 and is calculated based on photoelectric absorption and Compton-scattering losses (\texttt{zphabs $\times$ cabs} in \texttt{XSPEC}), XLF is the \mbox{X-ray} luminosity function with the $N_\mathrm{H}$ distribution included, and the summation runs over all the active dwarf galaxies. We leave the detailed derivation and explanation of $P_\mathrm{det}$ to Section~\ref{sec: agnfrac}. We adopt the XLF of \citet{Ueda14}, and \citetalias{Zou22} showed that XLFs from different works generally lead to consistent results given the XMM-SERVS depth. As in \citetalias{Zou22}, we limit the integration range of $N_\mathrm{H}$ to be below $10^{24}~\mathrm{cm^{-2}}$ because more heavily obscured active dwarf galaxies are generally undetectable. The above equations return $\hat{n}(>10^{23})=3.6$, 1.0, and 8.1 for W-CDF-S, ELAIS-S1, and XMM-LSS, respectively, which are significantly larger than our observed results in Figure~\ref{fig_zhr}. We find that the same conclusion still holds for our initial sample in Table~\ref{tbl_samplesize}, and thus the lack of heavily obscured sources is not caused by our selection biases.\par
Similar results are seen in COSMOS -- Figure~5 of \citet{Mezcua18} shows that, although mild absorption may sometimes exist, almost no sources below $z\approx1$ (above which sources may be unreliable, as we discussed in Section~\ref{sec: data_sample}) have sufficiently large HRs to indicate the existence of heavy obscuration. Indeed, heavily obscured active dwarf galaxies have almost not been reported even in the local universe, and \citet{Ansh22} report the first discovery of a type~2 dwarf galaxy showing heavy \mbox{X-ray} obscuration. Overall, this section indicates that the massive-AGN $N_\mathrm{H}$ distribution, and consequently, the XLF, does not appear to extend down to active dwarf galaxies. This may be explained as follows. The XLF encodes the obscuration through the inverse correlation between $L_\mathrm{X}$ and the obscuration fraction (e.g., \citealt{Brandt21} and references therein). However, \citet{Ricci17} argued that the $N_\mathrm{H}$ correlation with the mass-normalized accretion rate (i.e., Eddington ratio $\lambda_\mathrm{Edd}$; see Section~\ref{sec: ledd} for more details) is more fundamental than the correlation with $L_\mathrm{X}$. At a given $L_\mathrm{X}$, our sources have higher $\lambda_\mathrm{Edd}$ than for more massive SMBHs and thus should be less obscured. Unfortunately, we cannot reliably measure $\lambda_\mathrm{Edd}$ due to various challenges (see Section~\ref{sec: ledd}), and thus we cannot quantitatively revise the XLF predictions of the obscuration.\par
The right panel of Figure~\ref{fig_zhr} also shows that there is one source (XID = XMM02399) above $N_\mathrm{H}=10^{23}~\mathrm{cm^{-2}}$ in our XMM-LSS sample at $z=0.6$. We will discuss this source in greater detail in Appendix~\ref{sec: xmm02399}.

\subsection{Radio Properties}
\label{sec: radio}
Our fields also have deep radio coverage at 1.4~GHz from the Australia Telescope Large Area Survey (ATLAS; \citealt{Franzen15}),  the MeerKAT International GHz Tiered Extragalactic Exploration survey (MIGHTEE; \citealt{Heywood22}), and the Very Large Array (VLA) survey in the XMM-LSS field \citep{Heywood20}. ATLAS covers W-CDF-S and ELAIS-S1, and MIGHTEE and VLA cover XMM-LSS. These radio data have been compiled and analyzed in \citet{Zhu23}, wherein the radio sources have been matched to those in \citetalias{Zou22}.\par
To identify the origin of the radio emission of our sources, we define $q_\mathrm{24obs}$ as $\log(S_{24~\mu\mathrm{m}}/S_\mathrm{1.4~GHz})$, where $S_{24~\mu\mathrm{m}}$ and $S_\mathrm{1.4~GHz}$ are the observed Spitzer 24~$\mu\mathrm{m}$ and 1.4~GHz flux densities, respectively. We use $q_\mathrm{24obs}$ to identify radio-excess AGNs in our sample as outliers from the tight correlation between the FIR and radio emission. Such a FIR-radio correlation has been well constrained for star-forming galaxies over several decades (e.g., \citealt{Condon92}), and the physical reason behind it is that both the FIR and radio emission can trace star formation. That is, highly star-forming galaxies simultaneously produce strong FIR and radio emission, leading to a roughly constant $q_\mathrm{24obs}$. Therefore, outliers from the FIR-radio correlation with strong excess radio emission, which can be identified if their $q_\mathrm{24obs}$ values are below a given threshold, are thought to be powered by AGNs (e.g., \citealt{Ibar08}). Note, however, that $q_\mathrm{24obs}$ is not a useful indicator for other more detailed radio properties, such as morphology and radio spectral slope. We do not directly use the conventional radio-loudness parameter (e.g., \citealt{Kellermann89}), which is suitable for luminous AGNs, because the implicit assumption for appropriately calculating radio loudness is that the optical emission is dominated by AGN emission, while our sources generally do not satisfy this assumption. Therefore, classifications based on the FIR-radio correlation are often necessary and also have been shown to be currently the most effective method for selecting radio AGNs in deep surveys (e.g., \citealt{Zhu23}).\par
\begin{figure}
\centering
\resizebox{\hsize}{!}{\includegraphics{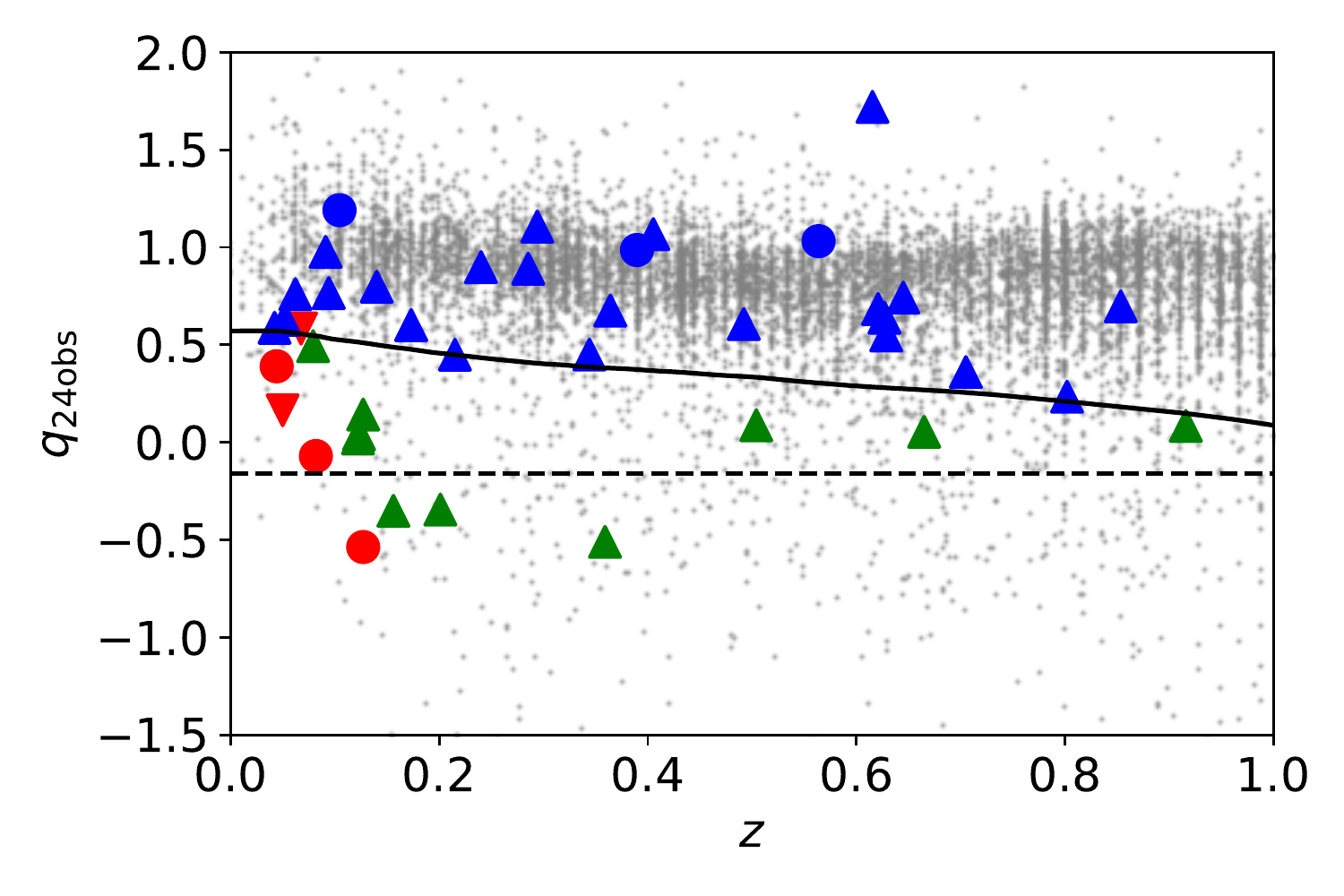}}
\caption{$q_\mathrm{24obs}$ versus $z$. The black solid and dashed lines are the $q_\mathrm{24obs}$ thresholds for radio-excess AGNs in \citet{Bonzini13} and \citet{Zhu23}, respectively. The colored points or triangles mark our active dwarf galaxies that have radio coverage and are detected in at least one of the 24~$\mu\mathrm{m}$ and 1.4~GHz bands. The points are those detected in both bands and thus have $q_\mathrm{24obs}$ values, the downward triangles represent $q_\mathrm{24obs}$ upper limits for sources only detected in the radio, and the upward triangles are $q_\mathrm{24obs}$ lower limits for sources only detected at 24~$\mu\mathrm{m}$. Sources are colored in red if they are classified as radio-excess AGNs according to the criterion in \citet{Bonzini13}, blue if they are above the black solid line (i.e., not radio-excess AGNs), and green if unable to be classified. The background grey points are general radio sources in our fields \citep{Zhu23}. Their photo-$z$s are mildly ``quantized'' because these photo-$z$s (same as those in \citetalias{Zou22}) are derived at a series of redshift grid points. This grid has a bin size of $\delta z/(1+z)<0.01$ and is thus sufficiently dense to avoid noticeable biases. Only a limited fraction of our sources are radio-excess AGNs.}
\label{fig_q24obs}
\end{figure}

We show our sources in the $z-q_\mathrm{24obs}$ plane in Figure~\ref{fig_q24obs}, where lower or upper limits of $q_\mathrm{24obs}$ are presented for sources only detected in one band. Figure~\ref{fig_q24obs} also presents the $q_\mathrm{24obs}$ threshold in \citet{Bonzini13} as the black solid curve, which is 0.7~dex below the expected relation based on the redshifted M82 SED. We follow \citet{Bonzini13} to classify a source as radio-excess if its $q_\mathrm{24obs}$ is below the curve, or its $q_\mathrm{24obs}$ upper limit (if undetected at 24~$\mu\mathrm{m}$) is no more than 0.35~dex higher than the curve. Similarly, we classify sources as not radio-excess if their $q_\mathrm{24obs}$ values or $q_\mathrm{24obs}$ lower limits (if undetected in the radio) are above the black solid curve. The remaining sources are unclassified. Five (XIDs = WCDFS0761, XMM00309, XMM00310, XMM00795, and XMM03004) of our active dwarf galaxies are radio-excess, and 28 are not radio-excess. However, the exact number of radio-excess sources depends on the adopted $q_\mathrm{24obs}$ threshold. If we adopt the threshold ($-0.16$; the black dashed line in Figure~\ref{fig_q24obs}) in \citet{Zhu23}, we would only have one radio-excess source (XID = XMM00309). This small incidence of radio-excess AGNs is generally consistent with those in \citet{Mezcua18}, where 3/40 active dwarf galaxies are radio-excess.\par

Simultaneously knowing the radio and \mbox{X-ray} luminosities can sometimes help measure $M_\mathrm{BH}$ through the so-called fundamental plane of black-hole activity, at least for some sources (e.g., \citealt{Gultekin19}). However, \citet{Gultekin22} showed that, for active dwarf galaxies, the $M_\mathrm{BH}$ inferred based on the fundamental plane is overestimated by several orders of magnitude. Our radio-excess sources have similar radio and \mbox{X-ray} luminosities as for the sample in \citet{Gultekin22} and are thus also not expected to follow the fundamental plane. Several causes that may lead to the discrepancy are detailedly discussed in \citet{Gultekin22}. Especially, the fundamental plane is suggested to hold only for low-$\lambda_\mathrm{Edd}$ sources, but our sources may have much higher $\lambda_\mathrm{Edd}$ (e.g., Section~\ref{sec: ledd}). In this case, our sources are more likely to follow a different relationship, i.e., being regulated by the corona-disk-jet connection in, e.g., \citet{Zhu20}. Therefore, the radio data cannot help constrain our sources' $M_\mathrm{BH}$.

\subsection{Host Environments}
\label{sec: host}
It is still unclear whether and how AGN activity can affect dwarf galaxies and their environments. Unfortunately, due to the strong selection effects, especially for sSFR (Section~\ref{sec: zphot_reliability}), we are unable to unbiasedly assess the star-formation activities of the active dwarf galaxy population. Nevertheless, the selection effects are not directly relevant to the galaxy environment, and thus we examine if our sources and normal dwarf galaxies reside in similar environments in this section.\par
\begin{figure*}
\centering
\resizebox{\hsize}{!}{\includegraphics{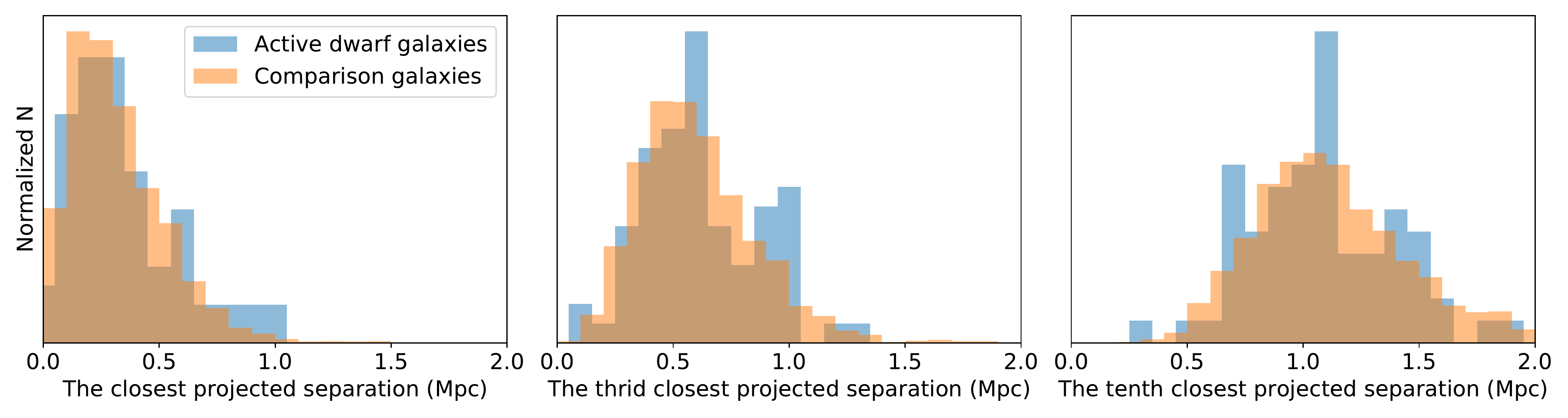}}
\caption{The normalized histograms of the first (left), third (middle), and tenth (right) closest projected separations of dwarf galaxies with respect to massive galaxies. Our active dwarf galaxies do not show clear differences from their comparison samples.}
\label{fig_compenv}
\end{figure*}

First, for each active dwarf galaxy with ($\log M_{\star, i}, z_i$), we locally construct a comparison galaxy sample from \citetalias{Zou22} by selecting galaxies with $z$ within $(1\pm0.075)\times z_i$ and $\log M_\star$ within $(1\pm0.1)\times\log M_{\star, i}$. We do not further apply any photo-$z$ quality cuts for these normal dwarf galaxies because their catastrophic outlier fraction is acceptably small (17.1\%), although much larger than that for massive galaxies (3.9\% for those with $M_\star>10^{10.5}~M_\odot$). There are typically 400 galaxies per dwarf satisfying the criterion, and we randomly pick out 100 comparison sources for each of our active dwarf galaxies. Following a similar method to \citet{Davis22}, we use projected distances to massive galaxies as an indication of the environment. We follow \citet{Yang18} to define the redshift slices for projections. We first calculate the dispersion of $(z_\mathrm{phot}-z_\mathrm{spec})/(1+z_\mathrm{spec})$, denoted as $\sigma_{\Delta z/(1+z)}$, as a function of $z$, from the photo-$z$ catalogs in \citet{Chen18} and \citet{Zou21b}. $\sigma_{\Delta z/(1+z)}$ is calculated within $z\pm0.2$ and is roughly 0.04 for the redshift range of our sources. For each given dwarf galaxy with redshift $z$, we select massive galaxies with $M_\star>10^{10.5}~M_\odot$ and redshifts within $z\pm1.5(1+z)\sigma_{\Delta z/(1+z)}$ and calculate the first, second, third, fifth, and tenth closest projected separations between the dwarf galaxy and the massive galaxies. These separations trace the environments on 100~kpc to 1~Mpc scales. We show the corresponding histograms of the first, third, and tenth closest projected separations in Figure~\ref{fig_compenv} as examples, and the histograms of our active dwarf galaxies do not visually show large differences from those of the comparison sample. Indeed, we found that our sources and the comparison samples do not show statistically significant differences for any of the separations, as also found in \citet{Davis22} and \citet{Siudek22}. Therefore, environmental effects are not significantly responsible for the presence of AGNs in dwarf galaxies. Similar conclusions are also drawn for massive galaxies in, e.g., \citet{Yang18}, where SMBH growth is found not to be correlated with the cosmic environment once $M_\star$ is controlled.\par

We also tried adding sources without \mbox{X-ray} excess in Section~\ref{sec: select_finalsample} because the exclusion of these sources may cause biases, e.g., those close to massive galaxies are more likely to be excluded. We repeated our analyses, and all the conclusions remain unchanged.

\subsection{AGN Accretion Distribution and Active Fraction}
\label{sec: agnfrac}
In this section, we explore the distribution of AGN accretion for dwarf galaxies, which can further return the active fraction ($\xi_\mathrm{AGN}$).\footnote{The term ``active fraction'' should be distinguished from the term ``fractional AGN contribution'' used in Section~\ref{sec: select_goodsample}. We used ``$f_\mathrm{AGN}$'' to denote the latter and will use ``$\xi_\mathrm{AGN}$'' to denote the former in this section.} This is important for at least two reasons. First, the accretion distribution provides clues for how MBHs coevolve with their dwarf hosts as a population. Second, although it is generally a consensus that most massive galaxies contain SMBHs, the BH occupation fraction of dwarf galaxies is largely unknown. The active fraction serves as a lower limit for the BH occupation fraction because active sources must contain BHs. Also, due to the strong selection effects, it is difficult to select sufficient sources in $z-M_\star-L_\mathrm{X}$ bins that are complete to calculate active fractions, and thus it is much more efficient to derive the active fraction from the accretion distribution, which is constrained by all the sources after accounting for the selection effects.\par
We quantify the accretion distribution as $p(L_\mathrm{X}\mid M_\star, z)$, the conditional probability density per unit $\log L_\mathrm{X}$ of a galaxy with ($M_\star$, $z$) hosting an AGN with intrinsic $2-10$~keV luminosity of $L_\mathrm{X}$ (e.g., \citealt{Aird12, Aird18, Yang18a}). We assume a power-law relation for $p(L_\mathrm{X}\mid M_\star, z)$:
\begin{align}
p(L_\mathrm{X}\mid M_\star, z)=A\left(\frac{M_\star}{M_0}\right)^{\gamma_M}\left(\frac{1+z}{1+z_0}\right)^{\gamma_z}\left(\frac{L_\mathrm{X}}{L_0}\right)^{\gamma_L},
\end{align}
where the normalization, $A$, and the power-law indices, $\gamma_M$, $\gamma_z$, and $\gamma_L$, are free parameters to be determined, and $M_0$, $z_0$, and $L_0$ are arbitrary scaling factors. This power-law form has been proven to be valid for massive galaxies (e.g., \citealt{Aird12}), and deviations from the power-law are minor and can only be revealed with sufficiently good data \citep{Aird18}. \citet{Birchall20, Birchall22} also showed that such a power-law is valid for dwarf galaxies. For easy comparison with \citet{Aird12}, we adopt the same scaling factors as theirs: $M_0=10^{11}~M_\odot$, $z_0=0.6$, and $L_0=10^{43}~\mathrm{erg~s^{-1}}$. In the following text, we consider selection effects and fit the data to constrain $p(L_\mathrm{X}\mid M_\star, z)$.\par
\begin{figure}
\centering
\resizebox{\hsize}{!}{\includegraphics{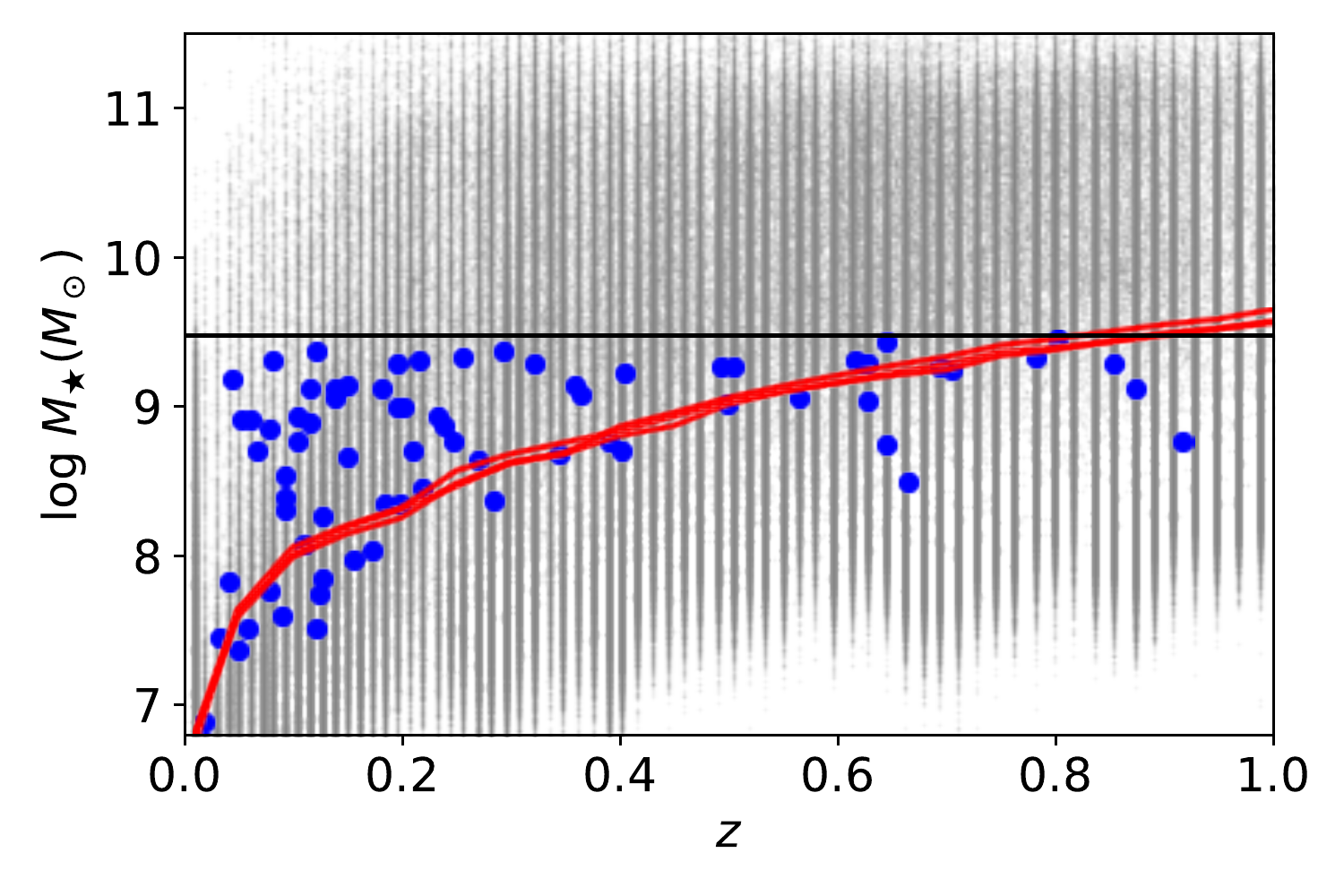}}
\caption{$M_\star$ versus $z$. The background grey points are all the sources in \citetalias{Zou22}, the blue points are our final active dwarf-galaxy sample, and the three red curves, which visually overlap with each other, are the $M_\star$ completeness curves for our three fields. Sources above the red curves are utilized to constrain the accretion distribution and active fraction of the dwarf-galaxy population.}
\label{fig_mstarlimit}
\end{figure}

We first construct a complete dwarf-galaxy sample by applying redshift-dependent $M_\star$ cuts. This is necessary because active dwarf galaxies may be subject to different incompleteness effects from inactive ones, and it is difficult to correct the incompleteness of the active population independently from the normal population. To estimate the $M_\star$ depth, we first adopt the limiting VIDEO $K_s$-band magnitude as $K_{s,\mathrm{lim}}=23.1$ mag, the $5\sigma$ $K_s$ depth in a $3''$ aperture. This limit is conservative for two reasons. First, the nominal magnitude depth becomes deeper with decreasing aperture size, and a $3''$ aperture may be large. As a comparison, the $K_s$-band depth of a $1''$ aperture is 1.5~mag deeper. Second, other VIDEO bands may reach deeper depths than $K_s$. Due to these reasons, less than half of the sources in \citetalias{Zou22} are more massive than the limiting $M_\star$ derived below. Nevertheless, this ensures that the samples are complete. Following \citet{Pozzetti10}, we then convert the magnitude depth to the expected limiting $M_\star$ for each $K_s$-detected galaxy in \citetalias{Zou22}: $\log M_\mathrm{lim}=\log M_\star+0.4(K_s-K_{s,\mathrm{lim}})$. At each redshift, we adopt the $M_\star$ completeness threshold as the value above which 90\% of the $M_\mathrm{lim}$ values lie. We derive the limiting $M_\star$ for all three XMM-SERVS fields independently, and Figure~\ref{fig_mstarlimit} presents the corresponding curves as functions of $z$. The curves in different fields are generally consistent. 54 out of the 73 active dwarf galaxies are above the completeness curves, and all the sources beyond $z\approx0.8$ are subject to incompleteness. As in Section~\ref{sec: host}, we do not apply photo-$z$ quality cuts for normal dwarf galaxies. The catastrophic outlier fraction is much smaller than the uncertainties of our parameters, as will be measured in the following text (Table~\ref{tbl_fitmodel}).\par

After settling the selection effects upon $M_\star$, we further quantify those from the XMM-SERVS survey, i.e., the probability that a source with observed FB flux of $f_\mathrm{X,obs}^\mathrm{FB}$ gets detected by the survey, which is denoted as $P_\mathrm{det}(f_\mathrm{X,obs}^\mathrm{FB})$. A common way of deriving $P_\mathrm{det}$ is to use the sensitivity curves and estimate $P_\mathrm{det}(f_\mathrm{X,obs}^\mathrm{FB})$ as the fraction of the total survey area with sensitivities deeper than $f_\mathrm{X,obs}^\mathrm{FB}$. However, this approach can cause biases because the sensitivity is derived for a given aperture ($20''\times20''$ for XMM-SERVS) while the real detection procedures are more complicated and cannot be accurately mimicked by a Poisson detection in a given constant aperture. We thus derive $P_\mathrm{det}$ by comparing the intrinsic $\log N-\log S$ relation in \citet{Georgakakis08} and the cataloged flux distribution, where the $\log N-\log S$ relation is the well-determined expected observed-flux distribution with the detection procedures deconvolved, given by $dN/df_\mathrm{X}^\mathrm{FB}$, the surface number density per unit $f_\mathrm{X}^\mathrm{FB}$. We assume a functional formula\footnote{We found it necessary to adopt a functional formula instead of estimating $P_\mathrm{det}$ in some $\log f_\mathrm{X,obs}$ bins because the limited number of high-$f_\mathrm{X,obs}$ sources cannot provide effective constraints to $P_\mathrm{det}$ when $f_\mathrm{X,obs}$ is high. We have also confirmed that this formula works well and is in perfect consistency with the binned estimations of $P_\mathrm{det}$ in $\log f_\mathrm{X,obs}$ bins with sufficient sources.} for $P_\mathrm{det}$:
\begin{align}
P_\mathrm{det}(f_\mathrm{X,obs}^\mathrm{FB})=\frac{1}{2}\left[\mathrm{erf}\left(b(\log f_\mathrm{X,obs}^\mathrm{FB}-a)\right)+1\right],
\end{align}
where $a$ and $b$ are free parameters to constrain, and the same functional formula has been adopted for optical surveys (e.g., \citealt{Bernstein04}). The log-likelihood (e.g., \citealt{Barlow90, Loredo04}) when comparing with the survey catalogs is
\begin{align}
\ln\mathcal{L}=-A_\mathrm{tot}\int\frac{dN}{df_\mathrm{X,obs}^\mathrm{FB}}P_\mathrm{det}df_\mathrm{X,obs}^\mathrm{FB}+\sum_i\ln P_\mathrm{det}(f_{i,\mathrm{obs}}^\mathrm{FB}),
\end{align}
where constants independent from $P_\mathrm{det}$ are discarded, $A_\mathrm{tot}$ is the survey area, $f_{i,\mathrm{obs}}^\mathrm{FB}$ is the observed FB flux of the $i^\mathrm{th}$ FB-detected source in the XMM-SERVS catalogs, and the summation runs over all the FB-detected sources. We only consider the regions overlapping with the VIDEO footprints and minimize the likelihood for each field independently to measure $a$ and $b$. The results are $(a, b)=(-14.36, 4.85), (-14.30, 3.80)$, and $(-14.42, 4.38)$ for W-CDF-S, ELAIS-S1, and XMM-LSS, respectively. We adopt the detection probability in the FB because most of our active dwarf galaxies are detected in the FB. Among the 54 sources above the $M_\star$ completeness, 5 are not detected in the FB. It is challenging to quantify the probability of not detecting a source in the FB but detecting it in another band because more detailed spectral shapes should be considered, and thus we remove the 5 FB-undetected sources from our following analyses for simplicity.\par
Overall, we use all the 49 active dwarf galaxies above the $M_\star$ completeness threshold and detected in the FB to constrain $p(L_\mathrm{X}\mid M_\star, z)$. The log-likelihood function (e.g., \citealt{Aird12, Yang18a}) is
\begin{align}
\ln\mathcal{L}=-n_\mathrm{mdl}+\sum_{i=1}^{n_\mathrm{AGN}}\ln p(L_{\mathrm{X}, i}\mid M_{\star, i}, z_i),
\end{align}
where the summation is for our 49 active dwarf galaxies, and $n_\mathrm{mdl}$ is the expected number of FB-detected active dwarf galaxies given a model parameter set. The log-likelihood is calculated in all three fields separately and added together to a merged log-likelihood function. $n_\mathrm{mdl}$ is
\begin{align}
n_\mathrm{mdl}=\sum_{i=1}^{n_\mathrm{gal}}\int_{-\infty}^{+\infty}p(L_\mathrm{X}\mid M_{\star, i}, z_i)P_\mathrm{det}(f_\mathrm{X}^\mathrm{FB})d\log L_\mathrm{X},\label{eq_nmdl}
\end{align}
where $f_\mathrm{X}^\mathrm{FB}=f_\mathrm{X}^\mathrm{FB}(L_\mathrm{X}, z)$ is the intrinsic FB flux for a source with $L_\mathrm{X}$ at redshift $z$ assuming a photon index of 1.8, as in Section~\ref{sec: hr}. Given our parametrization, the above equation can be solved analytically. The key is using the following integration \citep{Ng69}.
\begin{align}
\int e^{cx}\left[\mathrm{erf}(x)+1\right]dx=\frac{1}{c}\left\{e^{cx}\left[\mathrm{erf}(x)+1\right]-e^\frac{c^2}{4}\mathrm{erf}\left(x-\frac{c}{2}\right)\right\}.
\end{align}
Then the derivations become straightforward, and Equation~\ref{eq_nmdl} can be reduced to the following form when $\gamma_L<0$.
\begin{align}
n_\mathrm{mdl}=-\frac{\exp\left(\left(\frac{\gamma_L\ln10}{2b}\right)^2\right)}{\gamma_L\ln10}\sum_{i=1}^{n_\mathrm{gal}}p\left(L_\mathrm{X}(10^a, z_i)\mid M_{\star,i}, z_i\right),
\label{eq_nmdl_analytical}
\end{align}
where $L_\mathrm{X}(f_\mathrm{X}^\mathrm{FB}, z)$ is the inverse function of $f_\mathrm{X}^\mathrm{FB}(L_\mathrm{X}, z)$. Using Equation~\ref{eq_nmdl_analytical} instead of numerically solving the integration in Equation~\ref{eq_nmdl} increases the computational speed by several orders of magnitude.\par
Unlike \citet{Aird12}, who set a lower limit of $L_\mathrm{X}=10^{42}~\mathrm{erg~s^{-1}}$ for the integration in Equation~\ref{eq_nmdl}, we set the lower integration limit as $-\infty$. This effectively means that we regard all the dwarf galaxies that are detectable given the XMM-SERVS sensitivity to be powered by AGNs, which is supported in Section~\ref{sec: select_finalsample} -- the expected galaxy emission from all the \mbox{X-ray}-detected dwarf galaxies (not accounting for the emission from nearby sources) is far smaller than the observed fluxes. Note that we neglect the intrinsic obscuration because, as shown in Section~\ref{sec: hr}, active dwarf galaxies do not have a suitable a priori XLF available and generally are not heavily obscured. To quantitatively assess the possible impact of obscuration, we calculate the expected mean absorption factor predicted by the XLF in \citet{Ueda14} as follows.
\begin{align}
E\left\{\left<\log\beta_\mathrm{FB}\right>\right\}=\frac{1}{n_\mathrm{AGN}}\sum_{i=1}^{n_\mathrm{AGN}}\frac{\int_{20}^{24}\log\beta_\mathrm{FB}P_{\mathrm{det},i}\mathrm{XLF}_id\log N_\mathrm{H}}{\int_{20}^{24}P_{\mathrm{det},i}\mathrm{XLF}_id\log N_\mathrm{H}},
\end{align}
where $P_{\mathrm{det},i}(N_\mathrm{H})$ and $\mathrm{XLF}_i(N_\mathrm{H})$ are defined in Equations~\ref{eq_Pdet_i} and \ref{eq_XLF_i}, respectively. This returns $E\left\{\left<\log\beta_\mathrm{FB}\right>\right\}=-0.24$~dex, causing $\log A$ to shift by $\left|\gamma_L E\left\{\left<\log\beta_\mathrm{FB}\right>\right\}\right|\approx0.2$~dex, where $\gamma_L=0.9$, as will be measured in the following text (Table~\ref{tbl_fitmodel}). Such a difference is comparable to the $1\sigma$ uncertainty of $\log A$ in Table~\ref{tbl_fitmodel}. The actual absorption factor should be far smaller than the XLF-predicted value (Section~\ref{sec: hr}), and thus neglecting the intrinsic obscuration will not cause more than $1\sigma$ difference in our results.\par
We adopt a flat prior on $(\log A, \gamma_M, \gamma_z, \gamma_L)$ while restricting $\gamma_L<0$ and use \texttt{DynamicHMC.jl},\footnote{\url{https://www.tamaspapp.eu/DynamicHMC.jl/stable/}} a package implementing Hamiltonian Monte Carlo (e.g., \citealt{Betancourt17}), to sample the posterior. We apply the analyses to our final active dwarf galaxy sample, as presented above in the previous part of this section before this paragraph, and denote this case as Case~P, where ``P'' stands for ``purity''. However, this sample is designed to be pure instead of complete, and hence we may underestimate the active fraction. We thus regard the active fractions derived from this case as lower limits. We also conduct the same analyses on the initial sample in Table~\ref{tbl_samplesize}, which should be complete but not pure, and regard the corresponding active fractions as upper limits. We denote this case as Case~C, where ``C'' stands for ``completeness'', and there are 149 FB-detected active dwarf galaxies above the $M_\star$ completeness curve. In each case, we also conduct the same analyses in each field to check the consistency.\par
\begin{figure}
\centering
\resizebox{\hsize}{!}{\includegraphics{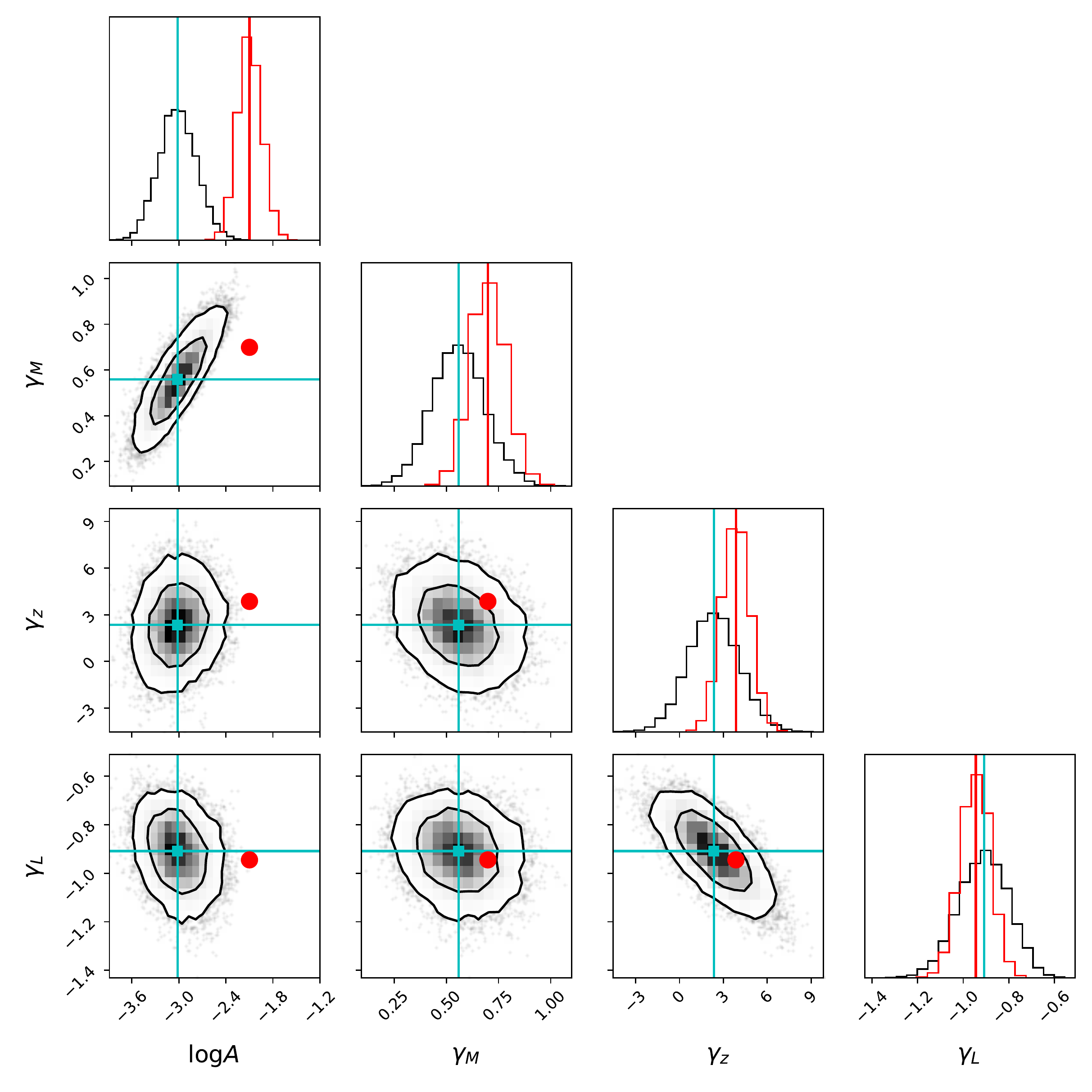}}
\caption{The sampling results of the accretion-distribution parameters. In the upper panels, the black and red histograms are for the sampling distributions in Cases~P and C, respectively; the cyan and red vertical lines mark the corresponding median values. In the other panels, the sampling points and contours are for Case~P, and those for Case~C are not shown for better visualization. The black contours represent 68\% and 95\% levels in Case~P, the grayscale pixels represent probabilities, and the rasterized points outside the 95\% contours are individual sampling points. The median sampling values in Case~P are shown as the cyan squares and lines, and those in Case~C are displayed as the large red points. The results in Cases~P and C mainly differ in $\log A$, and their $\gamma_z$ and $\gamma_L$ are consistent.}
\label{fig_mcmcsampling}
\end{figure}

We present the sampling results in Figure~\ref{fig_mcmcsampling}, based on which we report the fitted parameters in Table~\ref{tbl_fitmodel}. The contours for Case~C look similar aside from systematic differences in some parameters and are thus not plotted in Figure~\ref{fig_mcmcsampling} for clarity. The $\log A-\gamma_M$ and $\gamma_z-\gamma_L$ contours are highly tilted, indicating that their uncertainties are correlated. The same analyses are also applied to individual fields, and their results in Table~\ref{tbl_fitmodel} are statistically consistent with each other, indicating that there are no significant inter-field systematic biases that are larger than the statistical fluctuations. When comparing the merged results in Cases~P and C, their $\gamma_M$, $\gamma_z$, and $\gamma_L$ are in good agreement, while Case~C has a larger $\log A$ value than Case~P because more active dwarf galaxies are included.\par

\begin{table*}
\caption{Fitting results of the accretion-distribution model}
\label{tbl_fitmodel}
\centering
\begin{threeparttable}
\begin{tabular}{cccccc}
\hline
\hline
&& $\log A$ & $\gamma_M$ & $\gamma_z$ & $\gamma_L$\\
\hline
Case~P & Merged & $-3.02_{-0.24}^{+0.25}$ & $0.56_{-0.13}^{+0.13}$ & $2.35_{-1.74}^{+1.75}$ & $-0.91_{-0.11}^{+0.11}$\\
& W-CDF-S & $-3.50_{-0.56}^{+0.54}$ & $0.73_{-0.22}^{+0.23}$ & $-3.34_{-4.39}^{+4.33}$ & $-0.90_{-0.23}^{+0.20}$\\
& ELAIS-S1 & $-3.27_{-0.50}^{+0.48}$ & $0.46_{-0.27}^{+0.27}$ & $7.06_{-3.94}^{+3.90}$ & $-1.14_{-0.28}^{+0.26}$\\
& XMM-LSS & $-2.89_{-0.35}^{+0.37}$ & $0.49_{-0.19}^{+0.20}$ & $2.30_{-2.18}^{+2.28}$ & $-0.79_{-0.14}^{+0.13}$\\
\hline
Case~C & Merged & $-2.10_{-0.15}^{+0.16}$ & $0.70_{-0.08}^{+0.09}$ & $3.86_{-0.98}^{+0.95}$ & $-0.94_{-0.07}^{+0.06}$\\
& W-CDF-S & $-2.16_{-0.27}^{+0.28}$ & $0.70_{-0.15}^{+0.16}$ & $3.68_{-1.75}^{+1.74}$ & $-0.95_{-0.13}^{+0.12}$\\
& ELAIS-S1 & $-1.73_{-0.25}^{+0.26}$ & $0.91_{-0.14}^{+0.15}$ & $2.92_{-1.71}^{+1.69}$ & $-0.97_{-0.11}^{+0.11}$\\
& XMM-LSS & $-2.41_{-0.26}^{+0.27}$ & $0.50_{-0.15}^{+0.15}$ & $5.28_{-1.61}^{+1.67}$ & $-0.95_{-0.11}^{+0.11}$\\
\hline
\hline
\end{tabular}
\end{threeparttable}
\end{table*}

\begin{figure*}
\centering
\resizebox{\hsize}{!}{\includegraphics{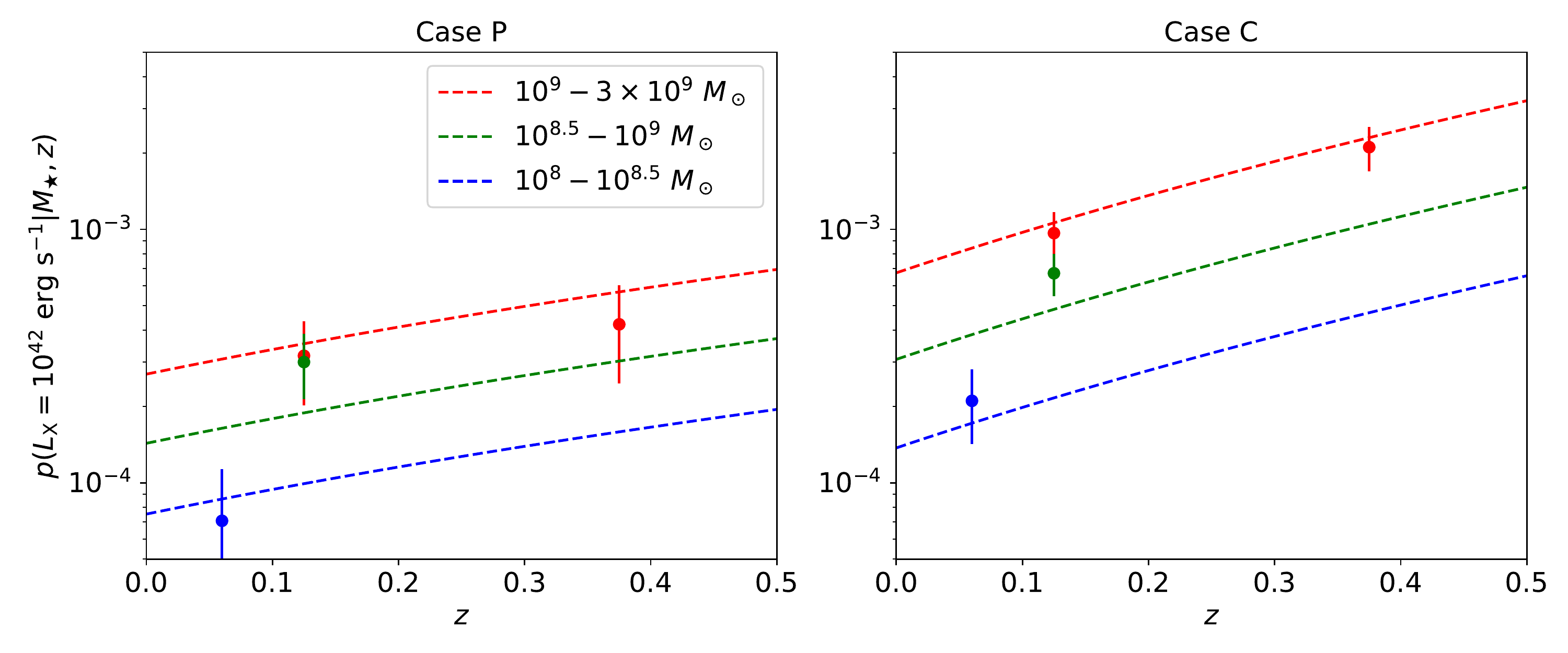}}
\caption{Our fitted $p(L_\mathrm{X}\mid M_\star, z)$ model (dashed lines) and the corresponding four binned estimators based on our data (points with error bars) as functions of $z$. The model and binned estimators are both evaluated at $L_\mathrm{X}=10^{42}~\mathrm{erg~s^{-1}}$ and the $M_\star$ bin centers. The left and right panels are for Cases~P and C, respectively. Our model is consistent with the binned estimators.}
\label{fig_p_z}
\end{figure*}

To help visualize the comparison between the model and the data, we use the $n_\mathrm{obs}/n_\mathrm{mdl}$ method to obtain binned estimators of $p(L_\mathrm{X}\mid M_\star, z)$, as outlined in \citet{Aird12}, which overcomes significant variations of the model within a bin and \mbox{X-ray} selection effects by applying model-dependent corrections. We select four bins in the $z-M_\star$ plane that are above the $M_\star$ completeness curves and below $M_\star=3\times10^9~M_\odot$: $z$ range $\times$ $\log M_\star$ range = $[0, 0.25]\times[9, 9.5]$, $[0.25, 0.5]\times[9, 9.5]$, $[0, 0.25]\times[8.5, 9]$, and $[0, 0.12]\times[8, 8.5]$. In each bin, we denote the number of observed active dwarf galaxies as $n_\mathrm{obs}$ and calculate the model-predicted number as $n_\mathrm{mdl}$ using Equation~\ref{eq_nmdl}. The binned estimator of $p(L_\mathrm{X}\mid M_\star, z)$ is then the fitted model evaluated at a given $(M_\star, z, L_\mathrm{X})$ scaled by $n_\mathrm{obs}/n_\mathrm{mdl}$, and its uncertainty is calculated from the Poisson error of $n_\mathrm{obs}$. We present our fitted model and the binned estimators in Figure~\ref{fig_p_z}, and they are consistent and both indicate that $p(L_\mathrm{X}\mid M_\star, z)$ increases with $M_\star$.\par

In both Cases~P and C, our $\gamma_M$ ($\gamma_L$) is positive (negative) above a $3\sigma$ confidence level,\footnote{Strictly speaking, $\gamma_L$ is always negative because any non-negative $\gamma_L$ value would cause the integration in Equation~\ref{eq_nmdl} to diverge. Here we just use the ratio between $\gamma_L$ and its uncertainty as its nominal significance level, and the significance here in fact means that the bulk of the $\gamma_L$ posterior is far from 0, as shown in Figure~\ref{fig_mcmcsampling}.} indicating that, at fixed $L_\mathrm{X}$, dwarf galaxies have larger probabilities of hosting AGNs with increasing $M_\star$, and the probability decreases as $L_\mathrm{X}$ increases. Although most of our fitted $\gamma_z$ values are positive, the evidence is not sufficiently strong to confirm positive redshift evolution in Case~P due to the large statistical uncertainties of $\gamma_z$. \citet{Aird12} conducted the same analyses for massive galaxies, and our fitted $\gamma_M$, $\gamma_z$, and $\gamma_L$ values are consistent with theirs, possibly indicating that the factors causing these dependencies in dwarf galaxies are similar to those in massive galaxies. However, our normalization, $A$, is significantly smaller than that in \citet{Aird12} in Case~P and only marginally consistent with \citet{Aird12} in Case~C.\par
\begin{figure*}
\centering
\resizebox{\hsize}{!}{\includegraphics{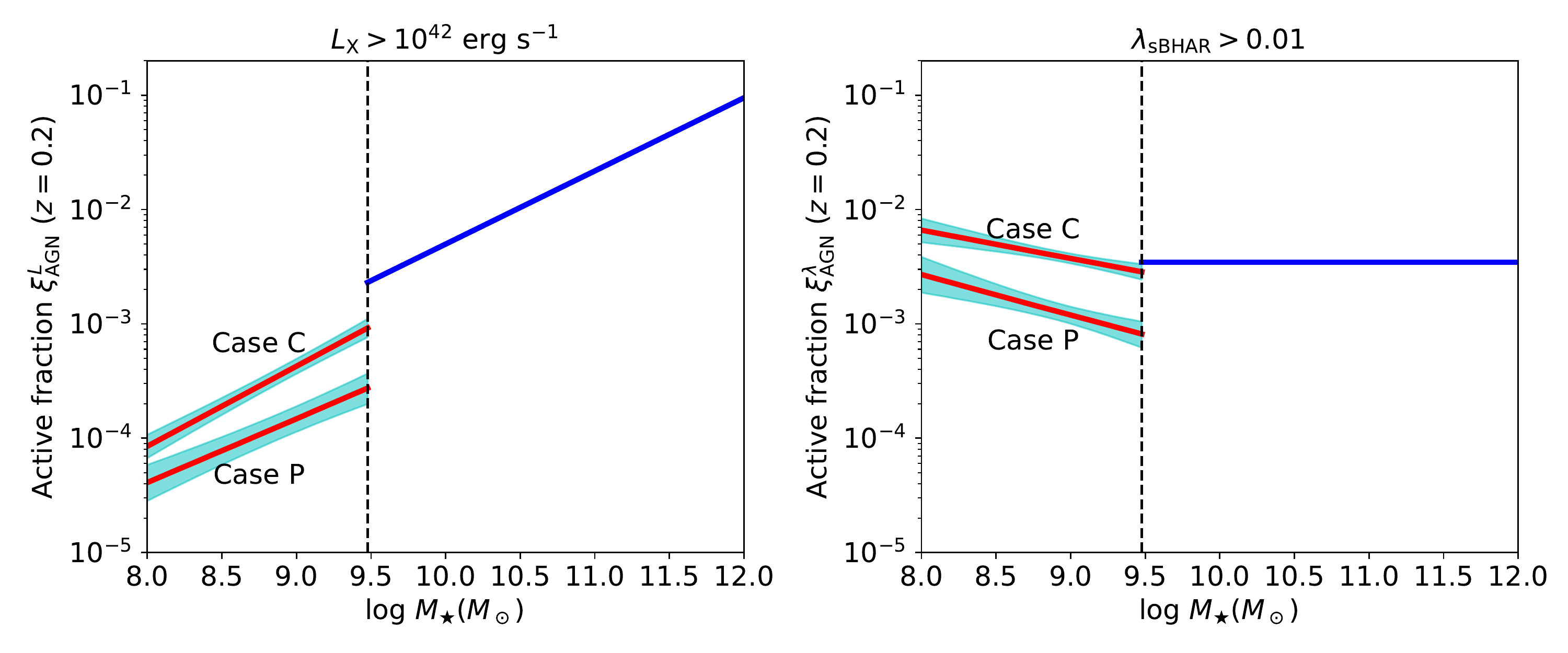}}
\caption{$\xi_\mathrm{AGN}$ as a function of $M_\star$, where the AGN definitions are presented as the panel titles, and the models are evaluated at our median redshift (0.2). The vertical dashed lines mark the $M_\star$ definitions ($3\times10^9~M_\odot$) of dwarf galaxies. The models in this work are presented as red lines with cyan regions showing $1\sigma$ confidence ranges, and blue lines are for massive galaxies, taken from \citet{Aird12}.}
\label{fig_agnfrac_mstar}
\end{figure*}

We further compare our $\xi_\mathrm{AGN}$ with previous work. $\xi_\mathrm{AGN}$ is generally poorly defined and often strongly depends on the AGN selection techniques and survey depths. However, under our accretion-distribution context, $\xi_\mathrm{AGN}$ has an unambiguous definition, which enables direct comparisons with previous work that also adopts the same definition. $\xi_\mathrm{AGN}$ within a given $L_\mathrm{X}$ range, $(L_\mathrm{X, low}, L_\mathrm{X, high})$, is defined as follows.
\begin{align}
\xi_\mathrm{AGN}^L(M_\star, z)=\int_{\log L_\mathrm{X, low}}^{\log L_\mathrm{X, high}}p(L_\mathrm{X}\mid M_\star, z)d\log L_\mathrm{X},\label{eq: xi_AGN_L}
\end{align}
where $L_\mathrm{X, high}$ is allowed to reach infinity, and we add a superscript ``$L$'' to indicate that the definition is based on luminosity. A common alternative definition is based on specific black-hole accretion rate ($\lambda_\mathrm{sBHAR}$; e.g., \citealt{Aird18, Birchall20, Birchall22}), which is defined as follows.
\begin{align}
\lambda_\mathrm{sBHAR}=\frac{25L_\mathrm{X}}{1.26\times10^{38}\times0.002M_\star},
\end{align}
where the multiplicative constants are chosen so that $\lambda_\mathrm{sBHAR}\approx\lambda_\mathrm{Edd}$ (Section~\ref{sec: ledd}) for massive galaxies that follow the relation between $M_\mathrm{BH}$ and the galaxy bulge mass. $p(L_\mathrm{X}\mid M_\star, z)$ can then be converted to $p(\lambda_\mathrm{sBHAR}\mid M_\star, z)$:
\begin{align}
p(\lambda_\mathrm{sBHAR}\mid M_\star, z)=A\left(\frac{M_\star}{M_0}\right)^{\gamma_M+\gamma_L}\left(\frac{1+z}{1+z_0}\right)^{\gamma_z}\left(\frac{\lambda_\mathrm{sBHAR}}{\lambda_0}\right)^{\gamma_L},
\end{align}
where $\lambda_0=\lambda_\mathrm{sBHAR}(L_\mathrm{X}=L_0, M_\star=M_0)$. Similar to Equation~\ref{eq: xi_AGN_L}, $\xi_\mathrm{AGN}$ within a given $\lambda_\mathrm{sBHAR}$ range, $(\lambda_\mathrm{sBHAR, low}, \lambda_\mathrm{sBHAR, high})$, is defined as follows.
\begin{align}
\xi_\mathrm{AGN}^\lambda(M_\star, z)=\int_{\log\lambda_\mathrm{sBHAR, low}}^{\log\lambda_\mathrm{sBHAR, high}}p(\lambda_\mathrm{sBHAR}\mid M_\star, z)d\log\lambda_\mathrm{sBHAR},
\end{align}
where we add a superscript ``$\lambda$'' to indicate that the definition is based on $\lambda_\mathrm{sBHAR}$. We show $\xi_\mathrm{AGN}$ as a function of $M_\star$ in Figure~\ref{fig_agnfrac_mstar}, where we evaluate it at our median redshift, 0.2. We adopt two different AGN definitions, $L_\mathrm{X}>10^{42}~\mathrm{erg~s^{-1}}$ \citep{Aird12} and $\lambda_\mathrm{sBHAR}>0.01$ \citep{Aird18}. The figure reveals that $\xi_\mathrm{AGN}^L$ increases with $M_\star$ and is systematically lower than the model in \citet[their Table~2]{Aird12}, though the difference is small in Case~C. However, $\xi_\mathrm{AGN}^\lambda$ is more independent of $M_\star$ and more consistent with the model in \citet[their Table~3]{Aird12}. The normalizations of $\xi_\mathrm{AGN}$ in the two panels of Figure~\ref{fig_agnfrac_mstar} also differ by $\gtrsim1$~dex -- $\xi_\mathrm{AGN}^L$ is roughly between $10^{-5}-10^{-3}$, while $\xi_\mathrm{AGN}^\lambda$ is roughly between $10^{-3}-10^{-2}$. These are consistent with the local-universe results in \citet{Birchall22}; the strong positive correlation between $\xi_\mathrm{AGN}^L$ and $M_\star$ is at least partly driven by the positive correlation between $M_\mathrm{BH}$ and $M_\star$, and the specific accretion rate does not necessarily strongly correlate with $M_\star$. Nevertheless, it is unclear if adopting $\xi_\mathrm{AGN}^\lambda$ can fully eliminate this factor, and Section~\ref{sec: ledd} will discuss related problems that may cause $\lambda_\mathrm{sBHAR}$ to systematically deviate from $\lambda_\mathrm{Edd}$ for dwarf galaxies.

\begin{figure}
\centering
\resizebox{\hsize}{!}{\includegraphics{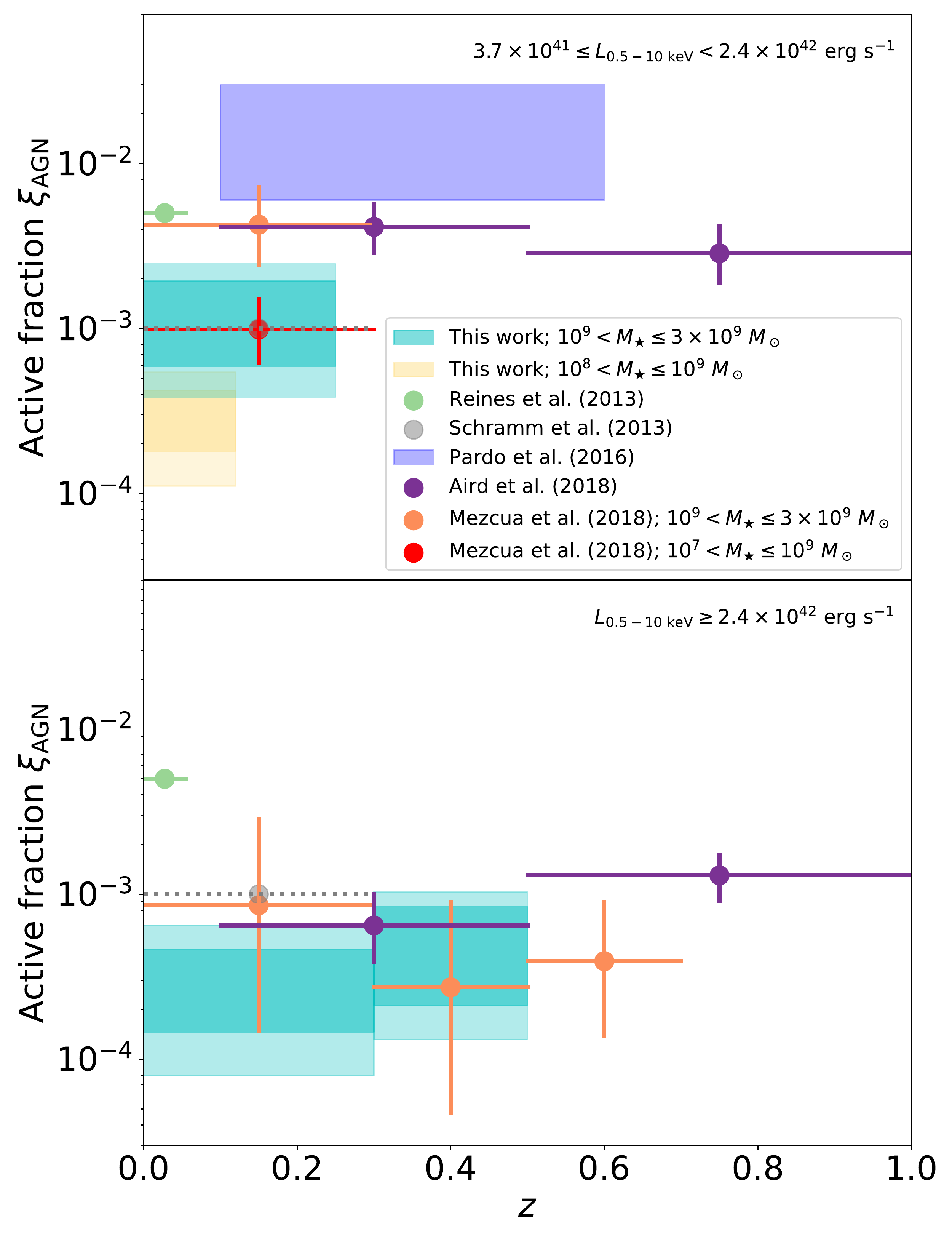}}
\caption{$\xi_\mathrm{AGN}$ in several $z-M_\star-L_\mathrm{X}$ bins compared with previous works, adapted from Figure~10 in \citet{Mezcua18}. The top and right panels are for $3.7\times10^{41}\leq L_\mathrm{0.5-10~keV}<2.4\times10^{42}~\mathrm{erg~s^{-1}}$ and $L_\mathrm{0.5-10~keV}\geq2.4\times10^{42}~\mathrm{erg~s^{-1}}$, respectively. Our measurements are shown as the cyan ($10^9<M_\star\leq3\times10^9~M_\odot$) and yellow ($10^8<M_\star\leq10^9~M_\odot$) boxes, whose redshift ranges are chosen so that the corresponding boxes are above the $M_\star$ completeness curves and if possible, match those in \citet{Mezcua18} for easy comparison. The parts with darker colors within the cyan and yellow boxes represent the $\xi_\mathrm{AGN}$ values calculated from the fitted model parameters in Case~P (lower boundaries) and Case~C (upper boundaries), and the outskirts with lighter colors are delimited by the lower $2\sigma$ sampling values of $\xi_\mathrm{AGN}$ in Case~P (lower boundaries) and upper $2\sigma$ values of $\xi_\mathrm{AGN}$ in Case~C (upper boundaries). Our $\xi_\mathrm{AGN}$ is generally consistent with those in previous works.}
\label{fig_agnfrac}
\end{figure}

We further present more comprehensive comparisons in Figure~\ref{fig_agnfrac}. First, we estimate $\xi_\mathrm{AGN}$ in a given mass-complete $z-M_\star$ bin as follows.
\begin{align}
\xi_\mathrm{AGN}=\frac{1}{n_\mathrm{gal}}\sum_{i=1}^{n_\mathrm{gal}}\xi_\mathrm{AGN}(M_i, z_i),
\end{align}
where the summation runs over all the sources in the bin. For easier comparisons with \citet{Mezcua18}, we estimate $\xi_\mathrm{AGN}$ in the same two $L_\mathrm{X}$ bins adopted in \citet{Mezcua18}: $3.7\times10^{41}\leq L_\mathrm{0.5-10~keV}<2.4\times10^{42}~\mathrm{erg~s^{-1}}$ and $L_\mathrm{0.5-10~keV}\geq2.4\times10^{42}~\mathrm{erg~s^{-1}}$, where $L_\mathrm{0.5-10~keV}$ is converted from $L_\mathrm{X}$ assuming a photon index of 1.8. We present our $\xi_\mathrm{AGN}$ estimations in Figure~\ref{fig_agnfrac}, where we also display previous estimations from \citet{Reines13}, \citet{Schramm13}, \citet{Pardo16}, \citet{Aird18}, and \citet{Mezcua18}. There are several more works presenting $\xi_\mathrm{AGN}$ (e.g., \citealt{Baldassare20b, Birchall20, Birchall22, Latimer21, Ward22}), and they are mainly limited to the local universe. We do not show them in Figure~\ref{fig_agnfrac} to avoid crowding, and interested readers can refer to these articles for more details. Note that due to different and sometimes unknown underlying definitions in most articles from ours, quantitative comparisons are meaningful only with the results in \citet{Mezcua18}. The figure shows that our binned $\xi_\mathrm{AGN}$ values are generally consistent with those in previous literature. Our constraints should be the best given our larger sample size and the fact that we have thoroughly identified multiple underlying issues in Section~\ref{sec: data_sample}. Both Table~\ref{tbl_fitmodel} and Figure~\ref{fig_agnfrac} do not indicate statistically significant redshift evolution of $\xi_\mathrm{AGN}$ for dwarf galaxies. Similar conclusions were drawn in previous works \citep{Aird18, Mezcua18, Birchall22}. However, the large statistical uncertainties may have hindered us from detecting any possible trend, and a larger sample with at least a few hundred objects is needed to provide more meaningful constraints.

\subsection{Bolometric Luminosity, Black-Hole Mass, Eddington Ratio, and The Underlying Challenges}
\label{sec: ledd}
We try to assess the AGN bolometric luminosities ($L_\mathrm{bol}^\mathrm{AGN}$), $M_\mathrm{BH}$, and $\lambda_\mathrm{Edd}$ of our sources in this section. $\lambda_\mathrm{Edd}$ quantifies the relative accretion power and is defined as follows.
\begin{align}
\lambda_\mathrm{Edd}=\frac{L_\mathrm{bol}^\mathrm{AGN}}{L_\mathrm{Edd}}=\frac{L_\mathrm{bol}^\mathrm{AGN}/(\mathrm{erg~s^{-1}})}{1.26\times10^{38}M_\mathrm{BH}/M_\odot}.
\end{align}
\par
\begin{figure}
\centering
\resizebox{\hsize}{!}{\includegraphics{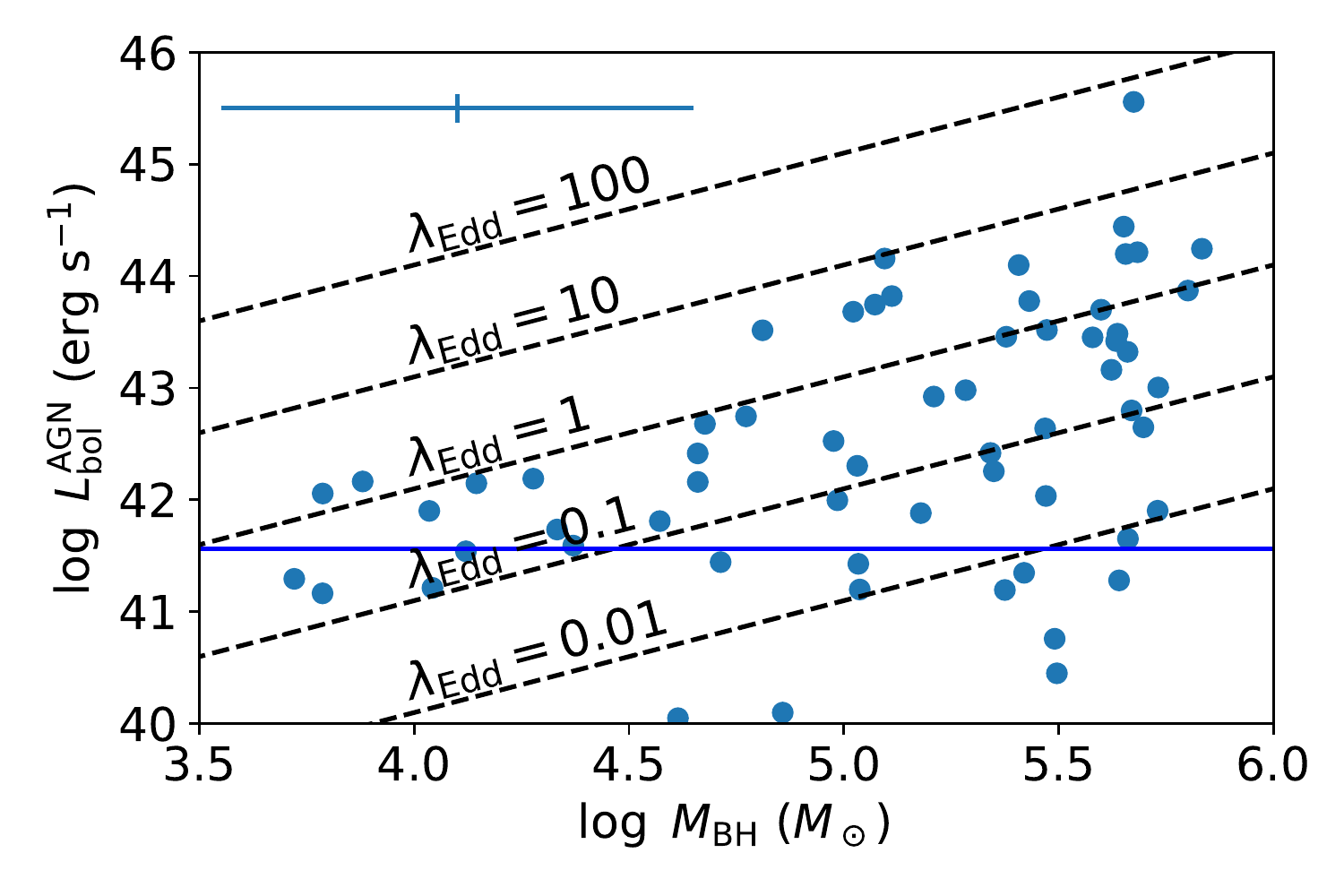}}
\caption{$M_\mathrm{BH}$ versus $L_\mathrm{bol}^\mathrm{AGN}$ for our final sample. Typical error bars are presented in the upper-left corner. The error of $M_\mathrm{BH}$ is 0.55~dex, i.e., the scatter of the scaling relation in \citet{Reines15}, compared to which the $M_\star$ uncertainty is negligible. The median error of $L_\mathrm{bol}^\mathrm{AGN}$ is 0.13~dex, which is from the SED fitting uncertainty and does not include possible systematic errors reported in Section~\ref{sec: ledd}. Several constant-$\lambda_\mathrm{Edd}$ lines are shown as the inclined dashed ones. The blue line presents a typical \mbox{X-ray} detection limit at our median redshift (0.2), assuming a typical $k_\mathrm{bol}^\mathrm{AGN}$ value of 16.75. There are around 11 sources with very small $L_\mathrm{bol}^\mathrm{AGN}$ ($<10^{40}~\mathrm{erg~s^{-1}}$) invisible in the figure, and their $L_\mathrm{bol}^\mathrm{AGN}$ estimations are highly unreliable and underestimated.}
\label{fig_mbh_lbol}
\end{figure}

We adopt $L_\mathrm{bol}^\mathrm{AGN}$ as the likelihood-weighted angle-averaged intrinsic AGN disk luminosity based on the SED fitting in \citetalias{Zou22}. We do not add the corona luminosity in \mbox{X-rays} into $L_\mathrm{bol}^\mathrm{AGN}$ because we usually expect that the disk luminosity dominates, and more importantly, as we will see later, only adopting the disk luminosity can help appreciate some important features. We then estimate $M_\mathrm{BH}$ based on the $M_\mathrm{BH}-M_\star$ scaling relation in \citet{Reines15}: $\log(M_\mathrm{BH}/M_\odot)=7.45+1.05\left[\log(M_\star/M_\odot)-11\right]$. The scatter of the relation is significant (0.55~dex). We plot $M_\mathrm{BH}$ versus $L_\mathrm{bol}^\mathrm{AGN}$ in Figure~\ref{fig_mbh_lbol}, in which several constant-$\lambda_\mathrm{Edd}$ lines are shown for comparison. We also plot a typical $L_\mathrm{bol}^\mathrm{AGN}$ limit corresponding to a typical FB \mbox{X-ray} detection limit of $10^{-14.4}~\mathrm{erg~cm^{-2}~s^{-1}}$ (see Section~\ref{sec: agnfrac}), a redshift of our median value (0.2), and a typical AGN bolometric correction ($k_\mathrm{bol}^\mathrm{AGN}=L_\mathrm{bol}^\mathrm{AGN}/L_\mathrm{X}$) of 16.75, the low-mass limit of $k_\mathrm{bol}^\mathrm{AGN}$ in \citet{Duras20}. The figure indicates that the inferred $M_\mathrm{BH}$ values range between $\approx10^4-10^6~M_\odot$ with a median value of $2\times10^5~M_\odot$ and thus at least some fraction are in the IMBH regime. Our sources, with a median $\lambda_\mathrm{Edd}$ value of 0.19, are often close to or are Eddington-limited. These high $\lambda_\mathrm{Edd}$ values are consistent with the findings in \citet{Mezcua18}, as largely expected because the involved \mbox{X-ray} surveys have similar depths and will preferentially select high-$\lambda_\mathrm{Edd}$ sources.\par

\begin{figure}
\centering
\resizebox{\hsize}{!}{\includegraphics{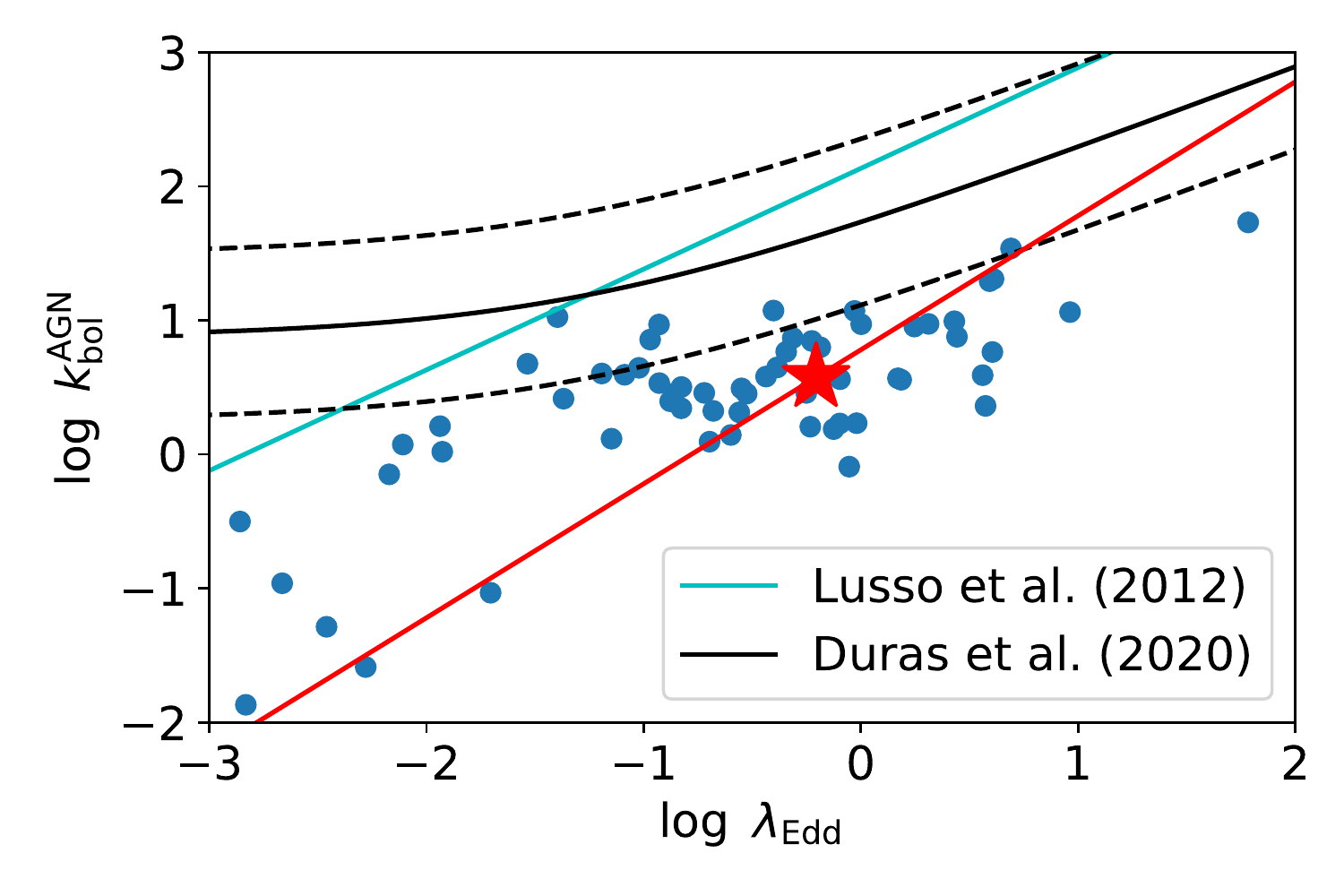}}
\caption{$\lambda_\mathrm{Edd}$ versus $k_\mathrm{bol}^\mathrm{AGN}$. Sources with $\lambda_\mathrm{Edd}\lesssim0.1$ and unphysically small $k_\mathrm{bol}^\mathrm{AGN}$ form a tail in the bottom-left corner and may have underestimated $L_\mathrm{bol}^\mathrm{AGN}$. The red star shows the median values among sources with $\lambda_\mathrm{Edd}>0.1$. The red line that has a slope of one and intersects with the red star is the trajectory when systematically shifting $L_\mathrm{bol}^\mathrm{AGN}$. The black lines are the $\lambda_\mathrm{Edd}-k_\mathrm{bol}^\mathrm{AGN}$ relation in \citet{Duras20} and the corresponding $2\sigma$ deviations, and the cyan line is the type~1 relation in \citet{Lusso12}. Our sample significantly deviates from the expected relations.}
\label{fig_ledd_kbol}
\end{figure}

To further check the reliability, we show our sources in the $\lambda_\mathrm{Edd}-k_\mathrm{bol}^\mathrm{AGN}$ plane in Figure~\ref{fig_ledd_kbol}. For SMBHs, $k_\mathrm{bol}^\mathrm{AGN}$ depends on $\lambda_\mathrm{Edd}$ because the corona emission becomes relatively weaker as $\lambda_\mathrm{Edd}$ increases. \citet{Lusso12} and \citet{Duras20} presented the calibrated correlation between the two quantities for massive SMBHs ($M_\mathrm{BH}\approx10^7-10^9~M_\odot$) with strong accretion ($\lambda_\mathrm{Edd}\approx0.01-1$), and these relations are also presented in Figure~\ref{fig_ledd_kbol}. Our sample significantly deviates from the expected relations, indicating that the overall estimations of the relevant parameters are problematic. We call this inconsistency the ``$\lambda_\mathrm{Edd}$ tension'' hereafter. Since $L_\mathrm{X}$ is largely robust, one or more of the following three factors must be wrong: $L_\mathrm{bol}^\mathrm{AGN}$ estimations, $M_\mathrm{BH}$ estimations, and the $\lambda_\mathrm{Edd}-k_\mathrm{bol}^\mathrm{AGN}$ relation. We will discuss these in the following text. As we will see, all three factors may have severe issues, and thus the ultimate goal of this section is not to accurately measure $\lambda_\mathrm{Edd}$ for our sample, but to point out the problems that should be solved before the measurement and subsequent scientific discussions.\par

\subsubsection{Challenges That May Lead to the $\lambda_\mathrm{Edd}$ Tension}
$L_\mathrm{bol}^\mathrm{AGN}$ is involved in both $\lambda_\mathrm{Edd}$ and $k_\mathrm{bol}^\mathrm{AGN}$, and changing $L_\mathrm{bol}^\mathrm{AGN}$ will cause the points in Figure~\ref{fig_ledd_kbol} to move along lines with slopes of unity in the $\log\lambda_\mathrm{Edd}-\log k_\mathrm{bol}^\mathrm{AGN}$ plane. The red star with an intersecting red line in Figure~\ref{fig_ledd_kbol} shows the median among our sources with $\lambda_\mathrm{Edd}\geq0.1$ and the corresponding trajectory when systematically varying $L_\mathrm{bol}^\mathrm{AGN}$. The lower left part of Figure~\ref{fig_ledd_kbol} is occupied by sources with $\lambda_\mathrm{Edd}\lesssim0.1$ and unphysically small $k_\mathrm{bol}^\mathrm{AGN}$ that roughly form a trend parallel to the red line, indicating that these sources may have underestimated $L_\mathrm{bol}^\mathrm{AGN}$ values. Figure~\ref{fig_ledd_kbol} indicates that the $\lambda_\mathrm{Edd}$ tension would be mitigated by systematically increasing $L_\mathrm{bol}^\mathrm{AGN}$. The first challenge of the $L_\mathrm{bol}^\mathrm{AGN}$ estimations is that, as pointed out in Section~3.2.4 of \citetalias{Zou22}, it is difficult to reliably constrain the strength of the AGN component unless the AGN emission dominates. \citetalias{Zou22} argued that this is a fundamental problem in SED fitting limited by the data and can hardly be solved merely with more sophisticated methods. The other problem arises from the fundamental fact that the accretion-disk temperature becomes higher as $M_\mathrm{BH}$ decreases ($\propto M_\mathrm{BH}^{-\frac{1}{4}}$). This causes a large amount of the disk emission to occur at extreme-UV (EUV) energies that are neither covered by our SED photometry nor the assumed disk SED model\footnote{\texttt{CIGALE} adopts a temperature-independent disk SED model constructed based on SMBH AGNs.} (e.g., \citealt{Cann18}).\par
\begin{figure}
\centering
\resizebox{\hsize}{!}{\includegraphics{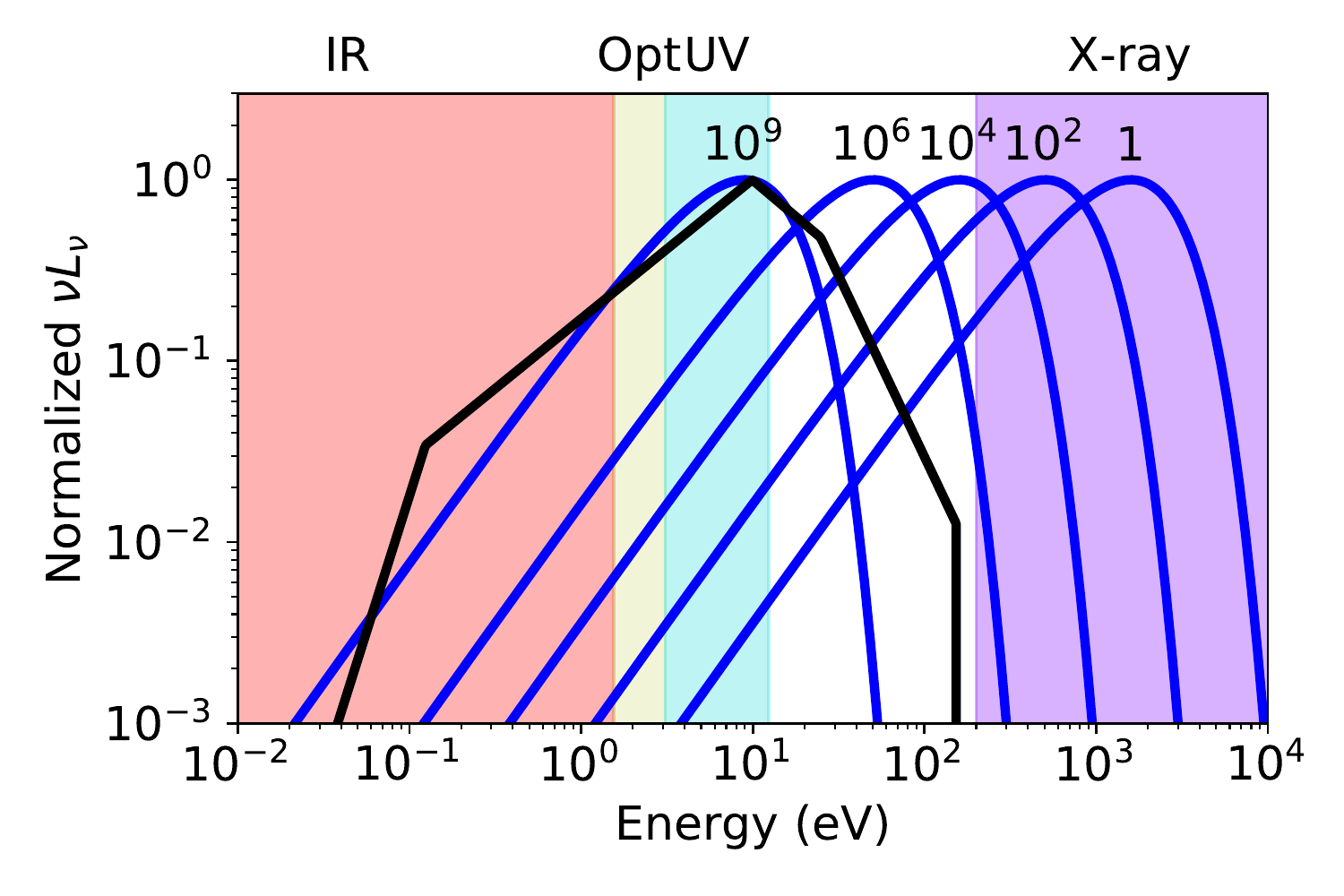}}
\caption{the blue curves show the predicted SS disk SEDs for different $M_\mathrm{BH}$, as labeled in units of $M_\odot$ at the SED peaks. The black curve is the adopted disk SED in \citetalias{Zou22}. The SEDs are normalized such that the peak value is unity. The background shaded regions show the typical band ranges: IR ($8000~\AA-1000~\mu\mathrm{m}$, $10^{-3}-1.5$~eV), optical ($4000-8000~\AA$, $1.5-3$~eV), UV (not including EUV; $1000-4000~\AA$, $3-12$~eV), and \mbox{X-ray} (only showing the typical XMM-Newton energy range; $0.2-12$~keV). The MBH disk SEDs may significantly deviate from the adopted \texttt{CIGALE} disk SED.}
\label{fig_sed_SSdisk}
\end{figure}

We use the Shakura-Sunyaev (SS) standard thin accretion disk \citep{Shakura73} to illustrate the second problem. Even if the SS disk is not exactly applicable, it illustrates the relevant trends; but also note that the standard accretion-disk theory faces significant challenges (e.g., \citealt{Lawrence18}). To calculate typical SS disk SEDs, we adopt a radiative efficiency of 10\%, $\lambda_\mathrm{Edd}=0.1$, and an inner disk radius of the innermost stable circular orbit radius for a non-spinning BH. We vary $M_\mathrm{BH}$ to calculate the corresponding SEDs. The results are shown in Figure~\ref{fig_sed_SSdisk}, including five representative $M_\mathrm{BH}$ values: $10^9~M_\odot$ is the typical SMBH mass for Sloan Digital Sky Survey (SDSS) quasars (e.g., \citealt{Shen11}), $10^6$ and $10^2~M_\odot$ are our MBH mass range, $10^4~M_\odot$ is the middle point of the MBH mass range and also roughly the lower boundary of our estimated $M_\mathrm{BH}$ distribution (Figure~\ref{fig_mbh_lbol}), and $1~M_\odot$ represents a typical stellar-mass BH. The figure shows that the $10^9~M_\odot$ SMBH disk SED largely shows similar patterns as for the \texttt{CIGALE} AGN disk model adopted in \citetalias{Zou22}. Their deviations are not surprising because our illustrative model may be over-simplified, and it is known that the actual SMBH AGN disk SED differs from the SS disk SED (e.g., \citealt{Davis11, Kubota18}). The disk SEDs of stellar-mass BHs are predicted to peak at \mbox{X-rays}, as known from XRBs. However, the MBH disk SEDs mainly peak in the EUV, where no direct data are available, though indirect probes such as high-excitation emission lines may help us understand the EUV emission (e.g., \citealt{Cann18, Timlin21}). Also note that the disk emission of our sources is expected to contribute little to the \mbox{X-rays}, especially above 2~keV, because our $M_\mathrm{BH}$ values ($\approx10^4-10^6~M_\odot$ with a median value of $2\times10^5~M_\odot$) are still much higher than those of stellar-mass BHs.\par

The MBH disk SEDs significantly deviate from the adopted AGN disk model in the SED fitting, which may cause strong biases. This problem is twofold. One is that the real MBH disk SED shape in the energy range covered by the data may be different, and the other is that the integrated $L_\mathrm{AGN}^\mathrm{bol}$ may be wrong. The first problem can cause the fitting to fail and consequently affect the derived host-galaxy properties. Although we currently still lack good observational constraints on intrinsic MBH SEDs, Figure~\ref{fig_sed_SSdisk} indicates that in the IR-to-optical range, both SMBH and MBH disk SEDs have similar shapes in their low-energy tails. We thus suspect that the shape difference of the disk SED should not cause major problems to the SED fitting quality and host-galaxy properties. The second problem, however, has potential for causing trouble. Figure~\ref{fig_sed_SSdisk} shows that the normalized MBH disk SEDs are around 1~dex below the SMBH disk SEDs in the optical-to-IR range. A large fraction of disk emission may not be accounted for in our SED fitting, and inferring the missing part would inevitably rely on the assumed disk model to conduct extrapolations. Overall, this fundamental issue may lead to large underestimations of $L_\mathrm{AGN}^\mathrm{bol}$. Nevertheless, this issue may be mitigated when considering the dust emission, especially for type~2 AGNs, whose observed AGN SEDs should be mainly from their dust emission instead of the disk. The \texttt{CIGALE} SED model may still be applicable if the dust temperature does not vary much with $M_\mathrm{BH}$, which is probably true because the dust temperature is bounded by dust sublimation.\par
We reiterate that our simple calculations are only for illustrative purposes and should not be taken as quantitative guidelines for correcting the bias. There are multiple factors not considered here. For example, more components besides the SS disk emission should be considered for more accurate SED calculations, and the model in \citet{Kubota18} predicts more UV emission than for the SS disk; the accretion disk may evolve to a slim disk when $\lambda_\mathrm{Edd}$ increases (e.g., \citealt{Abramowicz88}); it is unclear how the results would change after including the dust emission; the numerical underestimation factor should also depend on the adopted photometric bands; and even within our model, varying $\lambda_\mathrm{Edd}$ and inner disk radius, which depends on the BH spin, can also cause large shifts to the SEDs.\par
For the $M_\mathrm{BH}$ estimations, we used the empirical $M_\mathrm{BH}-M_\star$ scaling relationship. Even though this is largely calibrated from massive galaxies, it may also hold for dwarf galaxies (e.g., \citealt{Baldassare20a}), but possible biases may still exist (e.g., \citealt{Reines15}). Regardless of any possible systematic uncertainty of the relation itself, its large scatter (0.55~dex) is already notable. Our sample suffers from strong selection effects and thus may contain those objects with more massive MBHs so as to be detectable in \mbox{X-rays}. An example of a similar bias can be seen in Figure~18 of \citet{Burke22b}, where their $M_\mathrm{BH}$ values of variability-selected active dwarf galaxies are $2-3$~dex above the $M_\mathrm{BH}-M_\star$ scaling relation. Similarly, \citet{Mezcua23} also found a sample of distant active dwarf galaxies containing overmassive MBHs. It is difficult to quantify the underlying bias given the overall limited knowledge about the dwarf population, but it should be noted that even a $2\sigma$ underestimation of $M_\mathrm{BH}$ is more than 1~dex and can largely mitigate the $\lambda_\mathrm{Edd}$ tension. In this case, our sources may actually contain more massive SMBHs instead of MBHs. Also, given that both $L_\mathrm{bol}^\mathrm{AGN}$ and $M_\mathrm{BH}$ are likely to be underestimated, it is unclear if $\lambda_\mathrm{Edd}$ is underestimated or overestimated.\par
The $\lambda_\mathrm{Edd}-k_\mathrm{bol}^\mathrm{AGN}$ relation is also largely uncertain in at least two respects. First, this relation was calibrated mainly based on AGNs with $\lambda_\mathrm{Edd}\lesssim1$, and thus extrapolations toward the super-Eddington regime are inevitable in Figure~\ref{fig_ledd_kbol}. The relations in \citet{Lusso12} and \citet{Duras20} also deviate from each other in the super-Eddington regime in Figure~\ref{fig_ledd_kbol}. \citet{Liu21} found that candidate super-Eddington AGNs generally follow the same disk-corona correlation as for sub-Eddington AGNs, but their $\lambda_\mathrm{Edd}$ is still confined within 10. Given the aforementioned possible underestimations of $L_\mathrm{bol}^\mathrm{AGN}$, our sources may have larger $\lambda_\mathrm{Edd}$ beyond where there are good observational constraints, and thus the super-Eddington extrapolation is still uncertain. Second, the $\lambda_\mathrm{Edd}-k_\mathrm{bol}^\mathrm{AGN}$ relation may be different for MBHs. It has been shown that the \mbox{X-ray} emission becomes relatively stronger with decreasing $M_\mathrm{BH}$ for SMBHs (e.g., \citealt{Martocchia17, Duras20, Liu21}), and \citet{Desroches09} and \citet{Dong12} also found that type~1 active dwarf galaxies have higher ratios between their \mbox{X-ray} and UV emission compared to massive AGNs. Therefore, the real MBH $\lambda_\mathrm{Edd}-k_\mathrm{bol}^\mathrm{AGN}$ relation may be lower than that for massive AGNs.\par
Overall, this section highlights the difficulties in measuring $M_\mathrm{BH}$, $L_\mathrm{bol}^\mathrm{AGN}$, and consequently, $\lambda_\mathrm{Edd}$, for active dwarf galaxies. The $\lambda_\mathrm{Edd}$ tension may be caused by many factors, and future work is needed to solve the challenges discussed in this section.

\section{Summary}
\label{sec: summary}
In this work, we search for active dwarf galaxies in XMM-SERVS, and our main findings are summarized as follows.
\begin{enumerate}
\item{
After considering several contaminating factors, we found 73 active dwarf galaxies.\par
First, we highlight the importance of assessing the reliabilities of photo-$z$s and $M_\star$. Active dwarf galaxies present significant challenges for deriving reliable photo-$z$s, and the photo-$z$s of sources with high sSFR and/or $f_\mathrm{AGN}(0.36-4.5~\mu\mathrm{m}; \mathrm{obs})$ may be unreliable. This difficulty does not only exist in our sample, but is also likely present in previous literature. It should also be checked if the SEDs are dominated by AGNs. In such a case, the returned $M_\star$ from SED fitting may be unreliable, causing the misidentifications of massive galaxies as dwarfs. These problems become prevalent with increasing redshift for flux-limited samples, and all of our active dwarf galaxy candidates with $z>1.1$ are found to be unreliable.\par
We further estimate the expected \mbox{X-ray} flux if a given dwarf of interest does not contain an AGN, including the XRB and hot-gas emission that depend upon $M_\star$ and/or SFR. We found that given the XMM-Newton PSF aperture size, the flux from nearby sources usually dominates over the galaxy emission from the dwarf of interest itself. We require reliable active dwarf galaxies to have non-zero \mbox{X-ray} emission from their AGNs in a statistically significant manner. See Section~\ref{sec: data_sample}.
}
\item{
Based on our new implementation of HR measurements, we obtained the HRs of our sources as indicators of the \mbox{X-ray} spectral shapes. Our HRs indicate that nearly none of our sources is heavily obscured, which contradicts the predictions based on massive-AGN XLFs. This may be because our sources have lower $M_\mathrm{BH}$ than SMBHs and hence higher $\lambda_\mathrm{Edd}$, and the \mbox{X-ray} obscuration is fundamentally related to $\lambda_\mathrm{Edd}$ instead of $L_\mathrm{X}$. The massive-AGN XLFs thus overestimate the obscuration for active dwarf galaxies. See Section~\ref{sec: hr}.
}
\item{
We examined the radio properties of our sources. There are $\approx1-5$ radio-excess AGNs and 28 non-radio-excess sources. This small radio-excess incidence is consistent with previous works. See Section~\ref{sec: radio}.
}
\item{
We examined the host-galaxy environments of our sources. By measuring the projected separations on 100~kpc to $\gtrsim1$~Mpc scales between dwarf galaxies and massive galaxies, we found that our sources have similar values with inactive dwarf galaxies after matching $M_\star$ and $z$, indicating that dwarf AGN activity generally may not be triggered by environmental effects. See Section~\ref{sec: host}.
}
\item{
We explored the AGN accretion distribution of dwarf galaxies, as quantified by $p(L_\mathrm{X}\mid M_\star, z)$, after considering selection effects. A power-law formula with respect to $M_\star$, $z$, and $L_\mathrm{X}$ can provide good fits. The fitting results indicate that $p(L_\mathrm{X}\mid M_\star, z)$ decreases with $L_\mathrm{X}$ and increases with $M_\star$, but redshift evolution cannot be statistically confirmed. $p(L_\mathrm{X}\mid M_\star, z)$ can further be converted to the active fraction. Depending upon the exact AGN definition and $M_\star$ and $z$ ranges, the active fraction is generally $10^{-5}-10^{-2}$. As for $p(L_\mathrm{X}\mid M_\star, z)$, the $L_\mathrm{X}$-based active fraction increases with $M_\star$, but the $\lambda_\mathrm{sBHAR}$-based active fraction shows much less significant $M_\star$ evolution. See Section~\ref{sec: agnfrac}.
}
\item{
We estimated $L_\mathrm{bol}^\mathrm{AGN}$ from the SED fitting, $M_\mathrm{BH}$ from the $M_\mathrm{BH}-M_\star$ scaling relation, and consequently, $\lambda_\mathrm{Edd}$. However, our sources significantly deviate from the expected relation in the $\lambda_\mathrm{Edd}-k_\mathrm{bol}^\mathrm{AGN}$ plane. We discuss possible underlying uncertainties of $L_\mathrm{bol}^\mathrm{AGN}$, $M_\mathrm{BH}$, and the $\lambda_\mathrm{Edd}-k_\mathrm{bol}^\mathrm{AGN}$ relation.\par
Especially, even $L_\mathrm{bol}^\mathrm{AGN}$, which is often considered as robust, may suffer from strong biases inherent in the SED fitting. Our analyses raise the following questions that should be addressed in the future. How should one observationally constrain MBH SEDs? How large is the underlying bias when using SMBH AGN SED models to fit active dwarf galaxies? Do we need temperature-dependent AGN SED models, which currently have not been enabled in common SED fitting tools? See Section~\ref{sec: ledd}.
}
\end{enumerate}
Overall, our sample resides in a similar parameter space as the COSMOS sample in \citet{Mezcua18} because of similar multiwavelength (especially \mbox{X-ray}) depths. Several of our sample's characteristics echo those of the COSMOS sample, including the limited fractions of high-HR sources and radio-excess AGNs, the smaller $\xi_\mathrm{AGN}^L$ compared to massive galaxies, and the high $\lambda_\mathrm{Edd}$, as summarized above. Our results on, e.g., photo-$z$s and selection effects also highlight some likely limitations of the COSMOS sample. Our sample is also $\approx2-3$ times larger.\par
Our 73 \mbox{X-ray}-selected active dwarf galaxies can serve as a useful sample beyond the local universe. It is important to further enrich the sample in the same fields. Especially, LSST will enable variability selection, and the variability time scale is also an indicator of $M_\mathrm{BH}$. The unreliable photo-$z$ problem also causes strong selection biases and limits the sample size, and it may be necessary to obtain deep spectroscopic observations to measure spec-$z$s instead. Fortunately, several spectroscopic surveys will be available in our fields in the coming few years, as listed in Section~5 of \citetalias{Zou22}. Additionally, the \mbox{X-ray} sensitivity is the primary limitation for selecting more sources. Future missions such as STAR-X, AXIS, and Athena will push \mbox{X-ray} surveys on $\mathrm{deg}^2$ scales deeper and thus can provide much larger active dwarf galaxy samples.

\acknowledgments
We thank the anonymous referee for constructive suggestions and comments. We thank Joel Leja and Guang Yang for helpful discussions. F.Z., W.N.B., S.Z., and W.Y. acknowledge support from NASA grant 80NSSC19K0961, NSF grant AST-2106990, and the V.M. Willaman Endowment. D.M.A. acknowledges the Science Technology and Facilities Council (STFC) for support through grant code ST/T000244/1. F.E.B. acknowledges support from ANID-Chile BASAL CATA ACE210002 and FB210003, FONDECYT Regular 1200495 and 1190818, and Millennium Science Initiative Program -- ICN12\_009. C.-T.J.C. acknowledges support from Chandra X-ray Center grants AR0-21013A and AR0-21013B. B.L. acknowledges financial support from the National Natural Science Foundation of China grant 11991053. Y.X. acknowledges support from NSFC grants (12025303 and 11890693).

\appendix
\section{Estimation of Hardness Ratio}
\label{sec: myhr}
Our method is based on the framework of \citet{Park06} but with significant revisions and different algorithmic implementations. The main reasons for the revisions are to accelerate the computations and to implement the HR estimations in \texttt{Python} for the sake of convenience for \texttt{Python} users. For large \mbox{X-ray} surveys with thousands of sources, such as XMM-SERVS, a faster implementation than for \citet{Park06} becomes notably helpful. As we will show in the following, the computation is accelerated by avoiding numerical integrations; instead, all the integrations can be solved analytically with some special functions involved. Furthermore, we found that numerical overflows sometimes happen in \citet{Park06}, but our method overcomes this problem.\par
Throughout this appendix, SB and HB are regarded as any two nonoverlapping energy bands. We denote $S$ and $H$ as the total SB and HB counts in the source region, respectively, and they follow Poisson distributions:
\begin{align}
S\sim\mathrm{Poi}\left(e_S(\lambda_S+\xi_S)\right), H\sim\mathrm{Poi}\left(e_H(\lambda_H+\xi_H)\right),
\end{align}
where $e_\bullet$, $\lambda_\bullet$, and $\xi_\bullet$ are the exposure time, the true source count rate, and the expected background count rate in the source region, respectively. $e_\bullet$ is known from the observation, $\lambda_\bullet$ is unknown, and $\xi_\bullet$ is estimated from a background region. We assume that $\xi_\bullet$ is accurately known. This assumption is not adopted in \citet{Park06} but is appropriate in most cases because the total counts in the background region are usually much larger than for the source region, given that the background area is generally much larger than the source-region area. Besides, when this assumption fails, it is usually because the source is bright and many source counts are received. Such a case is even simpler -- it is sufficiently accurate to adopt Gaussian errors and apply classic error propagation, and there is no need for Bayesian estimations. Another reason is that, for \mbox{X-ray} surveys that construct background maps to measure $\xi_\bullet$, there is not an accurate definition of background regions. Although we adopt this assumption in this work, it is not always necessary in general cases, and we will also present the method without this assumption in the last part of this appendix.\par
Following \citet{Park06}, the priors on $\lambda_S$ and $\lambda_H$ are assumed to be
\begin{align}
\lambda_S\sim\gamma(\psi_{S1}, \psi_{S2}), \lambda_H\sim\gamma(\psi_{H1}, \psi_{H2}),\label{eq_hrprior}
\end{align}
where $\psi_\bullet$ are given in advance. The probability distribution function of $\gamma(\psi_1, \psi_2)$ is $\pi(x)=\psi_2^{\psi_1}x^{\psi_1-1}e^{-\psi_2x}/\Gamma(\psi_1)$. This prior family includes several common noninformative priors: $(\psi_1=1, \psi_2=0)$ is a flat prior in the linear scale, $(\psi_1=0.5, \psi_2=0)$ is a Jeffrey prior, and $(\psi_1\to0^+, \psi_2=0)$ is a flat prior in the logarithm scale.\par
The SB likelihood is
\begin{align}
p(S\mid\lambda_S)=\frac{\left[e_S(\lambda_S+\xi_S)\right]^Se^{-e_S(\lambda_S+\xi_S)}}{S!}=\frac{e_S^Se^{-e_S\xi_S}}{S!}\sum_{n=0}^{S}\binom{S}{n}\lambda_S^n\xi_S^{S-n}e^{-e_S\lambda_S}.
\end{align}
Therefore,
\begin{align}
p(\lambda_S\mid S)\propto\pi(\lambda_S)p(S\mid\lambda_S)\propto\sum_{n=0}^S\binom{S}{n}\lambda_S^{n+\psi_{S1}-1}\xi_S^{S-n}e^{-(e_S+\psi_{S2})\lambda_S}.
\end{align}
Similarly, for the HB,
\begin{align}
p(\lambda_H\mid H)\propto\sum_{m=0}^H\binom{H}{m}\lambda_H^{m+\psi_{H1}-1}\xi_H^{H-m}e^{-(e_H+\psi_{H2})\lambda_H}.
\end{align}
Under the assumption that the SB and the HB are independent, $p(\lambda_S, \lambda_H\mid S, H)=p(\lambda_S\mid S)p(\lambda_H\mid H)$. Denoting $w=\lambda_S+\lambda_H$, we can write $\lambda_S=(1-\mathrm{HR})w/2$ and $\lambda_H=(1+\mathrm{HR})w/2$. Thus, we have
\begin{align}
p(\mathrm{HR}, w\mid S, H)=p(\lambda_S, \lambda_H\mid S, H)\left|\frac{\partial(\lambda_S, \lambda_H)}{\partial(\mathrm{HR}, w)}\right|=\frac{1}{2}wp\left(\lambda_S=\frac{(1-\mathrm{HR})w}{2}\middle|S\right)p\left(\lambda_H=\frac{(1+\mathrm{HR})w}{2}\middle|H\right).
\end{align}
The HR posterior probability distribution function is
\begin{align}
p(\mathrm{HR}\mid S, H)=&\int_0^{+\infty}p(\mathrm{HR}, w\mid S, H)dw\\
\propto&\sum_{n=0}^S\sum_{m=0}^H\binom{S}{n}\binom{H}{m}\xi_S^{S-n}\xi_H^{H-m}2^{-(n+m+\psi_{S1}+\psi_{H1})}(1-\mathrm{HR})^{n+\psi_{S1}-1}(1+\mathrm{HR})^{m+\psi_{H1}-1}\times\\
&\int_0^{+\infty}w^{n+m+\psi_{S1}+\psi_{H1}-1}e^{-\frac{1}{2}\left[(e_S+\psi_{S2})(1-\mathrm{HR})+(e_H+\psi_{H2})(1+\mathrm{HR})\right]w}dw.
\end{align}
Using the fact that $\int_0^{+\infty}w^ae^{-bw}dw=\Gamma(1+a)/b^{1+a}$, the above equation can be expressed as follows.
\begin{align}
p(\mathrm{HR}\mid S, H)\propto\sum_{n=0}^S\sum_{m=0}^H\binom{S}{n}\binom{H}{m}\xi_S^{S-n}\xi_H^{H-m}\Gamma(n+m+\psi_{S1}+\psi_{H1})\frac{(1-\mathrm{HR})^{n+\psi_{S1}-1}(1+\mathrm{HR})^{m+\psi_{H1}-1}}{\left[(e_S+\psi_{S2})(1-\mathrm{HR})+(e_H+\psi_{H2})(1+\mathrm{HR})\right]^{n+m+\psi_{S1}+\psi_{H1}}}.
\end{align}
The HR posterior CDF is
\begin{align}
F(\mathrm{HR}\mid S, H)=&\int_{-1}^\mathrm{HR}p(\mathrm{HR}=h\mid S, H)dh\\
\propto&\sum_{n=0}^S\sum_{m=0}^H\binom{S}{n}\binom{H}{m}\xi_S^{S-n}\xi_H^{H-m}\Gamma(n+m+\psi_{S1}+\psi_{H1})\int_{-1}^\mathrm{HR}\frac{(1-h)^{n+\psi_{S1}-1}(1+h)^{m+\psi_{H1}-1}}{\left[(e_S+\psi_{S2})(1-h)+(e_H+\psi_{H2})(1+h)\right]^{n+m+\psi_{S1}+\psi_{H1}}}dh.\label{eq: cdf_G}
\end{align}
Denoting the integration as $G$ and substituting $x=(1+h)/2$, we have
\begin{align}
G=\frac{1}{2}\frac{1}{(e_S+\psi_{S2})^{n+m+\psi_{S1}+\psi_{H1}}}\int_0^\frac{1+\mathrm{HR}}{2}\frac{(1-x)^{n+\psi_{S1}-1}x^{m+\psi_{H1}-1}}{\left[1-\left(1-\frac{e_H+\psi_{H2}}{e_S+\psi_{S2}}\right)x\right]^{n+m+\psi_{S1}+\psi_{H1}}}dx.
\end{align}
The integration is of the form of $\int_0^b\frac{(1-x)^{z_1-1}x^{z_2-1}}{(1-ax)^{z_1+z_2}}dx$. We substitute $t=(1-a)x/(1-ax)$ and obtain
\begin{align}
\int_0^b\frac{(1-x)^{z_1-1}x^{z_2-1}}{(1-ax)^{z_1+z_2}}dx=\frac{1}{(1-a)^{z_2}}\int_0^\frac{(1-a)b}{1-ab}(1-t)^{z_1-1}t^{z_2-1}dt=\frac{1}{(1-a)^{z_2}}\frac{\Gamma(z_1)\Gamma(z_2)}{\Gamma(z_1+z_2)}I_\beta\left(z_2, z_1; \frac{(1-a)b}{1-ab}\right),
\end{align}
where $I_\beta$ is the incomplete Beta function. Therefore,
\begin{align}
G=\frac{\Gamma(n+\psi_{S1})\Gamma(m+\psi_{H1})}{2(e_S+\psi_{S2})^{n+\psi_{S1}}(e_H+\psi_{H2})^{m+\psi_{H1}}\Gamma(n+m+\psi_{S1}+\psi_{H1})}I_\beta\left(m+\psi_{H1}, n+\psi_{S1};\frac{(e_H+\psi_{H2})(1+\mathrm{HR})}{(e_S+\psi_{S2})(1-\mathrm{HR})+(e_H+\psi_{H2})(1+\mathrm{HR})}\right).
\end{align}
Substituting $G$ into Equation~\ref{eq: cdf_G}, we obtain
\begin{align}
F(\mathrm{HR}\mid S, H)\propto&\sum_{n=0}^S\sum_{m=0}^H\binom{S}{n}\binom{H}{m}\xi_S^{S-n}\xi_H^{H-m}\Gamma(n+m+\psi_{S1}+\psi_{H1})G\propto\sum_{n=0}^S\sum_{m=0}^H\Theta_{nm}(S, H, \xi_S, \xi_H)I_\mathrm{nm}\left(y(\mathrm{HR}, e_S, e_H)\right),\\
\Theta_{nm}(S, H, \xi_S, \xi_H)=&\frac{\Gamma(n+\psi_{S1})\Gamma(m+\psi_{H1})\left[(e_S+\psi_{S2})\xi_S\right]^{-n}\left[(e_H+\psi_{H2})\xi_H\right]^{-m}}{\Gamma(n+1)\Gamma(m+1)\Gamma(S-n+1)\Gamma(H-m+1)},\label{eq: theta_nm}\\
I_{nm}(y)=&I_\beta\left(m+\psi_{H1}, n+\psi_{S1};y\right),\label{eq: Inm}\\
y(\mathrm{HR}, e_S, e_H)=&\frac{(e_H+\psi_{H2})(1+\mathrm{HR})}{(e_S+\psi_{S2})(1-\mathrm{HR})+(e_H+\psi_{H2})(1+\mathrm{HR})}.
\end{align}
The above equations return unnormalized $F(\mathrm{HR}\mid S, H)$. The appropriate normalization such that $F(1\mid S, H)=1$ is calculated by substituting $\mathrm{HR}=1$, and we obtain the normalized HR CDF as follow.
\begin{align}
F(\mathrm{HR}\mid S, H)=\frac{\sum_{n=0}^S\sum_{m=0}^H\Theta_{nm}I_\mathrm{nm}}{\sum_{n=0}^S\sum_{m=0}^H\Theta_{nm}}.\label{eq: hrcdf_normed}
\end{align}
Therefore, the HR CDF can be calculated without numerically conducting integrations. The CDF can then be easily converted to point estimators and confidence intervals. Furthermore, we notice that $\Theta_{nm}$ does not depend on HR and thus only needs to be computed once for each source. $I_{nm}$ does not depend on $(S, H, \xi_S, \xi_H)$, and its dependence on $(e_S, e_H, \mathrm{HR})$ is all absorbed into the third parameter $y$ of $I_\beta$ in Equation~\ref{eq: Inm}; given a set of $(\psi_{S1}, \psi_{H1})$, the possible values of the first and second parameters of $I_\beta$ are countable. One can thus compute $I_{nm}$ on a grid of $y$ and $(n, m)$ from 0 up to reasonably large values (e.g., 1000) and save the results externally; then $I_{nm}$ does not need to be computed further when applying it to real sources. Note that $I_{nm}$ can be computed quickly if only once using \texttt{scipy.special.betainc}. Therefore, one only needs to compute $\Theta_{nm}$ once per source and do the summations to obtain the full HR CDF on the chosen $y$ grid, which can be converted to a HR grid. This implementation method allows fast and accurate HR estimations for a large number of sources. Practically, $\Theta_{nm}$ should be computed using a natural logarithm to avoid overflows (e.g., using \texttt{scipy.special.gammaln}).\par
In more general cases where $\xi_\bullet$ is not assumed to be accurately known, we need to estimate $\xi_\bullet$ from source-free background regions. We denote $B_S$ and $B_H$ as the SB and HB counts in the background region, respectively, and $r_\bullet$ as the ratio between the background-region area and the source-region area, where the term ``area'' here may include both the angular area and the sensitivity. That is, $B_S\sim\mathrm{Poi}(r_Se_S\xi_S)$ and $B_H\sim\mathrm{Poi}(r_He_H\xi_H)$. Similar to Equation~\ref{eq_hrprior}, the priors on $\xi_S$ and $\xi_H$ are assumed to be
\begin{align}
\xi_S\sim\gamma(\psi_{S3}, \psi_{S4}), \xi_H\sim\gamma(\psi_{H3}, \psi_{H4}).
\end{align}
Following the same procedures as above, except for an additional step to integrate out $\xi_\bullet$, we can prove that Equation~\ref{eq: hrcdf_normed} still holds, and the only difference is on $\Theta_{nm}$. Equation~\ref{eq: theta_nm} should be modified as follows.
\begin{align}
\Theta_{nm}=&\frac{\Gamma(S+B_S+\psi_{S3}-n)\Gamma(H+B_H+\psi_{H3}-m)\Gamma(n+\psi_{S1})\Gamma(m+\psi_{H1})}{\Gamma(n+1)\Gamma(m+1)\Gamma(S-n+1)\Gamma(H-m+1)}\left[\frac{(1+r_S)e_S+\psi_{S4}}{e_S+\psi_{S2}}\right]^n\left[\frac{(1+r_H)e_H+\psi_{H4}}{e_H+\psi_{H2}}\right]^m.
\end{align}\par
We release \texttt{Python} codes implementing this algorithm at \url{https://github.com/fanzou99/FastHR}.

\section{Detailed Analyses of XMM02399}
\label{sec: xmm02399}
Section~\ref{sec: hr} suggests that XMM02399 (R. A. =  02:21:56.53, Decl. = $-$04:07:58.0, and SDSS spec-$z$ = 0.615) is the only source in our sample with clear evidence of $N_\mathrm{H}\gtrsim10^{23}~\mathrm{cm^{-2}}$. As far as we know, there has not been any Compton-thick (CT) active dwarf galaxy beyond the local universe confirmed yet. We thus analyze this candidate in detail in this appendix.\par

\subsection{\mbox{X-ray} Analyses}
\label{sec: xray_xmm02399}
XMM02399 has been observed by XMM-Newton six times and by Chandra once. We list them as follows: XMM-Newton ObsIDs = 0037982201 (16.4~ks), 0404960601 (11.9~ks), 0785101501 (22.0~ks), 0785102001 (22.0~ks), 0793581001 (9.0~ks), 0793581301 (9.0~ks); Chandra ObsID = 20538 (9.9~ks). We reduce these observations and extract the corresponding spectra using the XMM-Newton Science Analysis System (SAS; v20.0) and the Chandra Interactive Analysis of Observations (CIAO; v4.12). We group these spectra to one count per bin. To help visualize the spectral fitting, we further merge all the XMM-Newton spectra into a single spectrum using the SAS task \texttt{epicspeccombine}. This merging procedure may produce slight biases, and thus we derive our numerical results by jointly fitting all the individual spectra instead of from the single merged spectrum.\par
\begin{figure*}
\centering
\resizebox{\hsize}{!}{\includegraphics{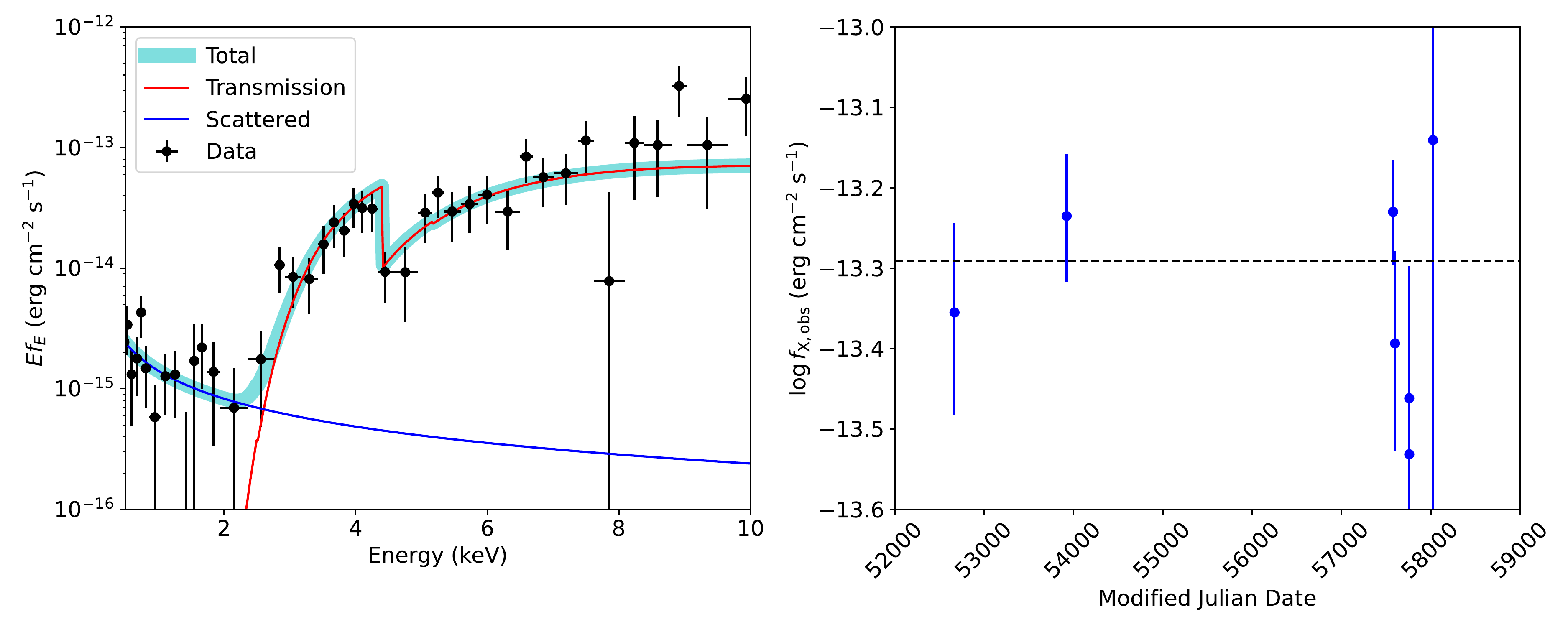}}
\caption{Left: the unfolded merged XMM-Newton spectrum (the points with error bars) fitted with the \textit{AbsPow} model (the thick cyan curve). The model includes transmission and scattered components, as labeled by the legend. The spectrum is rebinned for visualization only, and this merged spectrum is presented only for illustration, while our scientific analyses are based on jointly fitting all the individual spectra. Right: $f_\mathrm{X,obs}$ versus time. Each data point represents one \mbox{X-ray} observation. The horizontal black dashed line marks our $f_\mathrm{X,obs}$ based on the original non-variable \textit{Pow} model. No variability is visible for XMM02399.}
\label{fig_xray_xmm02399}
\end{figure*}

We use \texttt{sherpa} to fit the \mbox{X-ray} spectra, in which the W statistic is adopted. We limit our spectral fitting range to $0.5-10$~keV for XMM-Newton and $0.5-7$~keV for Chandra and use the same model and parameters to fit all the spectra simultaneously. Using a simple power-law absorbed by the Galactic absorption, to which we refer as the \textit{Pow} model, we obtain a hard effective photon index of $-0.04_{-0.16}^{+0.15}$. To measure $N_\mathrm{H}$, we phenomenologically adopt an absorbed power-law model. We also found the existence of an apparent scattered component below 2~keV, and hence we use another power-law to account for this. We find no evidence indicating iron K lines and thus do not add them to the model. The resulting model, which will be called \textit{AbsPow}, is \texttt{phabs $\times$ (zphabs $\times$ cabs $\times$ zpowerlw + constant $\times$ zpowerlw)}, where \texttt{phabs} models the Galactic absorption, the $N_\mathrm{H}$ values of \texttt{zphabs} and \texttt{cabs} are linked to account for the intrinsic obscuration, the two \texttt{zpowerlw} components are set to be the same, and \texttt{constant} describes the scattered fraction ($f_\mathrm{sc}$). We obtain a nearly CT-level $N_\mathrm{H}=9.6_{-1.6}^{+1.8}\times10^{23}~\mathrm{cm^{-2}}$ and a soft $\Gamma=2.77_{-0.37}^{+0.39}$ with a good fit (W-stat/d.o.f. = 428/420). Physically, there should be reprocessed emission, and thus we try the following model: \texttt{phabs $\times$ (borus + zphabs $\times$ cabs $\times$ cutoffpl + constant $\times$ cutoffpl)}, where \texttt{borus} is the reprocessed torus emission model in \citet{Balokociv18}. We name this model \textit{Torus} hereafter. The $N_\mathrm{H}$ values of \texttt{borus}, \texttt{zphabs}, and \texttt{cabs} are all set the same, and the \texttt{cutoffpl} parameters are linked to those of \texttt{borus}. The best-fit statistic is W-stat/d.o.f. = 432/418, and the fitting returns $N_\mathrm{H}=9.5_{-1.6}^{+1.8}\times10^{23}~\mathrm{cm^{-2}}$ and $\Gamma=2.60_{-0.27}^{+u}$, where $+u$ means that the best-fit $\Gamma$ reaches the upper limit of the $\Gamma$ domain, $1.4-2.6$, allowed by the \texttt{borus} component. These results are similar to those from the \textit{AbsPow} model. In fact, we found the reflection component in the best-fit \textit{Torus} model is negligible, and thus the \textit{Torus} model is effectively similar to the \textit{AbsPow} model. Given the fact that the \textit{AbsPow} model allows $\Gamma$ to reach beyond 2.6 and provides a smaller best-fit statistic, we will adopt this model as our final one. We also refer to the observed flux ($f_\mathrm{X,obs}$) as the \textit{Pow} result. We summarize our fitting results in Table~\ref{tbl_fitres}. To visualize the spectrum, we show the \textit{Torus} model fitting results in Figure~\ref{fig_xray_xmm02399}. XMM02399 has a very high $\Gamma$ value, which is rare and may indicate a high $\lambda_\mathrm{Edd}$ (see also Section~\ref{sec: ledd} and Appendix~\ref{sec: discuss_xmm02399}) because $\Gamma$ is positively correlated with $\lambda_\mathrm{Edd}$ (e.g., \citealt{Shemmer08, Brightman13}). Sub-Eddington AGNs generally have $\Gamma\lesssim2.2$, and even the $\Gamma$ of reported super-Eddington AGN candidates ($1\lesssim\lambda_\mathrm{Edd}\lesssim10$) rarely exceeds 2.5 (e.g., \citealt{Huang20, Liu21}). Nevertheless, our $\Gamma$ uncertainty is large, and thus the high $\Gamma$ value is still statistically consistent with those of general AGNs.\par

\begin{table*}
\caption{\mbox{X-ray} spectral fitting results of XMM02399}
\label{tbl_fitres}
\centering
\begin{threeparttable}
\begin{tabular}{ccccccc}
\hline
\hline
Model & W-stat/d.o.f. & $\Gamma$ & $N_\mathrm{H}$ & $f_\mathrm{sc}$ & $\log f_\mathrm{X}$ & $\log L_\mathrm{X}$\\
& & & ($\mathrm{cm^{-2}}$) & & ($\mathrm{erg~cm^{-2}~s^{-1}}$) & ($\mathrm{erg~s^{-1}}$)\\
\hline
\textit{Pow} & 484/422 & $-0.04_{-0.16}^{+0.15}$ & --- & --- & $-13.29_{-0.05}^{+0.05}$ & $43.50_{-0.03}^{+0.03}$\\
\textit{AbsPow} & 428/420 & $2.77_{-0.37}^{+0.39}$ & $9.6_{-1.6}^{+1.8}\times10^{23}$ & $1.4_{-0.9}^{+2.2}\times10^{-3}$ & $-12.22_{-0.20}^{+0.24}$ & $45.12_{-0.23}^{+0.34}$\\
\textit{Torus} & 432/418 & $2.60_{-0.27}^{+u}$ & $9.5_{-1.6}^{+1.8}\times10^{23}$ & $2.2_{-0.7}^{+2.6}\times10^{-3}$ & $-12.32_{-0.32}^{+0.19}$ & $45.01_{-0.33}^{+0.19}$\\
\hline
\hline
\end{tabular}
\begin{tablenotes}
\item
\textit{Notes.} $f_\mathrm{X}$ is the \mbox{X-ray} flux in the observed-frame $2-10$~keV band, and $L_\mathrm{X}$ is the \mbox{X-ray} luminosity in the rest-frame $2-10$~keV band. For the \textit{Pow} model, the reported $f_\mathrm{X}$ and $L_\mathrm{X}$ are observed ones without absorption corrections, while the $f_\mathrm{X}$ and $L_\mathrm{X}$ of the \textit{AbsPow} and \textit{Torus} models are intrinsic ones. The results from the \textit{AbsPow} model are adopted as the fiducial ones throughout this appendix.
\end{tablenotes}
\end{threeparttable}
\end{table*}

The \mbox{X-ray} observations span a time range of 14~yr, which enables us to check if XMM02399 presents \mbox{X-ray} variability. We first check if $f_\mathrm{X,obs}$ is variable. We use the \textit{Pow} model to simultaneously fit all the observations, linking their $\Gamma$ while allowing their normalizations to vary freely. Using a universal $\Gamma$ reduces the number of parameters and helps improve the variability significance (if any). The resulting $f_\mathrm{X,obs}$ as a function of time is plotted in Figure~\ref{fig_xray_xmm02399}, and no significant variability is visible. More rigorously, we compare the Bayesian information criteria (BICs) between the original \textit{Pow} model (i.e., the normalizations are linked) and the variable \textit{Pow} model (i.e., the normalizations are freed), and the latter has a much larger BIC than the former (differ by 30), which strongly disfavors the variable model because more free parameters are introduced while nearly not improving the fitting. Similarly, when using the \textit{AbsPow} model, we always detect no variability, no matter if we free or link the $\Gamma$ and $N_\mathrm{H}$ parameters. We thus conclude that XMM02399 is not a variable source in \mbox{X-rays}.

\subsection{Optical Spectrum}
\begin{figure*}
\centering
\resizebox{\hsize}{!}{\includegraphics{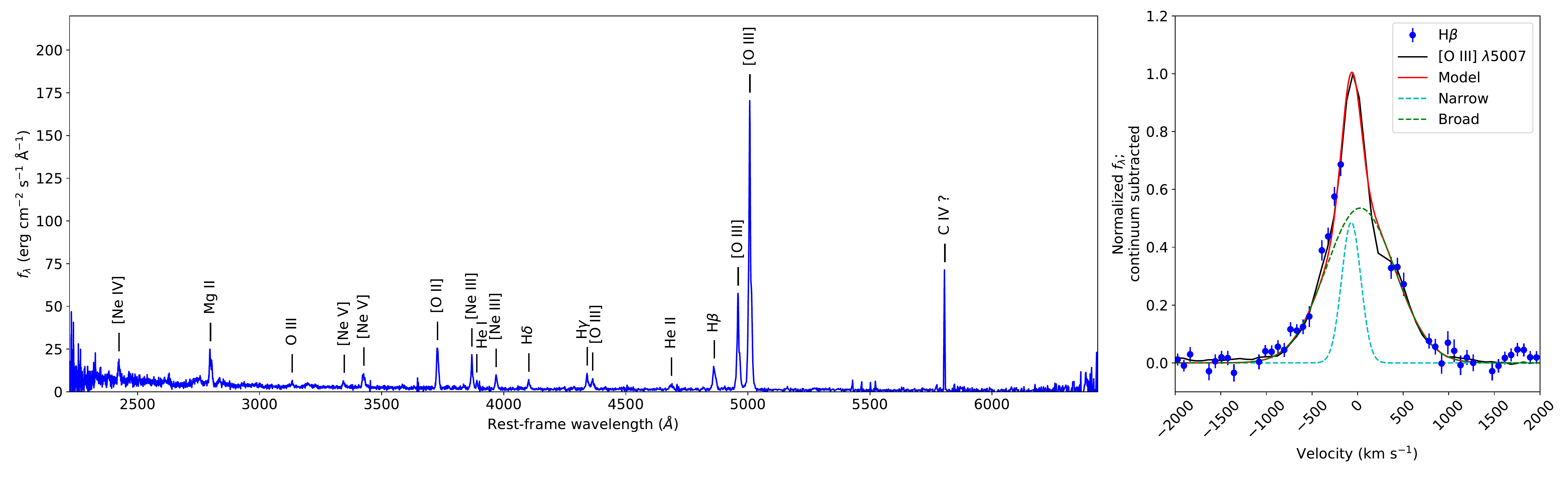}}
\caption{Left: SDSS rest-frame spectrum of XMM02399 after correcting for Galactic extinction. Apparent emission lines are marked explicitly. The \iona{C}{iv}~$\lambda5808$ line is uncertain due to large uncertainties in the corresponding pixels. Right: Normalized H$\beta$ and [\iona{O}{iii}]~$\lambda5007$ profiles in the velocity space. Negative velocity indicates blueshift. H$\beta$ is shown as points with error bars, while [\iona{O}{iii}]~$\lambda5007$ is shown as the black solid line. The best-fit model for [\iona{O}{iii}]~$\lambda5007$ is shown as the red solid line, whose narrow and broad components are plotted as the cyan and green dashed lines, respectively. H$\beta$ is consistent with the [\iona{O}{iii}]~$\lambda5007$ profile.}
\label{fig_opt_xmm02399}
\end{figure*}

XMM02399 was spectroscopically observed by the Sloan Digital Sky Survey (SDSS). We show its spectrum in the left panel of Figure~\ref{fig_opt_xmm02399} after correcting the Galactic extinction based on the $E(B-V)$ map in \citet{Schlegel98} and the extinction law in \citet{Cardelli89}. We explicitly mark emission lines identified through visual inspection and matching wavelengths with those in \citet{Vanden_Berk01}. There is a visually strong \iona{C}{iv}~$\lambda5808$ line, but the corresponding pixel errors are large, and thus its existence is uncertain. The velocity shifts of these emission lines (e.g., \iona{Mg}{ii}, [\iona{O}{ii}], and [\iona{O}{iii}]) relative to each other and the SDSS pipeline spec-$z$ are generally a few tens of $\mathrm{km~s^{-1}}$, and such small shifts are within intrinsic uncertainties of their systemic redshifts (e.g., \citealt{Shen16}).\par

Several high-ionization lines are present, including the [\iona{Ne}{v}] and \iona{He}{ii}~$\lambda4686$ lines, which require ionizing photon energies above 97.1 and 54.4~eV, respectively. They cannot be explained by normal stellar populations. To examine the general excitation mechanism, we rely on the emission-line diagnostics in \citet{Lamareille10} and \citet{Juneau14}, where the former uses [\iona{O}{iii}]~$\lambda5007$/H$\beta$ versus [\iona{O}{ii}]~$\lambda3727$/H$\beta$, and the latter uses [\iona{O}{iii}]~$\lambda5007$/H$\beta$ versus $M_\star$. To model these emission lines, we fit the continuum locally with a power-law. [\iona{O}{ii}]~$\lambda3727$ is then fitted using a narrow Gaussian line. [\iona{O}{iii}]~$\lambda5007$ and H$\beta$ are fitted together with [\iona{O}{iii}]~$\lambda4959$, and each of them is fitted with two Gaussian lines to represent a narrow component and a broad component (see the next paragraph). We then measure their narrow-component equivalent widths to calculate the line ratios and obtain [\iona{O}{iii}]~$\lambda5007$/H$\beta$ = 8.3 and [\iona{O}{ii}]~$\lambda3727$/H$\beta$ = 3.0. These place our source into the AGN locus under both criteria in \citet{Lamareille10} and \citet{Juneau14}. Therefore, the primary excitation source of these lines is the AGN.\par
We further check its \iona{Mg}{ii} and H$\beta$ lines. \iona{Mg}{ii} can be explained by two narrow components accounting for its doublet nature. For H$\beta$, we show its normalized profile together with [\iona{O}{iii}]~$\lambda5007$ in the right panel of Figure~\ref{fig_opt_xmm02399}. The [\iona{O}{iii}]~$\lambda5007$ profile has a narrow core and a broad component, and the broad component is expected to be from outflows and has a FWHM of $891~\mathrm{km~s^{-1}}$. The H$\beta$ data points are in perfect consistency with the [\iona{O}{iii}]~$\lambda5007$ profile. Several H$\beta$ data points are missing around its peak because they fail the SDSS \texttt{and\_mask}, but they are also consistent with the [\iona{O}{iii}]~$\lambda5007$ profile if added. Similarly, H$\beta$ is also consistent with the [\iona{O}{iii}]~$\lambda4959$ profile. These similarities in line profiles are generally not expected if the broad H$\beta$ component is from the broad-line region (e.g., \citealt{Zou20}), and we hence suspect that H$\beta$ is also broadened by outflows. Therefore, XMM02399 is actually a narrow-line AGN in terms of its optical spectrum.\par

\subsection{SED and $M_\star$}
\label{sec: sed_xmm02399}
\begin{figure*}
\centering
\resizebox{\hsize}{!}{\includegraphics{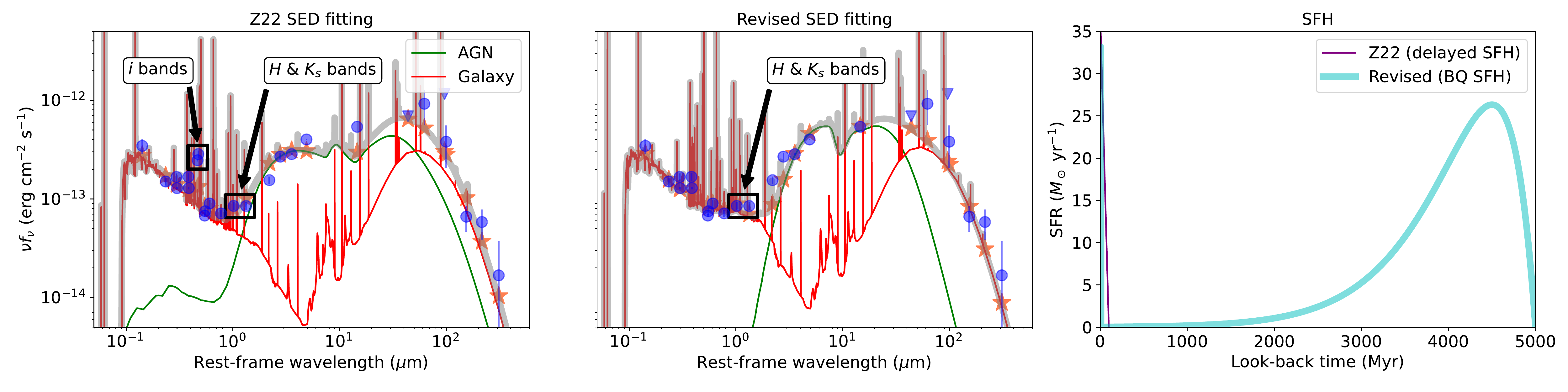}}
\caption{The best-fit SEDs of XMM02399 in \citetalias{Zou22} (left) and this appendix (Middle) and the corresponding SFHs (Right). These SEDs are plotted in the same manner as in Figure~\ref{fig_example_sed}, and their abscissa axes are truncated in the UV to help focus on their differences in the IR. The $i$ bands, which are contaminated by strong [\iona{O}{iii}] lines, are marked explicitly in the left panel and excluded in our revised SED fitting. The $H$ and $K_s$ bands are also marked to highlight the different explanations of their origins in the two different SED models -- the galaxy component dominates in the $H$ and $K_s$ bands in the middle panel but not in the left panel.}
\label{fig_sedsfh_xmm02399}
\end{figure*}

We revise the bulk SED fitting in \citetalias{Zou22} for XMM02399. There are three main reasons that the SED fitting can be improved. 
First, \texttt{CIGALE} requires the input \mbox{X-ray} photometry to be de-absorbed. The empirical \mbox{X-ray} absorption 
correction in \citetalias{Zou22} is not appropriate for Compton-thick AGNs, and we will instead update the \mbox{X-ray} photometry using our results in Appendix~\ref{sec: xray_xmm02399}. Second, the strong [\iona{O}{iii}]~$\lambda4959$ and [\iona{O}{iii}]~$\lambda5007$ lines, as can be clearly seen in Figures~\ref{fig_opt_xmm02399} and \ref{fig_sedsfh_xmm02399}, contaminate the $i$-band photometry. \texttt{CIGALE} is currently incapable of fitting AGN-produced emission lines (but it can fit emission lines from H~II regions; e.g., \citealt{Villa-Velez21}), and thus we exclude the $i$-band photometry in the SED. Third, the reported sSFR of our source in \citetalias{Zou22} is $10^{-7.67}~\mathrm{yr^{-1}}$, almost reaching the maximum sSFR allowed by the normal-galaxy SFH in \citetalias{Zou22}. This indicates that the SFH of XMM02399 may need to be revised to a bursting one, and we will use the BQ SFH in \citetalias{Zou22}. The first two problems actually do not have substantial impacts on $M_\star$, but the third problem regarding its SFH is notable. We apply all three changes and show the best-fit SED and SFH, together with the original ones in \citetalias{Zou22}, in Figure~\ref{fig_sedsfh_xmm02399}. Their best-fit SFHs are significantly different -- the \citetalias{Zou22} SFH only includes a young stellar population, while our revised SFH includes a strong old stellar population and a very young bursting population. Consequently, the revised SED fitting returns a much larger $M_\star$ -- the \citetalias{Zou22} (best-fit and likelihood-weighted) $M_\star=2\times10^9~M_\odot$, while the revised best-fit $M_\star=4\times10^{10}~M_\odot$, and the revised likelihood-weighted $M_\star=8\times10^9~M_\odot$.\par

The SEDs reveal that two main underlying uncertainties undermine $M_\star$ measurements. First, both best-fit SEDs in Figure~\ref{fig_sedsfh_xmm02399} indicate that the galaxy component dominates below rest-frame $\sim0.8~\mu\mathrm{m}$, and the AGN component dominates above rest-frame $\sim2.2~\mu\mathrm{m}$, but the relative AGN contribution to the part between rest-frame $\sim0.8-2.2~\mu\mathrm{m}$ is unclear. The VIDEO $H$ and $K_s$ bands reside in this wavelength range, and SED models with either significant or negligible AGN contributions to these two bands can return reasonably good results. However, this wavelength range is where the old stellar emission peaks, and thus the AGN contamination prevents us from constraining the old stellar emission. Second, XMM02399 is a strong starburst galaxy, in which old stars may also be outshined by recently-formed young stars, and thus the SED-fitted $M_\star$ strongly depends upon how many old stars are allowed in the adopted SFH. When using galaxy-only models, the delayed SFH and BQ SFH return very different $M_\star$ by up to 1~dex. In fact, the second problem is aggravated by the first one -- it is necessary to obtain the amount of galaxy emission at $\sim1~\mu\mathrm{m}$ to constrain the old stellar emission, but this becomes infeasible given our data because of plausible AGN contamination.\par
Therefore, the real $M_\star$ is upper-bounded by the extreme case that the rest-frame $\sim0.8-2.2~\mu\mathrm{m}$ is dominated by the galaxy component, and many old stars exist; similarly, $M_\star$ is lower-bounded in the conversely extreme case. The revised best-fit SED (middle panel in Figure~\ref{fig_sedsfh_xmm02399}) happens to be the first extreme case, and the \citetalias{Zou22} result corresponds to the second. Therefore, the real $M_\star$ should be roughly between $2\times10^9$ and $4\times10^{10}~M_\odot$ (the statistical uncertainties are much smaller than these systematic uncertainties). Any value between this range is possible, which explains why the revised likelihood-weighted $M_\star$ is much smaller than the revised best-fit $M_\star$. Therefore, XMM02399 is still a low-mass galaxy candidate, but we cannot determine its actual $M_\star$. To further narrow the plausible $M_\star$ range, it is essential to directly constrain its galaxy and AGN contribution at rest-frame $\sim0.8-2.2~\mu\mathrm{m}$ (i.e., observed-frame $\sim1.3-3.5~\mu\mathrm{m}$). Hubble or JWST IR imaging should be able to spatially resolve the galaxy and AGN components, as done for SDSS quasars (e.g., \citealt{Li21}).\par
We note that the problem discussed in this appendix is not prevalent in our whole sample. XMM02399 happens to be the most extreme case in several respects (e.g., high $N_\mathrm{H}$ and sSFR). The last paragraph of Section~\ref{sec: mstar_reliability} indicates that XMM02399 is the only source with $M_\star^\mathrm{bqgal}>10^{10}~M_\odot$, and our other sources generally have much more reliable $M_\star$.

\subsection{Further Discussion}
\label{sec: discuss_xmm02399}
Regardless of the $M_\star$ uncertainty, XMM02399 is an interesting object. Its several properties are the same as the high-$\lambda_\mathrm{Edd}$ dust-obscured galaxy (DOG) population in \citet{Zou20}: heavy obscuration in \mbox{X-rays}, moderate [\iona{O}{iii}] outflow (FWHM $\lesssim1000~\mathrm{km~s^{-1}}$), and starburst host galaxy. This may indicate that XMM02399 is physically at the same evolutionary stage with high-$\lambda_\mathrm{Edd}$ DOGs, i.e., at the peak of both BH and galaxy growth in a merger-driven event (see Section~5.1 of \citealt{Zou20} for more discussions).\par
If XMM02399 is indeed similar to high-$\lambda_\mathrm{Edd}$ DOGs, it should lie around the boundary between the allowed and forbidden regions in the $\lambda_\mathrm{Edd}-N_\mathrm{H}$ plane (e.g., \citealt{Ishibashi18}; Figure~7 of \citealt{Zou20}), and this boundary returns $\lambda_\mathrm{Edd}\approx1$ at our $N_\mathrm{H}$ in Appendix~\ref{sec: xray_xmm02399}. Independently, its soft $\Gamma$ also indicates that its $\lambda_\mathrm{Edd}$ may be high, and we obtain $\lambda_\mathrm{Edd}=11$ using the super-Eddington $\Gamma-\lambda_\mathrm{Edd}$ relation in \citet{Huang20}. As another approach, following Section~\ref{sec: ledd}, we use the $M_\star$ range in Appendix~\ref{sec: sed_xmm02399} to estimate the $\lambda_\mathrm{Edd}$ range and obtain $\lambda_\mathrm{Edd}$ to be between 3 and 71. These $\lambda_\mathrm{Edd}$ estimations are crude, but all lead to the same conclusion that XMM02399 is an Eddington-limited AGN with $\lambda_\mathrm{Edd}$ between unity and a few tens.

\bibliography{citations}

\end{document}